\documentclass[authoryear,3p,twocolumn,nofootinbib]{emulateapj-rtx4}
\usepackage[english]{babel}
\usepackage{graphicx}% Include figure files
\usepackage{dcolumn}% Align table columns on decimal point
\usepackage{bm}% bold math
\usepackage{longtable}
\usepackage{amsmath}
\usepackage{amssymb}
\usepackage{latexsym}
\usepackage{mathrsfs}
\usepackage{amsfonts}
\usepackage[usenames]{color}
\usepackage{syntonly}
\usepackage{natbib}

\usepackage{epstopdf}
\usepackage{rotating}
\usepackage{latexsym}
\usepackage{verbatim}
\usepackage{hyperref}
\usepackage{fancyheadings}
\usepackage{fancyhdr}

\newcommand{\be} {\begin{equation}}
\newcommand{\ee} {\end{equation}}

\newcommand{\bea} {\begin{eqnarray}}
\newcommand{\eea} {\end{eqnarray}}
\newcommand{\bdm} {\begin{displaymath}}
\newcommand{\edm} {\end{displaymath}}
\newcommand{\ba} {\begin{array}}
\newcommand{\ea} {\end{array}}

\newcommand{\bfg}  {\begin{figure}}
\newcommand{\efg}  {\end{figure}}
\newcommand{\bfgd}  {\begin{figure*}}
\newcommand{\efgd}  {\end{figure*}}
\newcommand{\incgr} {\includegraphics}
\newcommand{\mbf} {\mathbf}

\newcommand{\btb} {\begin{table}}
\newcommand{\etb} {\end{table}}
\newcommand{\ben} {\begin{enumerate}}
\newcommand{\een} {\end{enumerate}}

\newcommand{\bit} {\begin{itemize}}
\newcommand{\eit} {\end{itemize}}

\newcommand{\ns}{N_{\rm side}}
\newcommand{\commander}{{\tt Commander}}

\newcommand{\Ainu}{A^\nu_i}

\def\aap  {A\&A}

\def\apjl {ApJ}
\def\apjs {ApJ}

\newcommand{\BA}{\mathbf{A}}

\newcommand{\Bc}{\mathbf{c}}
\newcommand{\Bs}{\mathbf{s}}
\newcommand{\BS}{\mathbf{S}}
\newcommand{\Bn}{\mathbf{n}}
\newcommand{\BN}{\mathbf{N}}

\newcommand{\Bf}{\mathbf{f}}

\newcommand{\Bd}{\mathbf{d}}

\newcommand{\Bt}{\mathbf{t}}

\newcommand{\muK}{\mu\textrm{K}}
\def\L2{\ifmmode L_2\else $L_2$\fi}

\def\DeltaT{\ifmmode \Delta T\else $\Delta T$\fi}
\def\deltat{\ifmmode \Delta t\else $\Delta t$\fi}
\def\fknee{\ifmmode f_{\rm knee}\else $f_{\rm knee}$\fi}
\def\Fmax{\ifmmode F_{\rm max}\else $F_{\rm max}$\fi}
\def\solar{\ifmmode{\rm M}_{\mathord\odot}\else${\rm M}_{\mathord\odot}$\fi}
\def\Msolar{\ifmmode{\rm M}_{\mathord\odot}\else${\rm M}_{\mathord\odot}$\fi}
\def\Lsolar{\ifmmode{\rm L}_{\mathord\odot}\else${\rm L}_{\mathord\odot}$\fi}

\def\inv{\ifmmode^{-1}\else$^{-1}$\fi}
\def\mo{\ifmmode^{-1}\else$^{-1}$\fi}
\def\sup#1{\ifmmode ^{\rm #1}\else $^{\rm #1}$\fi}
\def\expo#1{\ifmmode \times 10^{#1}\else $\times 10^{#1}$\fi}
\def\,{\thinspace}
\def\lsim{\mathrel{\raise .4ex\hbox{\rlap{$<$}\lower 1.2ex\hbox{$\sim$}}}}
\def\gsim{\mathrel{\raise .4ex\hbox{\rlap{$>$}\lower 1.2ex\hbox{$\sim$}}}}

\def\simprop{\mathrel{\raise .4ex\hbox{\rlap{$\propto$}\lower 1.2ex\hbox{$\sim$}}}}
\def\deg{\ifmmode^\circ\else$^\circ$\fi}
\def\pdeg{\ifmmode $\setbox0=\hbox{$^{\circ}$}\rlap{\hskip.11\wd0 .}$^{\circ}
          \else \setbox0=\hbox{$^{\circ}$}\rlap{\hskip.11\wd0 .}$^{\circ}$\fi}
\def\arcs{\ifmmode {^{\scriptstyle\prime\prime}}
          \else $^{\scriptstyle\prime\prime}$\fi}
\def\arcm{\ifmmode {^{\scriptstyle\prime}}
          \else $^{\scriptstyle\prime}$\fi}
\newdimen\sa  \newdimen\sb
\def\parcs{\sa=.07em \sb=.03em
     \ifmmode \hbox{\rlap{.}}^{\scriptstyle\prime\kern -\sb\prime}\hbox{\kern -\sa}
     \else \rlap{.}$^{\scriptstyle\prime\kern -\sb\prime}$\kern -\sa\fi}
\def\parcm{\sa=.08em \sb=.03em
     \ifmmode \hbox{\rlap{.}\kern\sa}^{\scriptstyle\prime}\hbox{\kern-\sb}
     \else \rlap{.}\kern\sa$^{\scriptstyle\prime}$\kern-\sb\fi}
\def\ra[#1 #2 #3.#4]{#1\sup{h}#2\sup{m}#3\sup{s}\llap.#4}
\def\dec[#1 #2 #3.#4]{#1\deg#2\arcm#3\arcs\llap.#4}
\def\deco[#1 #2 #3]{#1\deg#2\arcm#3\arcs}
\def\rra[#1 #2]{#1\sup{h}#2\sup{m}}
\def\dots{\relax\ifmmode \ldots\else $\ldots$\fi}
\def\WHzsr{\ifmmode $W\,Hz\mo\,sr\mo$\else W\,Hz\mo\,sr\mo\fi}
\def\mHz{\ifmmode $\,mHz$\else \,mHz\fi}
\def\GHz{\ifmmode $\,GHz$\else \,GHz\fi}
\def\mKs{\ifmmode $\,mK\,s$^{1/2}\else \,mK\,s$^{1/2}$\fi}
\def\muKs{\ifmmode \,\mu$K\,s$^{1/2}\else \,$\mu$K\,s$^{1/2}$\fi}
\def\muKRJs{\ifmmode \,\mu$K$_{\rm RJ}$\,s$^{1/2}\else \,$\mu$K$_{\rm RJ}$\,s$^{1/2}$\fi}
\def\muKHz{\ifmmode \,\mu$K\,Hz$^{-1/2}\else \,$\mu$K\,Hz$^{-1/2}$\fi}
\def\MJysr{\ifmmode \,$MJy\,sr\mo$\else \,MJy\,sr\mo\fi}
\def\MJysrmK{\ifmmode \,$MJy\,sr\mo$\,mK$_{\rm CMB}\mo\else \,MJy\,sr\mo\,mK$_{\rm CMB}\mo$\fi}
\def\microns{\ifmmode \,\mu$m$\else \,$\mu$m\fi}

\def\muK{\ifmmode \,\mu$K$\else \,$\mu$\hbox{K}\fi}
\def\microK{\ifmmode \,\mu$K$\else \,$\mu$\hbox{K}\fi}
\def\muW{\ifmmode \,\mu$W$\else \,$\mu$\hbox{W}\fi}
\def\kms{\ifmmode $\,km\,s$^{-1}\else \,km\,s$^{-1}$\fi}
\def\kmsMpc{\ifmmode $\,\kms\,Mpc\mo$\else \,\kms\,Mpc\mo\fi}

\begin{document}

\bibliographystyle{apj}

\title{Analysis of WMAP 7-year Temperature Data: Astrophysics of the Galactic Haze}
\author{Davide Pietrobon\altaffilmark{1}}
\email{davide.pietrobon@jpl.nasa.gov}
\altaffiltext{1}{Jet Propulsion Laboratory, California Institute of Technology, 4800 Oak Grove Drive, Pasadena, CA 91109-8099, U.S.A.}

\author{Krzysztof M.~  G\'orski\altaffilmark{1,2}}
\altaffiltext{2}{
%Jet Propulsion Laboratory, California Institute of Technology\\
%4800 Oak Grove Dr.~ 91109 Pasadena CA\\
Warsaw University Observatory, Aleje Ujazdowskie 4, 00478 Warszawa, Poland}

\author{James Bartlett\altaffilmark{1,3}}
\altaffiltext{3}{Laboratoire AstroParticule \& Cosmologie (APC), Universit\`e Paris Diderot, CNRS/IN2P3, CEA/lrfu, Observatoire de Paris, Sorbonne Paris Cit\'e, 10, rue Alice Domon et L\'eonie Duquet, 75205 Paris Cedex 13, France}

\author{A.~ J.~ Banday\altaffilmark{4,5}}
\altaffiltext{4}{Universit\`e de Toulouse; UPS-OMP; IRAP; Toulouse, France}
\altaffiltext{5}{CNRS; IRAP; 9 Av.~Colonel Roche, BP 44346, F-31028 Toulouse Cedex 4, France}

\author{Gregory Dobler\altaffilmark{6}}
\altaffiltext{6}{Kavli Institute for Theoretical Physics, University of California, Santa Barbara Kohn Hall, Santa Barbara, CA 93106 USA}

\author{Loris P.~L.~ Colombo\altaffilmark{7,1}}
\altaffiltext{7}{USC Dana and David Dornsife College of Letters, Arts and Sciences, University of Southern California, University Park Campus, Los Angeles, CA 90089}

\author{Sergi R.~ Hildebrandt\altaffilmark{8}}
\altaffiltext{8}{Division of Physics, Mathematics and Astronomy, California Institute of Technology, 1200 East California Blvd.~, 91125 Pasadena, CA}

\author{Jeffrey B.~ Jewell\altaffilmark{1}}

\author{Luca Pagano\altaffilmark{1}}
%\altaffiltext{8}{Physics Department and INFN, Universit\'a di Roma ÒLa SapienzaÓ, Ple Aldo Moro 2, 00185, Rome, Italy}

\author{Graca Rocha\altaffilmark{1}}

\author{Hans~ Kristian~ Eriksen\altaffilmark{9,10}}
\altaffiltext{9}{Institute of Theoretical Astrophysics, University of Oslo,
P.O. Box 1029 Blindern, N-0315 Oslo, Norway}
\altaffiltext{10}{4 Centre of Mathematics for Applications, University of Oslo,
P.O. Box 1053 Blindern, N-0316 Oslo, Norway}

\author{Rajib Saha\altaffilmark{11}}
\altaffiltext{11}{Physics Department, Indian Institute of Science Education and Research Bhopal, Bhopal, M.P, 462023, India}

\author{Charles R.~ Lawrence\altaffilmark{1}}

\begin{abstract}
We perform a joint analysis of the cosmic microwave background (CMB) and Galactic emission
from the WMAP 7-year temperature data.  Using the Commander code, based on Gibbs sampling, 
we simultaneously derive the CMB and Galactic components on scales larger than 1\deg\ with
improved sensitivity over previous work.  We conduct a detailed study of the low-frequency 
Galactic foreground, focussing on the ``microwave haze" emission around the Galactic center.
We demonstrate improved  performance in quantifying the diffuse Galactic emission when including Haslam 
408MHz data and when jointly modeling  the spinning and thermal 
dust emission. We examine whether the hypothetical Galactic haze can be explained 
by a spatial variation of the synchrotron spectral index, and find that the excess of emission around the Galactic 
center is stable with respect to variations of the foreground model. Our results demonstrate that 
the new Galactic foreground component - the microwave haze - is indeed present.
\end{abstract}


\maketitle

\section{Introduction}
\label{sec:intro}

While data from the \emph{Wilkinson Microwave Anisotropy Probe} (WMAP) \citep[see][and references therein]{Jarosik:2010,Komatsu:2010wmap7} has enabled unprecedented advances in the understanding of cosmology over the past decade, it has also opened a unique window into the fundamental physical processes of the interstellar medium (ISM).  The choice of observing bands for WMAP insured that multiple emission mechanisms would be observed across the frequency coverage.  In particular, there are at least three distinct physical processes at low frequencies (23-33-41 GHz),: free-free, synchrotron, and anomalous microwave emission (AME), which falls with frequency and is highly correlated with 100 $\mu$m thermal dust emission. At high frequencies (61-94 GHz), Galactic emission is completely dominated by thermal dust emission \citep{gold11}.

Free-free emission (or thermal bremsstrahlung) is generated by scattering of ionized electrons off the proton nuclei in hot ($\sim5000$ K) gas, and has a brightness temperature which scales as $T \propto \nu^{-2.15}$ through the WMAP bands, where $\nu$ represents frequency. The bulk of the synchrotron emission observed by WMAP 
%is caused to cosmic-ray electrons/positrons, accelerated by supernova (SN) shocks, that spiral in the Galactic magnetic field.  For the 1st-order Fermi acceleration in shocks, the resultant spectrum should be $T\propto\nu^{-2.5}$ at the injection site, softening to $T\propto\nu^{-3}$ after diffusing through the ISM. Remarkably, this spectrum is almost exactly what 
is seen to closely follow a power-law $T\propto\nu^{-3}$ \citep{kogut07,dobler11b}. Lastly, spinning dust (whose presence has been observed in small, dusty clouds \citep{casassus08,planck11XX}, as well as hotter diffuse regions \citep{dobler08b,dobler09}), is the likely cause of the anomalous microwave emission. Small dust grains with non-zero dipole moments are spun up by a variety of mechanisms such as ion collisions, plasma density fluctuations, photon fields, etc.,\ and produce spinning dipole radiation \citep[see][for the original theoretical realization of this idea]{Erickson1957,draine1998}.

Using simple template regression techniques (see Section~\ref{sec:wanalysis}), \cite{Bennett:2003ca} and \cite{Haze_Finkbeiner2004ApJ...614..186F} showed that the Galactic emissions are highly spatially correlated with maps at other frequencies. Free-free emission is morphologically correlated with H$\alpha$ recombination line emission, synchrotron with low frequency radio emission (e.g., at 408 MHz), and spinning (and thermal) dust with total dust column density \citep[e.g., \cite{schlegel98} evaluated using models for the thermal emission to 94 GHz by][]{Finkbeiner1999ApJ524}.  After removing emission correlated with these templates, \cite{Haze_Finkbeiner2004ApJ...614..186F} found that there was an excess signal centered on the Galactic center (GC) and extending out roughly $\sim30$ degrees. A more detailed study of this Galactic ``haze'' by \citep{Haze_Dobler2008ApJ...680.1222D} \citep[and more recently with the 7-year data by][]{dobler11b} showed that its spectrum ($T\propto\nu^{\beta}$ with $\beta\sim-2.5$) was too soft to be free-free emission and too hard to be synchrotron emission associated with acceleration by super novae (SN) shocks (after taking into account cosmic-ray diffusion). The origin of this residual component
%e haze electrons
remains a mystery and has been the object of intense theoretical scrutiny \citep[e.g.,][]{Haze_Hooper2007PhRvD..76h3012H,Haze_Cholis2009PhRvD..80l3518C,biermann10,dobler11,crocker11,guo11}.

However, the subject of the haze has not been without controversy, most notably due to the claim by the WMAP team \citep[as well as others, e.g.,][]{Dickinson2009ApJ705} that the haze is not detected in their analyses. In particular, \cite{gold11} claimed lack of evidence for the haze based on WMAP polarization data, though it is unclear that a polarization signal would be detectable \citep[see][]{dobler11b}. A comprehensive study of component separation (and CMB cleaning) was performed by \cite{Eriksen2006ApJ641,Eriksen2007ApJ656,Dickinson2009ApJ705} utilizing Bayesian inference of foreground amplitudes and spectra via Gibbs sampling. These studies have never claimed a significant excess of emission towards the GC. Nevertheless, with the release of gamma-ray data from the \emph{Fermi Gamma-Ray Space Telescope} in 2009, a corresponding feature at high energies was discovered \citep{Haze_Dobler2010ApJ...717..825D}, likely generated by the haze electron inverse Compton (IC) scattering starlight, infrared, and CMB photons up to \emph{Fermi} energies.

%With some exceptions \citep[see for example][]{Haze_Bottino2010MNRAS.402..207B}, regression techniques suffer one additional complication, which is related to the CMB subtraction, normally performed by fitting some linear combination of WMAP channels, approach which seems circular. In this work we overcome this specific issue, solving jointly CMB map
With few exceptions \citep[see for example][]{Haze_Bottino2010MNRAS.402..207B},
foreground studies using template fitting pre-subtract a CMB estimate
which imprints a bias in the foreground spectra \citep[see][]{Haze_Dobler2008ApJ...680.1222D} since no CMB estimate is completely clean of foregrounds.  In this work, we attempt to minimize this effect by solving jointly for the CMB map and angular power spectrum and Galactic emission parameters, within a Bayesian framework. We fit foreground parameters in every pixel, thus taking into account spatial variation of foreground spectra, and we test the stability of the solution against several Galactic emission models. Seeking greater independence from correlation with external templates, we instead treat other data sets as input channel maps and process them through the Gibbs sampler: to this end we add Haslam at 408 MHz \citep{Haslam1982A&AS47} to the WMAP data set.

In \S\ref{sec:data} we describe the data and our foreground model, while in \S\ref{sec:wanalysis} we describe our its application in our \commander\ analysis.  Our approach enables us to improve upon previous studies.  In \S\ref{sec:discussion} we discuss our results and refine the methodology by performing a joint analysis of the WMAP 7-year data set and Haslam at 408 MHz. Lastly in \S\ref{sec:conclusions} we summarize and draw our conclusions.

\section{Datasets and Foreground Model}
\label{sec:data}
Our approach uses the Gibbs sampling algorithm  introduced by \citet{Jewell2004ApJ609} and \citet{Wandelt2004PhRvD70}, and further developed by \cite{Eriksen2004ApJS155, ODwyer2004ApJ617L, Eriksen2007ApJ656, Chu2005PhRvD71, Jewell2009ApJ697, Rudjord2009ApJ692, Larson2007ApJ656}. The method has been numerically implemented in a computer program called  \commander, and successfully applied to previous releases of the WMAP data as reported in \citet{Eriksen2008ApJ672L}, \citet{Eriksen2007ApJ665L}, and \citet{Dickinson2009ApJ705}. \commander\ is a maximum likelihood method, which generates samples from the joint posterior density for the CMB map and angular power spectrum, as well as foreground components for the chosen sky model. A detailed description of the algorithm and its validation on simulated data is provided by \citet[][and references therein]{Eriksen2007ApJ656}. For the interested reader, we summarize the technical aspects of the sampling procedure in Appendix~\ref{app:commander}.

The main advantage of the approach is full characterization of the (posterior) probability distribution of the parameter space spanned by the adopted sky model. Moreover, we can evaluate the goodness-of-fit of each sample, i.e., of a given set of model parameters.  Any parametric model can be encoded, leaving the freedom to vary parameters pixel-by-pixel as well as  to fit templates.

In the following we describe the data used in our analysis and their processing.

\subsection{WMAP 7-year dataset}
\label{subsec:wmap}

Our aim is to perform a comprehensive study of the WMAP 7-year temperature data \citep{Jarosik:2010}, focusing not only on estimating the CMB signal and corresponding power spectrum, but also on characterizing the properties of the foreground emission, with particular attention to evidence for the haze signal.

The WMAP data set comprises data from ten Differential Assemblies (DAs) covering five frequencies from 23 to 94~GHz. The WMAP team provided maps centered at frequencies of 23, 30, 40, 60 and 90~GHz [K,Ka,Q,V,W bands] resulting from the coadition of the DAs.  We smooth the frequency maps to an effective 60 arcminute resolution by deconvolving the maps with the appropriately coadded transfer functions provided by the WMAP team and convolving with a Gaussian beam of 1 degree FWHM.  We then downgraded the sky maps to a working resolution of $\ns=128$ in the HEALPix\footnote{http://healpix.jpl.nasa.gov/} scheme \citep{Gorski2005Healpix}. The choice of angular resolution, and subsequently of the pixelization, was dictated by both the angular resolution of the available foreground templates and by the need to resolve the first acoustic peak of the CMB power spectrum. This represents a novel element of our analysis  compared to previous work \citep{Eriksen2008ApJ672L,Dickinson2009ApJ705}.

In a further break with earlier analyses, we do not add uniform white noise to regularize the maps \citep{Eriksen2007ApJ665L}, but instead add a noise component proportional to the actual noise
variance in the smoothed, processed maps. To do so, we compute the RMS noise per pixel of the maps at the lower resolution via a Monte Carlo approach. We generate 1,000 non-uniform white noise realizations for each channel following the prescription given by the WMAP
team.
%The values of the effective standard deviation quoted by the WMAP team for each channel are reported in Table \ref{tab:w7sigma}. 
In practice, we draw random Gaussian noise maps for each frequency band with zero mean and a variance given by $\sigma_\nu^2/N_{\rm obs}$ at the native WMAP resolution $N_{\rm side}=512$, where $N_{\rm obs}$ represents the number of observations in a given pixel. We then smooth and downgrade the noise maps as above and finally recompute the variance per pixel averaged over 1,000 simulations. Noise is then added to each frequency based on the computed noise variance, and chosen so that the signal-to-noise ratio of the masked smoothed map at $\ell\simeq2N_{\rm side}$ is of order unity. 
%The final RMS maps from which the noise realizations are drawn are shown in Figure~\ref{fig:wmap_maps}.
The scaling factors applied are [16,12,12,6,6] for the  [K,Ka,Q,V,W]  channels respectively. One of the advantages of this approach is that it preserves the noise structure imposed by the scanning strategy and describes at least the diagonal part of the instrumental noise, which has been correlated by the smoothing procedure.

%\begin{table}[h]
%\begin{center}
%\caption{Effective noise standard deviation for the WMAP 7-year channels.\label{tab:w7sigma}}
%\begin{tabular}{ccc}
%\tableline\tableline
% {\bf Band} & {\bf $\nu$ GHz} & {\bf $\sigma_\nu$ [$\mu$K CMB]} \\
%%\multicolumn{1}{c}{$P$\tablenotemark{a}} & $P R_{maj}$ & $P R_{min}$ &
%%\multicolumn{1}{c}{$\Theta$\tablenotemark{b}} \\
%\tableline
% K & 23GHz &1437 \\
% Ka & 33GHz & 1470 \\
% Q & 41GHz & 2197 \\
% V & 61GHz & 3137 \\
% W & 94GHz & 6549\\
%\tableline
%\end{tabular}
%\end{center}
%\end{table}

\subsection{Foregrounds}

The diffuse Galactic emission consists of three contributions
from well understood foregrounds --
synchrotron radiation from cosmic ray electrons losing energy in the
Galactic magnetic field, free-free emission in the diffuse ionised
medium, and radiation from dust grains heated by the interstellar
radiation field. In addition, there is a strong contribution at
frequencies in the range 20 -- 70~GHz referred to as 
anomalous microwave emission (AME)
that is strongly correlated with the thermal dust emission and which has
been explained by rotational excitation of small
grains - the so-called spinning dust emission \citep[see for example][for recent studies on the subject and references therein]{Hoang2010ApJ,Hoang2011ApJ}. In the frequency range
spanned by WMAP, the synchrotron emission is reasonably well described by
a power law brightness temperature emissivity with spectral index
$\beta\sim-3$.  The free-free
emission also follows an approximate power law, well described
by a spectral index $\alpha\sim-2.15$. The thermal dust component is
usually modeled as a gray body with emissivity of the form $T(\nu)\propto\nu^\epsilon
B(\nu,T_{\rm d})$ with typical values for the parameters given by 
$\epsilon\simeq1.6-2.0$ and $T_{\rm d}\approx18$K. Since the
highest frequency WMAP band W has a nominal central frequency of $94$~GHz, the
overall thermal dust contribution is small and can be approximated by 
a simpler power law model with a spectral index
$\epsilon\simeq1.7$. The AME is not completely well-characterised as
yet, but falls rapidly below 20 GHz. Fits to the high latitude sky
suggest that it may be reasonably well approximated by a power-law emissivity over the WMAP range of wavelengths, although this may reflect the combination of multiple
spinning dust populations in different physical conditions
along a given line-of-sight.

The total foreground intensity observed by the $\nu$ channel in a given direction $p$ can then be summarized as follows:
% --------------------------------------------------------------------------------------------------------
\be
\begin{split}
T_\nu(p) &= M + \sum_{d=x,y,z} D_{\nu, d}(p) + \Big(\frac{\nu}{\nu_0}\Big)^{\beta(p)}A_{\rm synch}(p) \\
&+   \Big(\frac{\nu}{\mu_0}\Big)^{\alpha(p)}A_{\rm f-f}(p) +  \Big(\frac{\nu}{\lambda_0}\Big)^{\epsilon}A_{\rm dust}(p) \\
&+ {\rm AME}(\nu)\,,
\end{split}
\label{eq:fg}
\ee
% --------------------------------------------------------------------------------------------------------
where M and D represent monopole and dipole residuals, respectively,
$A_i$ is the amplitude in antenna temperature units of the $i^{\rm th}$ foreground component at the
reference frequency ($\nu_0$, $\mu_0$, $\lambda_0$), and $\beta$,
$\alpha$ and $\epsilon$ describe the spectral response of synchrotron,
free-free and thermal dust emission, and an AME contribution has also
been included. 

In principle, there is no reason why these parameters
should be constant over the sky, so we should allow them to vary pixel by
pixel and let \commander\ solve for the most likely value. In practice,
we are limited by the number of frequency bands, five in the case of WMAP,
from which we also wish to determine the CMB contribution. Therefore,
there remain only 4 maps that can be used to infer the foreground emission. Addressing the full
problem as posed in Eq.~\ref{eq:fg} in an independent and
self-consistent way is therefore not possible.
Since our present analysis is motivated by a desire to assess the presence of a hard synchrotron component in the Galactic center, the microwave haze, we prefer to disentangle the foreground contributions in the low frequency range.

\begin{figure}[!h]
\begin{center}
\incgr[width=.5\columnwidth, angle=90]{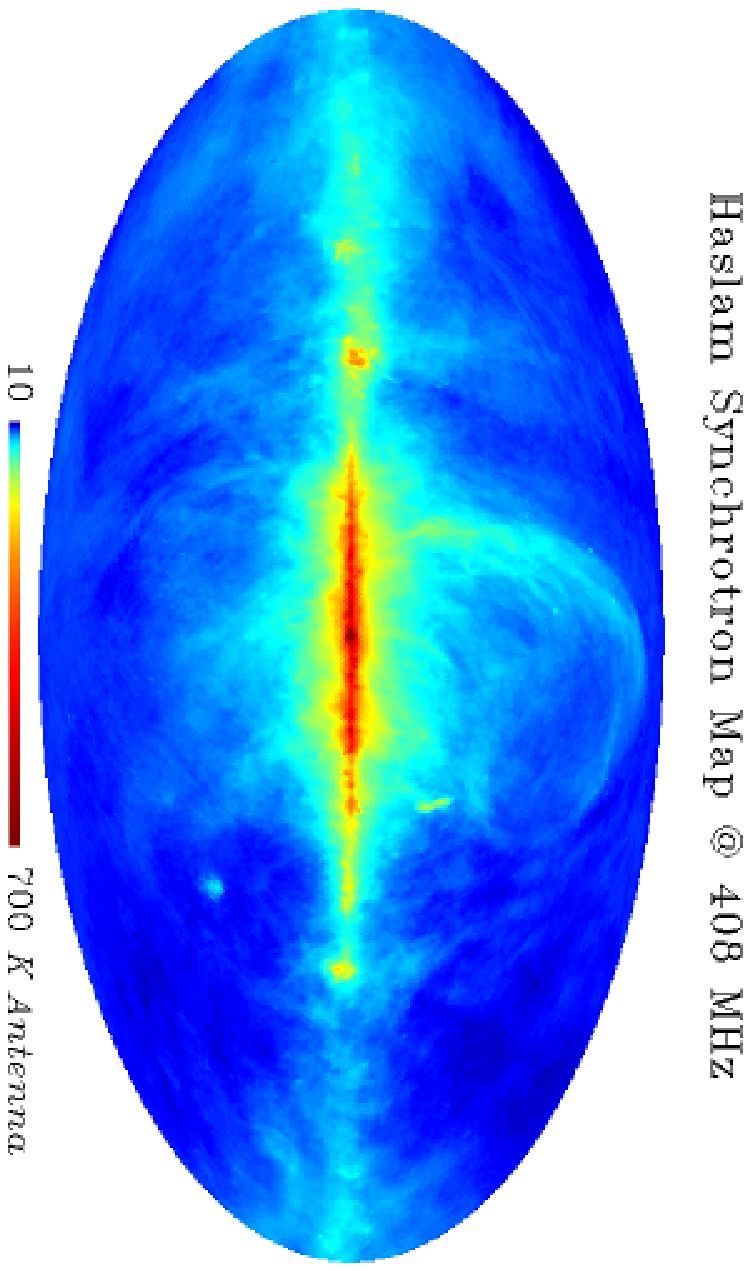}
\incgr[width=.5\columnwidth, angle=90]{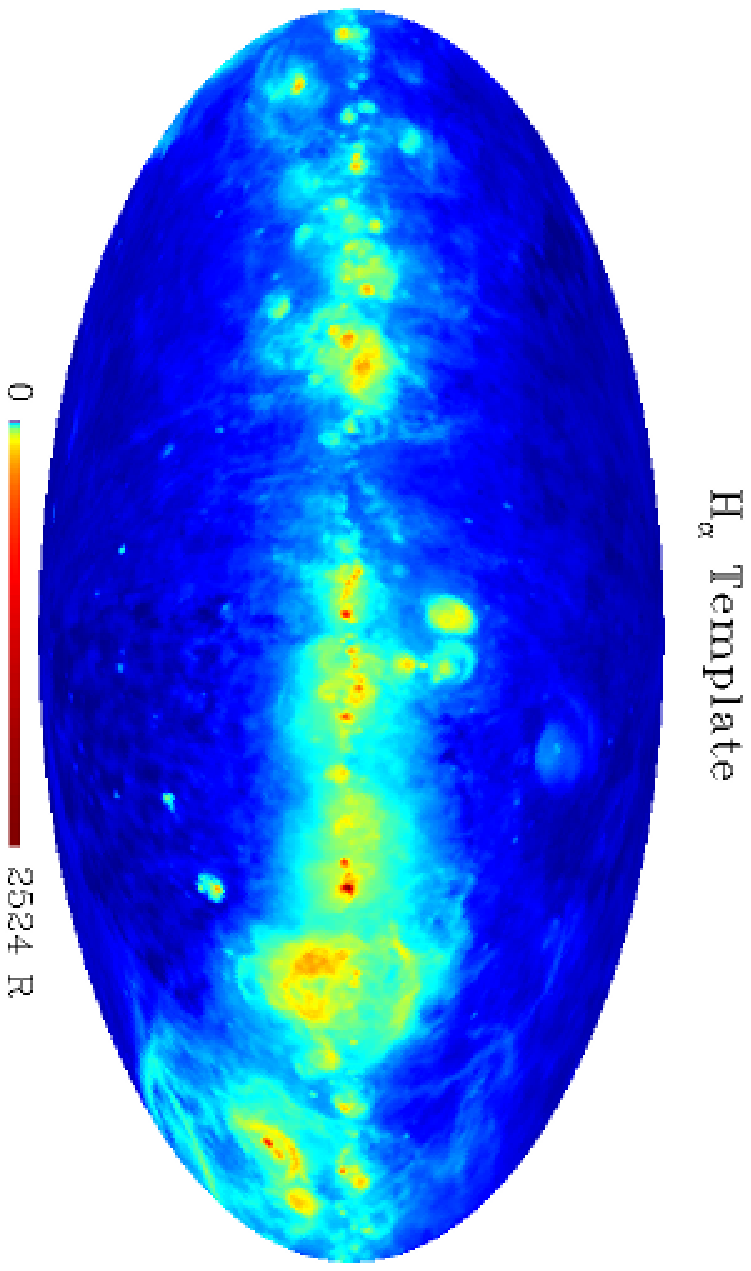}
\incgr[width=.5\columnwidth, angle=90]{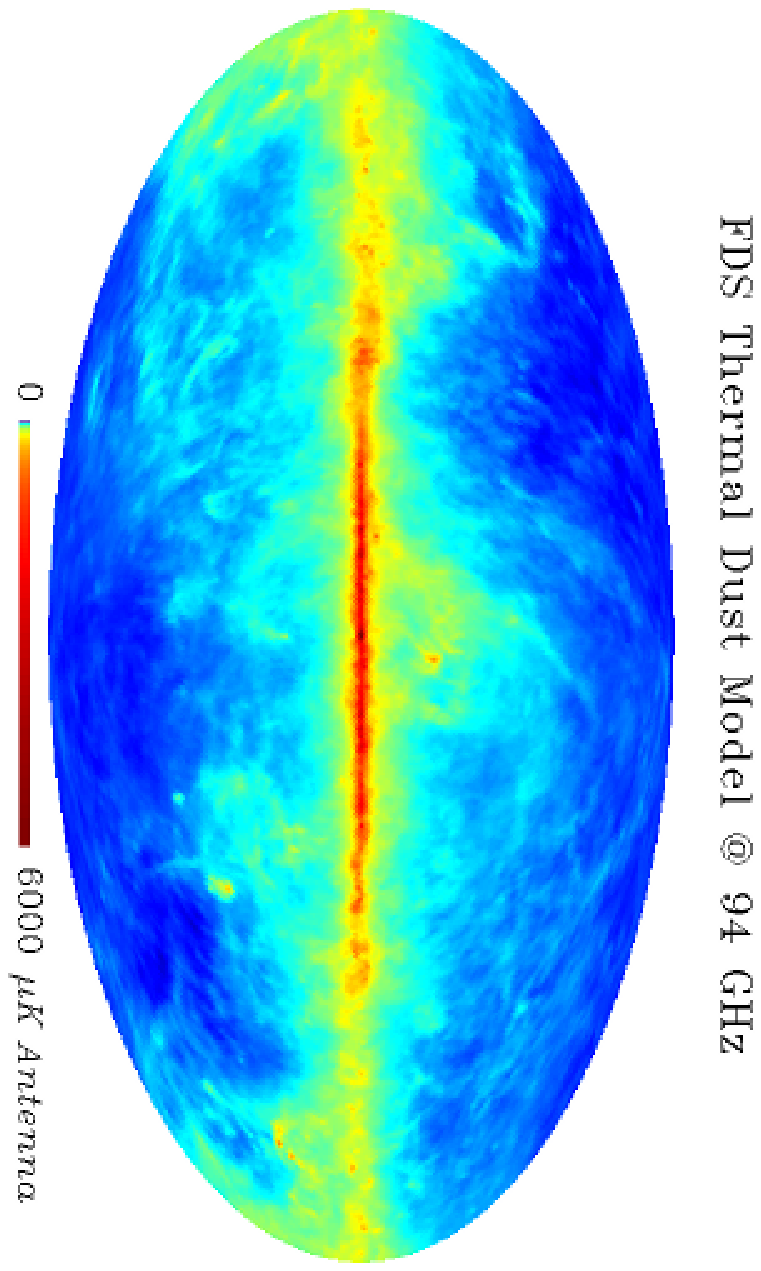}
\caption{Top to bottom, the three templates used to trace synchrotron (Haslam 408 MHz, top panel), free-free (H$\alpha$, middle panel) and dust, both thermal and spinning components (FDS, lower panel).}
\label{fig:templates}
\end{center}
\end{figure}

 A plausible and widely used alternative, which allows for more realistic foreground modeling, is the use of external
templates.  
At very low frequency ($<$ 1~GHz), the observed sky signal is dominated by the
synchrotron emission from our Galaxy,  and is little contaminated by
free-free emission, at least away from the Galactic plane. 
A full-sky map of the synchrotron
emission is provided by the 408 MHz map of \citet{Haslam1982A&AS47}.
The optical H$\alpha$ line is known to be a tracer of the
free-free continuum emission at microwave wavelengths. \cite{Finkbeiner2003ApJS..146..407F}
produced a full-sky H$\alpha$ map
as a composite of various surveys in both the northern 
and southern hemispheres. 
Finally, \citet{Finkbeiner1999ApJ524} predicted the thermal
dust contribution at microwave frequencies from a series of models
based on the \emph{COBE}-DIRBE 100 and 240 $\mu m$ maps tied
to \emph{COBE}-FIRAS spectral data.
We use predicted emission at 94~GHz from the preferred Model 8 (FDS8) as our reference
template for dust emission. These templates, smoothed to 60
arcminutes and downgraded to $N_{\rm side}=128$, are shown in
Figure~\ref{fig:templates}.
% -----------------------------------------------------------------------------------------
% -----------------------------------------------------------------------------------------

Since the thermal dust contribution to the WMAP bands is small compared to synchrotron and free-free emission, we can describe it by means of the FDS template with a fixed spectral index $\beta=1.7$, allowing for an overall amplitude, $b$. The same can be done for the bremsstrahlung emission, assuming the H$\alpha$ template as a sufficiently accurate description at 23 GHz and rescaling the amplitude according to a power law with index $\alpha=-2.15$. The synchrotron emission is expected to vary across the sky, and the template we have is at 408 MHz, quite a large stretch from the first band of WMAP. A wise choice is to let \commander\ solve for an amplitude and spectral index at every pixel, choosing 23 GHz as pivot. It should be noted that variations in the gas temperature will be imprinted onto the synchrotron component, as will deviations in the spectrum of thermal dust (which is a much smaller contribution).

A detailed study of the foreground emission in the frequency range 23-94 GHz employing external templates can be found in \cite{Dickinson2009ApJ705}. Notice that the value of the spectral index of the dust, $1.7$, is somewhat dependent on the specific method used to derive it. A direct fit to the predicted dust templates at WMAP frequencies using the FDS8 model yields a lower value of $\beta=1.55$, which is in agreement with what we find when applying template fitting procedure to WMAP maps (see Figure \ref{fig:dust_sed}).

\section{WMAP Analysis}
\label{sec:wanalysis}
Previous work on the Galactic Haze relied heavily on template regression techniques, raising a debate as to whether the 
assumption of constant spectral behavior across the sky is the cause of the excess emission near the Galactic Center.
To address this issue, we follow an alternate approach and solve instead for a simple model describing Galactic emission
with two power laws. One is a low-frequency component with a falling spectrum to account for synchrotron and free-free emission, 
as well as AME.  The second is a higher frequency component with a rising spectrum to represent thermal dust emission. This appears to be well-motivated in studies of the WMAP Maximum Entropy Method (MEM) foreground solutions, as in \cite{Park2007ApJ...660..959P}.  While we can solve for the amplitude and spectral index at every pixel at of the low-frequency component, where the variability of the signal is large, we fix the dust emissivity at high frequency to $\epsilon=1.7$, consistent with previous applications of the Gibbs sampling technique to the WMAP data. We choose $22.8$~GHz and $94$~GHz as pivot frequency for the low-frequency and thermal dust component respectively. The drawback of this approach is that the three low-frequency emission mechanisms are combined into a single power law component  that may not provide an fully adequate description of their physical complexity.  However, we will attempt to separate these components post-sampling in Section~ \ref{subsec:regression}.

We stress that our approach differs from that followed by \cite{Haze_Cumberbatch2009arXiv0902.0039C}, where \commander\ was only used to extract a CMB map while modeling the foreground components with templates. The main focus of that work was comparison between the WMAP Internal Linear Combination (ILC) CMB map, previously adopted in studies of the Haze, and the posterior mean CMB map derived from Gibbs sampling.  The presence of the Haze was found to be stable with respect to the CMB map choice.  In the present paper, we take full advantage of \commander\ to jointly sample from foregrounds and CMB.  In addition, we choose a less aggressive smoothing of the WMAP maps and mask, both factors that are likely to enhance the signal-to-noise ratio of the Haze component.
%\subsection{Sky Maps and CMB results}

% ---------------------------------------------------------------------------------------------------
\begin{figure}[!h]
\begin{center}
\incgr[width=.6\columnwidth, angle=90]{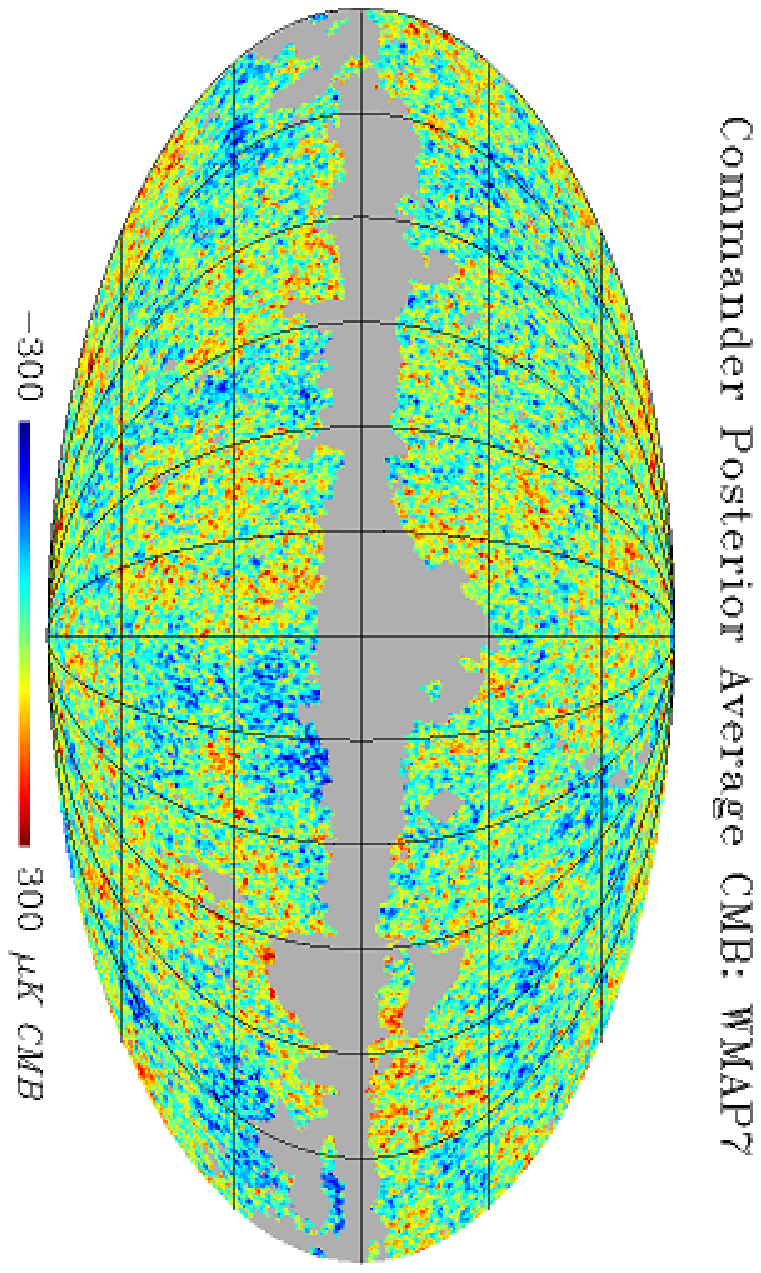}
\incgr[width=.6\columnwidth, angle=90]{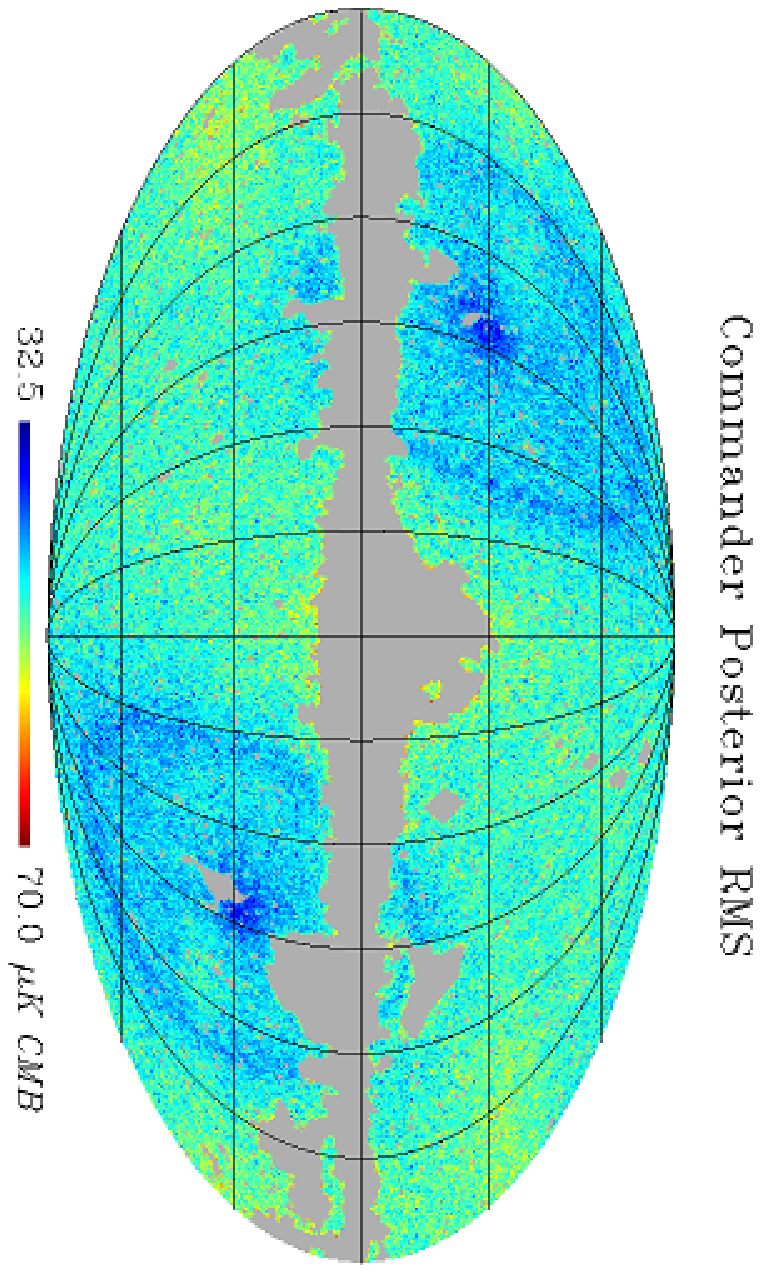}
\incgr[width=.6\columnwidth, angle=90]{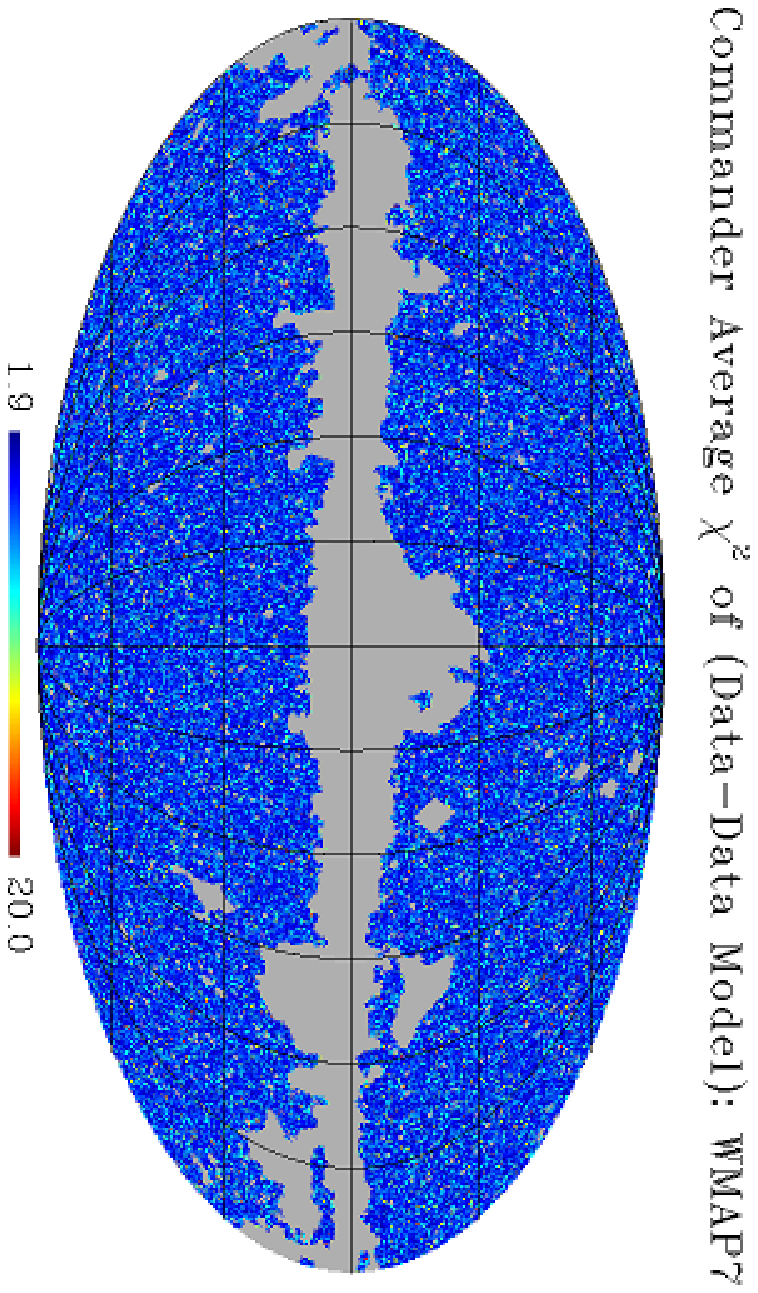}
\caption{CMB \commander\ posterior average, rms and mean $\chi^2$ map. No particular features are present, meaning that the model is a very good fit to the WMAP 7-year data. {\bf The lower limit results 1.9, whereas} the upper limit, 20, corresponds to $0.001$ probability for a $\chi^2$ distribution with 5 degrees of freedom. 
Within the Gibbs sampling framework, we do not search for the maximum-likelihood (ML) point, but rather sample from the posterior: in practice, we add a scatter term around the the ML point, drawn from the noise model of the data. This results into a number that is distributed with 5 degrees of freedom.}
\label{fig:cmb_map}
\end{center}
\end{figure}
% ---------------------------------------------------------------------------------------------------

The \commander\ result for the CMB map is shown in
Figure~\ref{fig:cmb_map}. As a diagnostic, we show the mean $\chi^2$ map in the lower panel of Figure~\ref{fig:cmb_map}. The number of pixels with $\chi^2 >20$, corresponding to the 99.9\% confidence level for 5 degrees of freedom, is 0.7\%. Most of these pixels lie near point sources and we conclude that they correspond to point source bleeding outside the mask after convolution with a large beam. The overall lack of features in the $\chi^2$ map is an indication of the goodness-of-fit of the model and that the residuals at each frequency are compatible with our noise description.

%Figures~\ref{fig:cmd_spectrum} and \ref{fig:cmd_spectrum_diff} show that
We checked that the \commander\ power spectrum is consistent with the best
estimate provided by the WMAP team \citep{Larson2010} at the 1$\sigma$ level up to $\ell=200$.
Beyond that, the \commander\ solution has significantly more scatter
due to the regularizing noise added to the smoothed, binned maps (see Sec.~\ref{subsec:wmap}).
In addition, while the WMAP team also used Gibbs sampling
techniques for low multiples, their high $\ell$ power spectrum was
obtained by applying a quadratic estimator to the cross-spectra of V
and W bands. We emphasize that the
agreement between the two power spectra below $\ell=200$ is what is
important for our purposes since we are concentrating on features in
the map that are much larger than 1$^{\circ}$. The possible presence of a residual monopole and dipole in the $\Delta T$ data has been taken into account and the mean values obtained are displayed in Table ~\ref{tab:mondip}. They are small and compatible with those found by \cite{Dickinson2009ApJ705} in the WMAP 5-year data, though it is important to keep in mind that monopole and dipole features become strongly coupled to foregrounds in CMB analyses, as discussed by \cite{Eriksen2008ApJ676}.
% ---------------------------------------------------------------------------------------------------
\begin{table}[!htb]
\begin{center}
\caption{Mean values ($\mu$K) for the monopole and dipole residuals at
  every frequency.\label{tab:mondip}}
\begin{tabular}{lccccc}
   \tableline\tableline
   & K-band & Ka-band & Q-band & V-band & W-band \\ 
   & 23 GHz & 33 GHz & 41 GHz & 61 GHz & 94 GHz \\ 
   \tableline
   M & $2.4\pm0.9$ & $5.6\pm1.0$ & $3.5\pm1.1$ & $2.7\pm1.0$ & $3.9\pm1.0$ \\
   ${\rm D_x}$ & $-2.6\pm1.2$ & $-3.4\pm1.2$ & $-2.7\pm1.2$ & $-2.7\pm1.2$ & $-3.0\pm1.2$ \\ 
   ${\rm D_y}$ & $-3.6\pm0.9$ & $-4.3\pm0.9$ & $-4.4\pm0.9$ & $-3.5\pm0.9$ & $-4.0\pm0.9$ \\ 
   ${\rm D_z}$ & $1.9\pm0.2$ & $1.9\pm0.2$ & $2.4\pm0.2$ & $1.8\pm0.2$ & $2.0\pm0.2$ \\
   \tableline
\end{tabular}
\end{center}
\end{table}
% ---------------------------------------------------------------------------------------------------

\begin{figure*}[hb]
\begin{center}
\incgr[width=.5\columnwidth, angle=90]{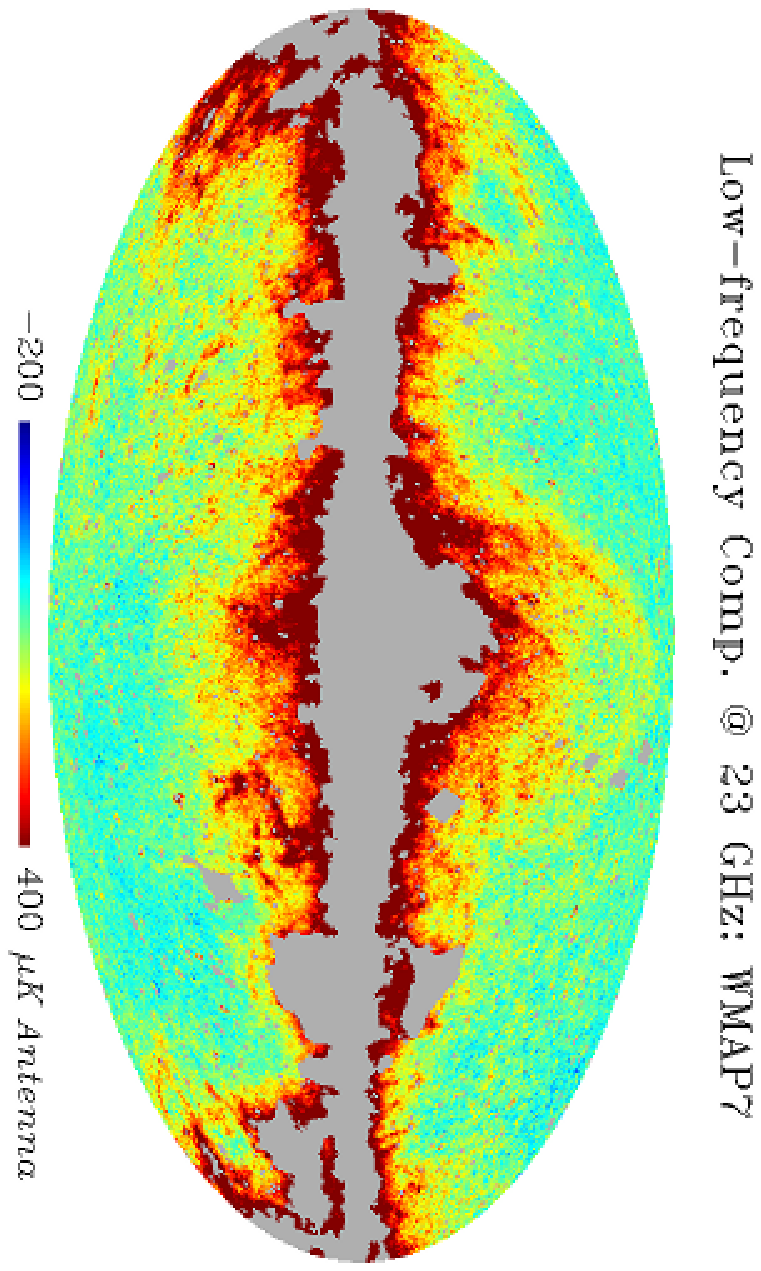}
\incgr[width=.5\columnwidth, angle=90]{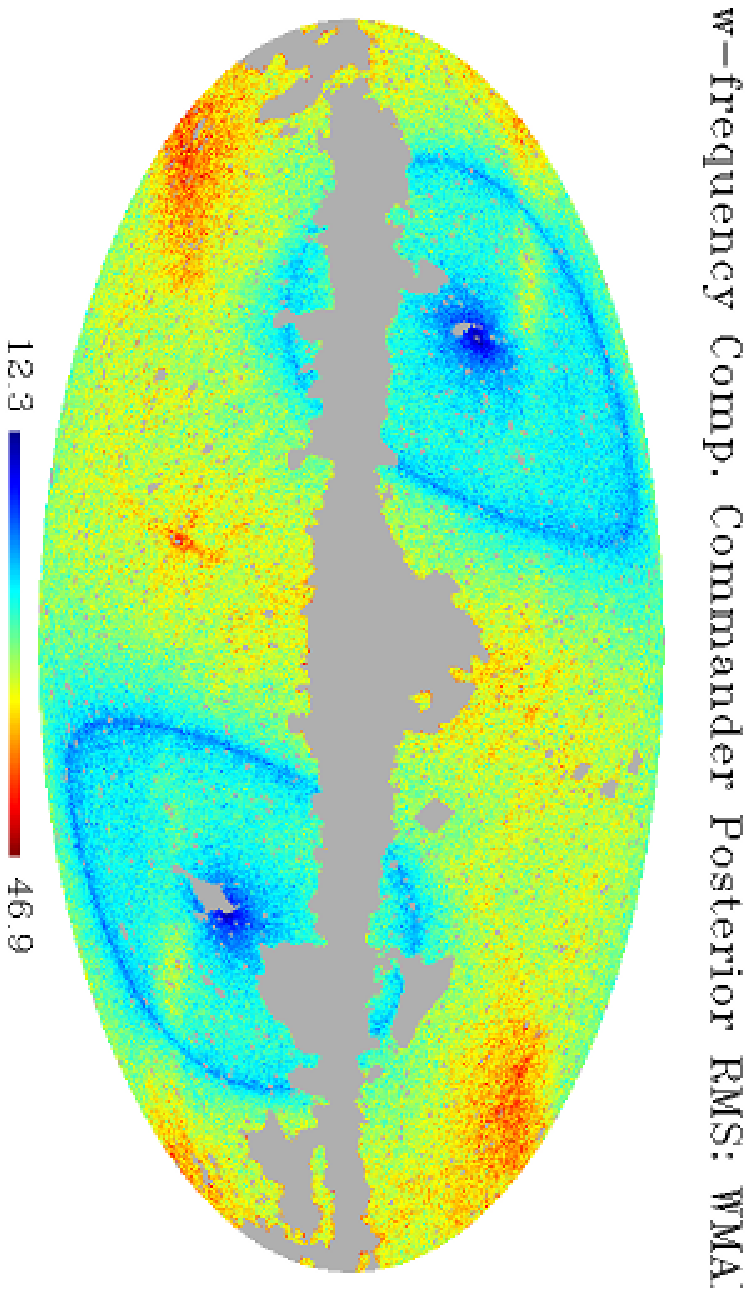}
%\caption{Mean field map (top panel) and rms (bottom panel) for low-frequency component in the case the two power law foreground model}.
%\label{fig:standard_wmap_fga1}
%\end{center}
%\end{figure}
%% ---
%\begin{figure}[h]
%\begin{center}
\incgr[width=.5\columnwidth, angle=90]{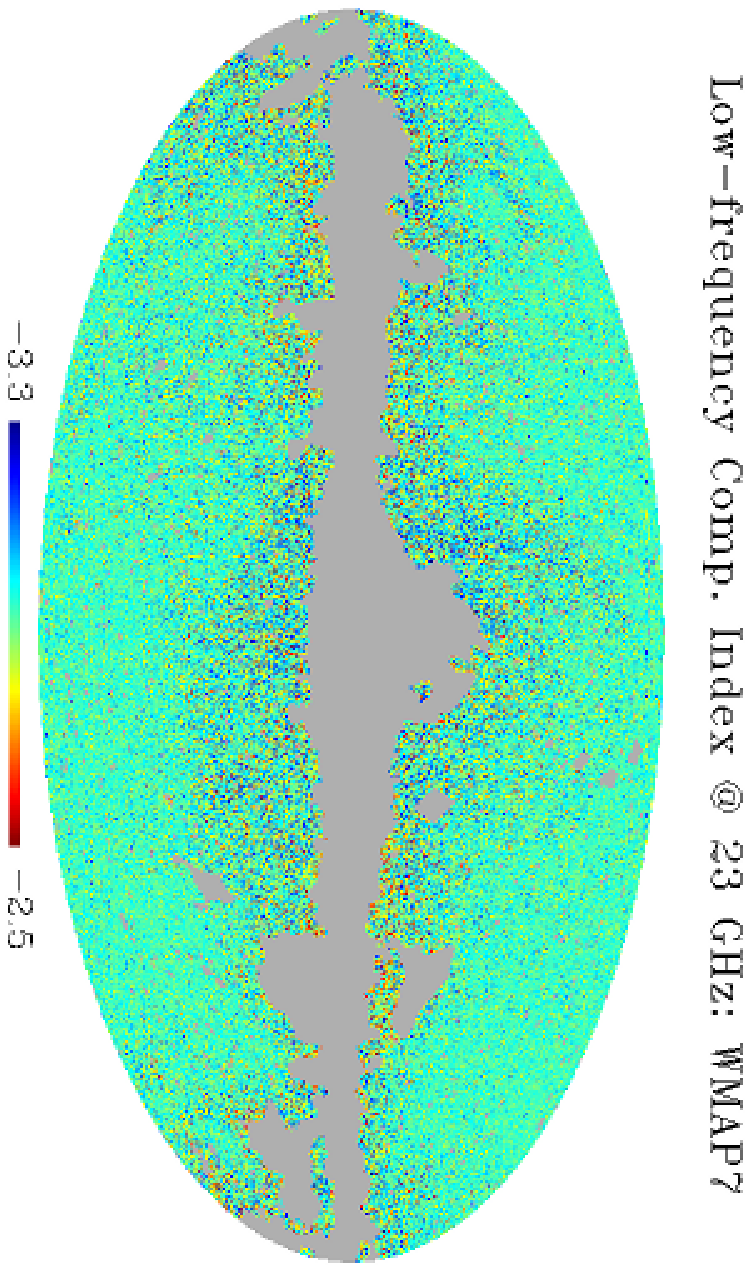}
\incgr[width=.5\columnwidth, angle=90]{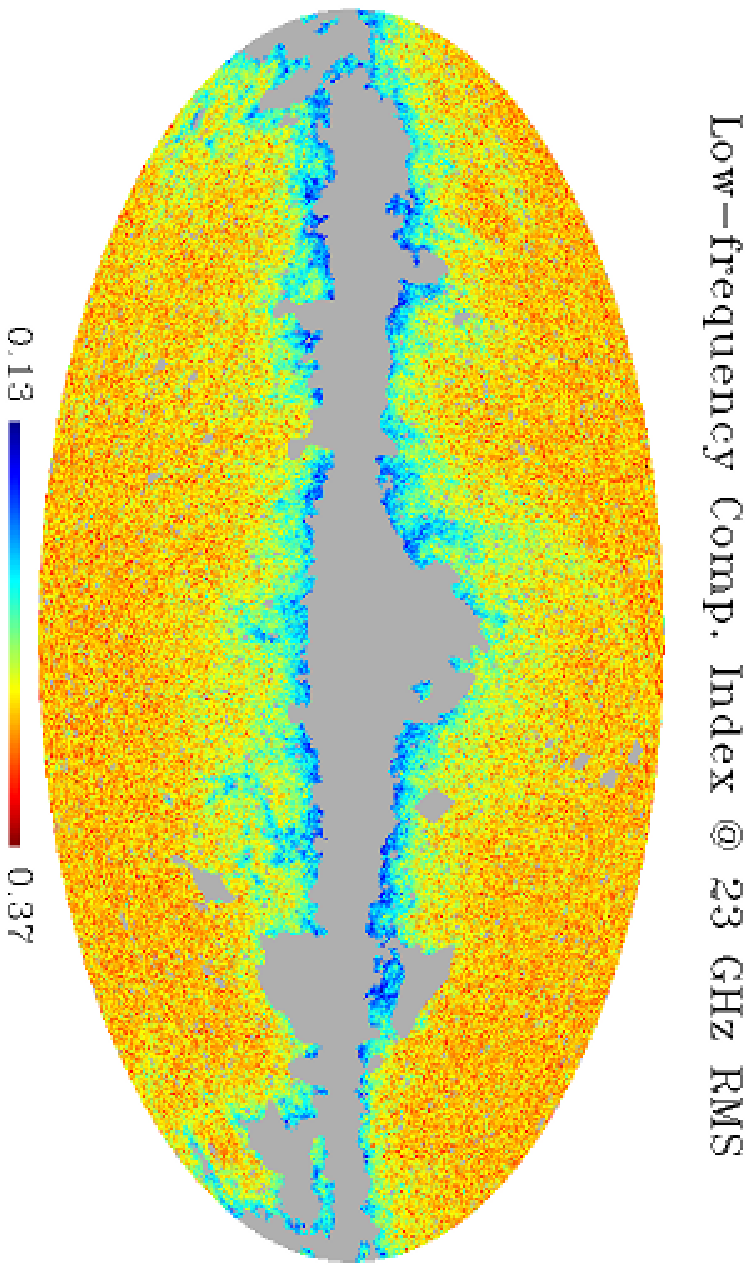}
%\caption{Mean field map (top panel) and rms (bottom panel) for synchrotron spectral index for the two power laws foreground model}.
%\label{fig:standard_wmap_fgi1}
%\end{center}
%\end{figure}
%% ---
%\begin{figure}[h]
%\begin{center}
%##\incgr[width=.5\columnwidth, angle=90]{wmap-pl_foreg_amp_94.eps}
\incgr[width=.5\columnwidth, angle=90]{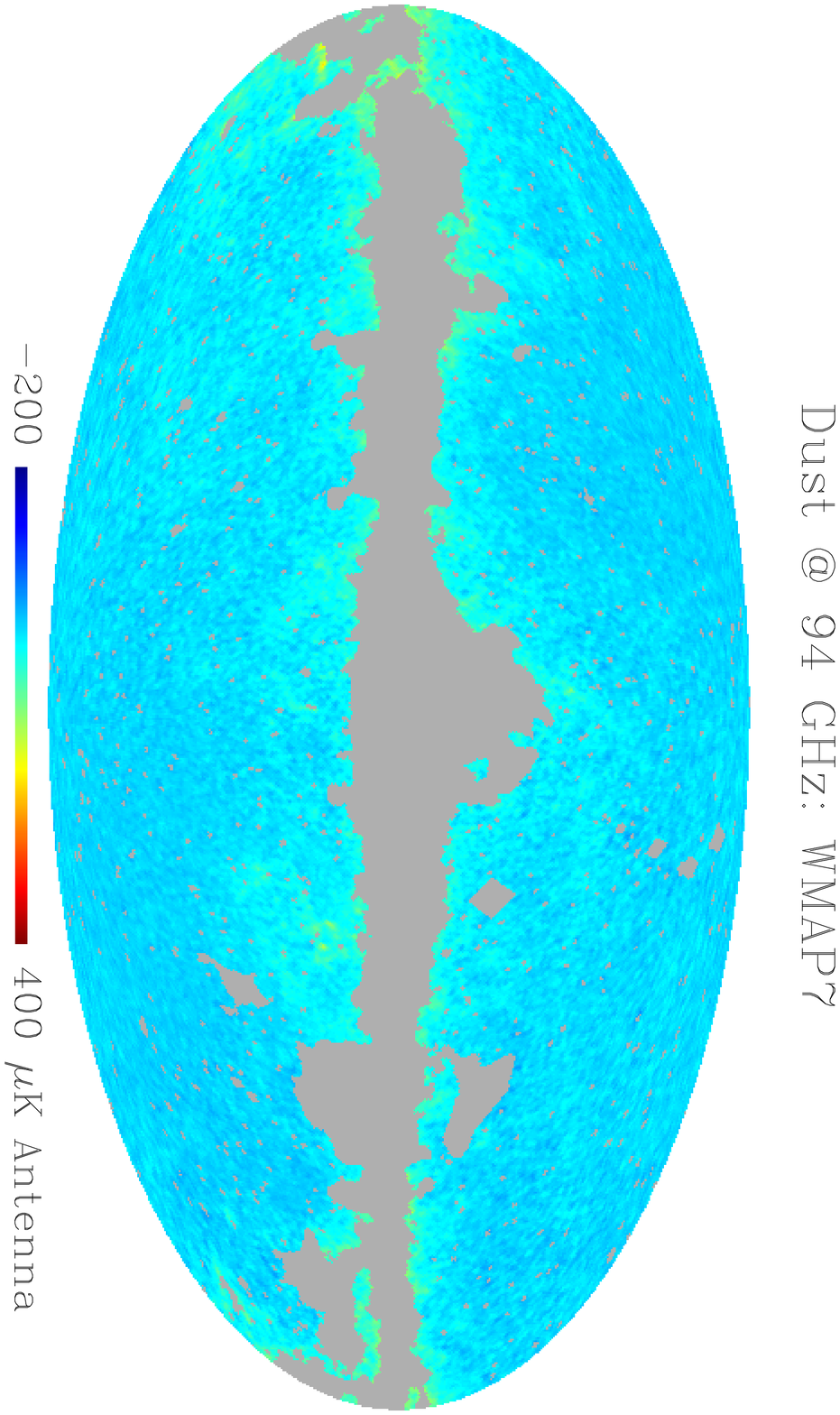}
\incgr[width=.5\columnwidth, angle=90]{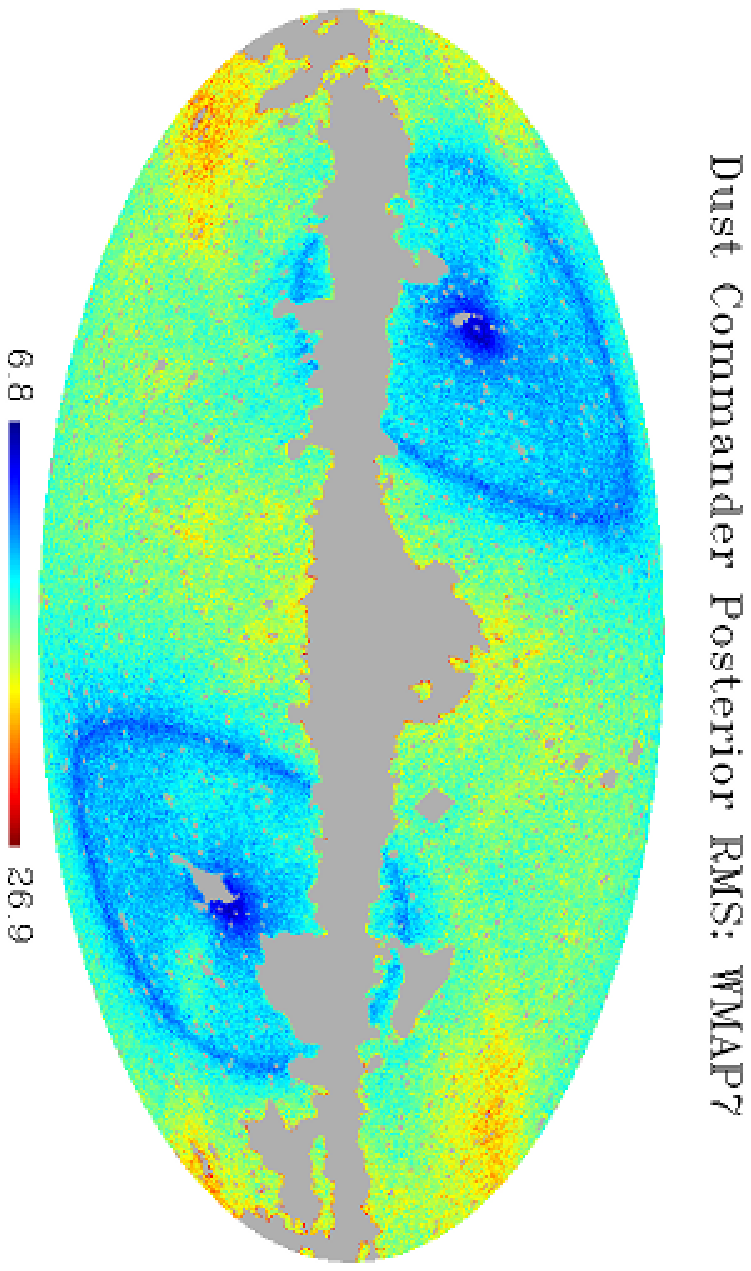}
\caption{Mean field map (left column) and rms (right column) for low-frequency component amplitude (upper panel) and spectral index (middle panel) and dust amplitude (lower panel) for the foreground model with two power laws.}
\label{fig:standard_wmap}
\end{center}
\end{figure*}

Foreground maps (left column) and associated errors (right column) for low frequency amplitude and spectral index measured at 23 GHz and the thermal dust contribution at 94 GHz are shown in Figure~\ref{fig:standard_wmap}. 
% ---------------------------------------------------------------------------------------------------
% ---------------------------------------------------------------------------------------------------

\begin{figure}[htb]
\begin{center}
\incgr[width=.5\columnwidth, angle=90]{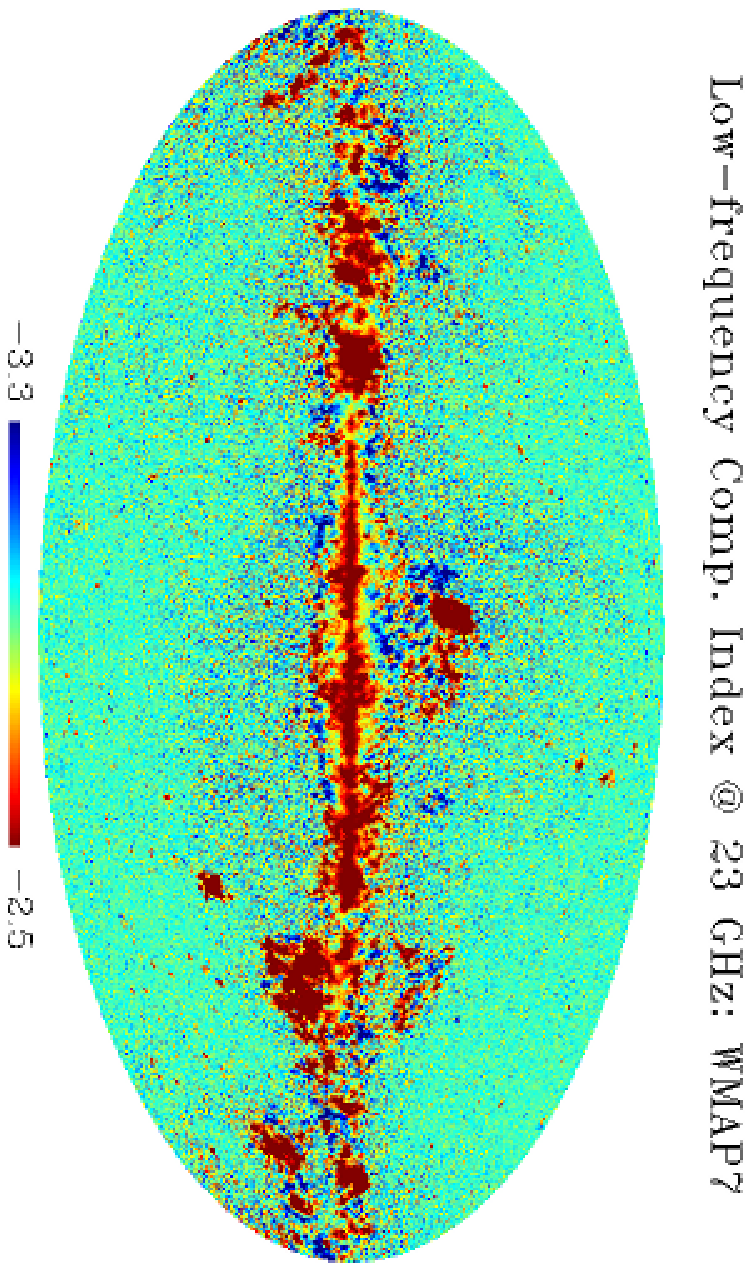}
%\incgr[width=.6\columnwidth, angle=90]{wmap-pl_foreg_ind_23_rms_unmask.eps}
\caption{Full sky spectral index mean field map. The Galactic plane shows strong variation corresponding to regions where one component among synchrotron, free-free and spinning dust emission dominates.}
\label{fig:unmask_ind}
\end{center}
\end{figure}

It is clear that our derived low frequency component represents a combination of multiple emission mechanisms, which we now attempt to disentangle. The low frequency component shows a strong correlation with thermal dust emission at 94 GHz as modeled by FDS, correlation which is interpreted as signature of a spinning dust contribution. The spectral index map, second row in Figure~\ref{fig:standard_wmap}, is particularly informative in this respect since departures from the prior assumed in \commander, $-3.0\pm0.3$ are driven by the data. The mean value at high Galactic latitudes turns out to be a bit higher than the prior ($\simeq-2.9$) as consequence of the superposition along the line of sight of multiple components. It is remarkable how very bright free-free regions show up clearly, requiring a spectral index close to $-2.15$, which saturates the scale in Figure~\ref{fig:unmask_ind}. The visual correlation of these diffuse gas clouds with those present in the H$\alpha$ map is striking. Spectral index values lower than the prior are visible in regions where both synchrotron and dust correlated emission are present, and distinguishing between the two is difficult.
%% ---------------------------------------------------------------------------------------------------
%% ---------------------------------------------------------------------------------------------------

\subsection{Template regression}
\label{subsec:regression}
Following a common approach to the problem \citep[see for example][and references therein]{dobler08b,Haze_Bottino2010MNRAS.402..207B}, we regress our amplitude maps against the three templates shown in Figure~\ref{fig:templates}.

Assuming the amplitude map to be a linear combination of these templates, the coefficients $\Ainu$ are computed from a $\chi^2$ minimization of the form:
\bea
&& \chi^2 = \sum_\nu \sum_{\rm p} \Big(S_\nu(p) - \sum_i \Ainu T_i(p) \Big) \\ %\sum_\nu \sum_{\rm p} \Big(M_\nu(p) - \sum_i A_i T_i(p) \Big) N_\nu^{-1}(p) \nonumber \\
&& \qquad N_\nu^{-1}(p) \Big( S_\nu(p) - \sum_i \Ainu T_i(p) \Big);\nonumber \\
&& \Ainu : \frac{\partial\chi^2}{\partial \Ainu} = 0; \nonumber \\
&& \Delta^2 \Ainu = \left( \frac{\partial^2\chi^2}{\partial \left[\Ainu\right]^2} \right)^{-1}. \nonumber\label{eq:Amp}
\eea
where $T_i$ are the foreground templates and $S_\nu$ is the \commander\ amplitude solution.
%The CMB map can be indeed considered as an external template we simply subtract from the data. As a cross-check we tried to include it in the templates and fitting for an amplitude. The results were consistent with $A_{\rm CMB}=1$ at all frequencies. The CMB map we adopt in the analysis comes from the \commander\ sample processing. Notice that the uncertainty on the correct CMB map was the main source of error in the analysis by \cite{Haze_Dobler2008ApJ...680.1222D}.
The explicit expression for the regression solution is given by:
\bea
&& \Ainu = \sum_j \mbf{T^\nu}^{-1}_{ij} B^\nu_j, \nonumber \\
&& T^\nu_{ij} = \sum_p \frac{T_i(p)T_j(p)} {N_\nu^2(p)}, \nonumber \\
&& B^\nu_j = \sum_p \frac{S^\nu_i(p)T_j(p) }{N_\mu^2(p)}.
\label{eq:regression_coef}
\eea
where $T^\nu_{ij}$ and $B^\nu_j$ describe the noise weighted correlation between foreground templates and frequency maps and templates, respectively. %Notice that the sum includes pixels outside the mask only. 
%The actual formula for the correlation is computed on zero-mean arrays, operation which has been advocated by \cite{Haze_Dobler2008ApJ...680.1222D} to improve the amplitude determination. 
The error $\Delta A_i^\mu$ is given by $\sqrt{(2 \mbf{T^\mu})^{-1}_{ii}}$. See also \citep{Sergi2006,Hildebrandt2007}. Since templates are not reliable in the Galactic plane, the fit is performed outside the Kq85 mask \citep{Larson2010}, which removes the plane and detected point sources. In addition, the mask covers regions of high dust column density where extinction makes our H$\alpha$ template a poor tracer of free--free template emission.

\begin{figure}[htb]
\begin{center}
\incgr[width=.5\columnwidth, angle=90]{wmap-pl_foreg_amp_23.eps}
\incgr[width=.5\columnwidth, angle=90]{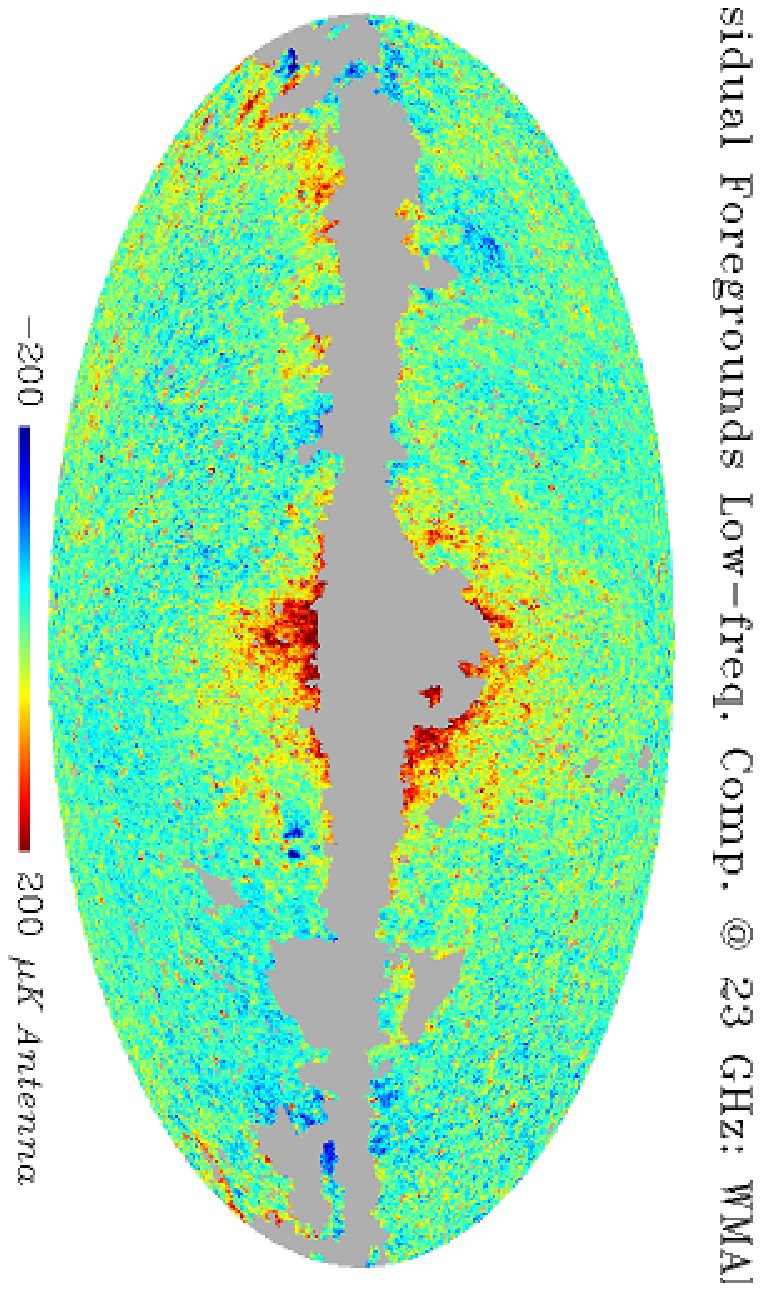}
\incgr[width=.5\columnwidth, angle=90]{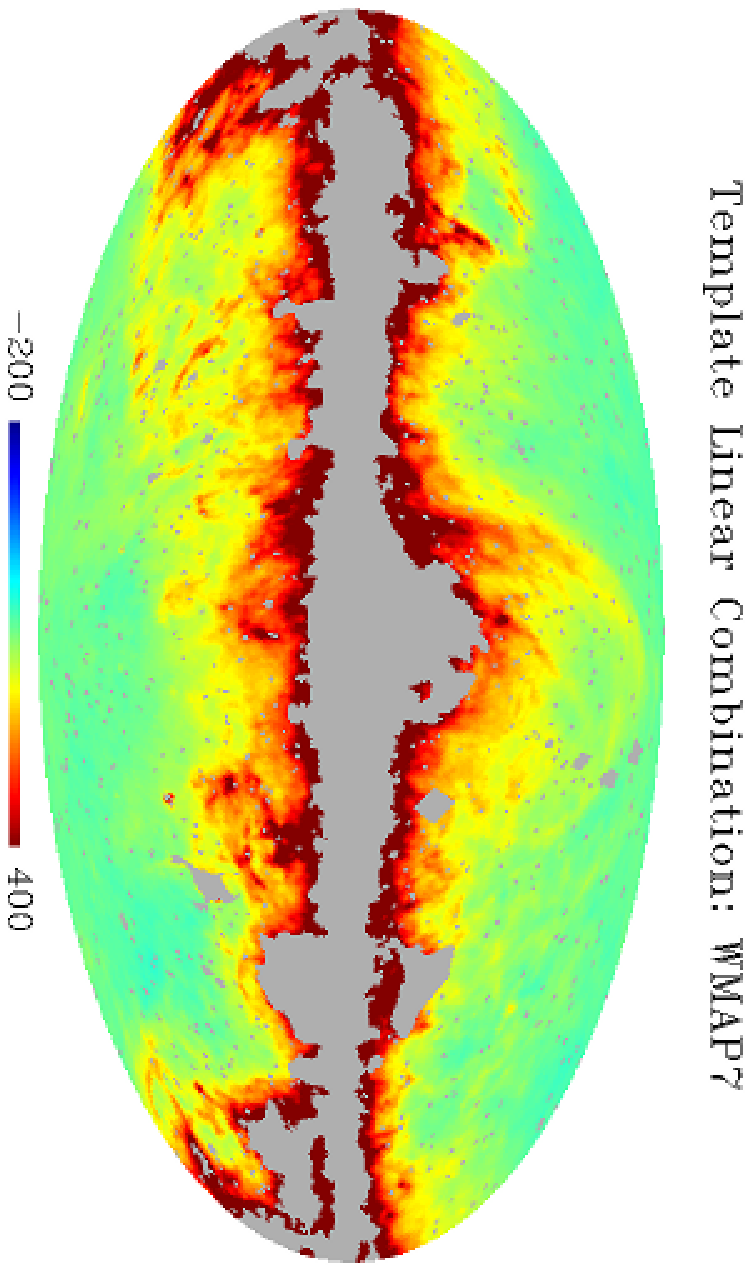}
\caption{Low-frequency foreground amplitude maps (top) compared to the linear combination of templates (bottom). The difference (i.~e.~ residuals) is shown in the middle panel.}
\label{fig:wmap-pl_regression_fg1}
\end{center}
\end{figure}

\begin{figure}[hbt]
\begin{center}
\incgr[width=.5\columnwidth, angle=90]{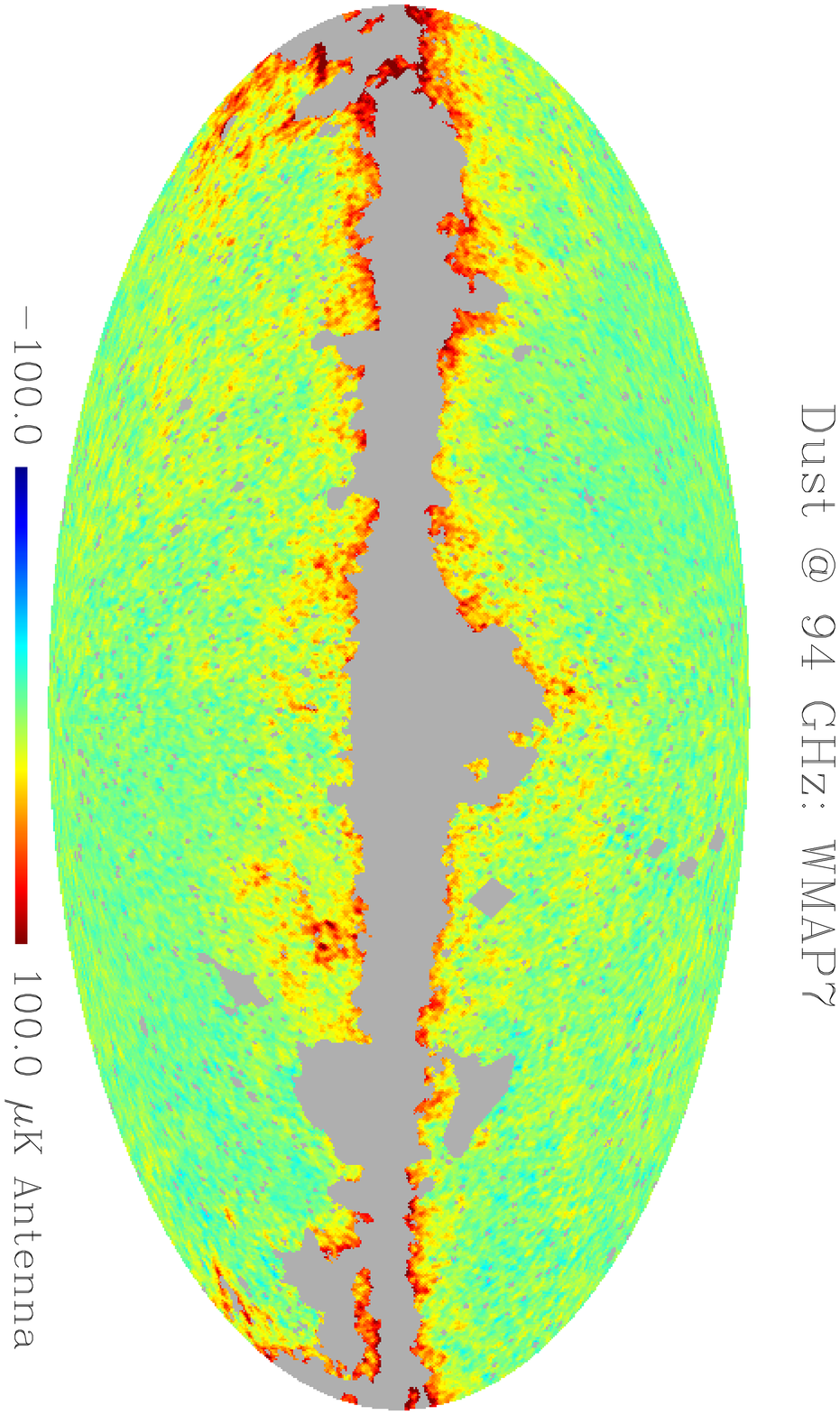}
\incgr[width=.5\columnwidth, angle=90]{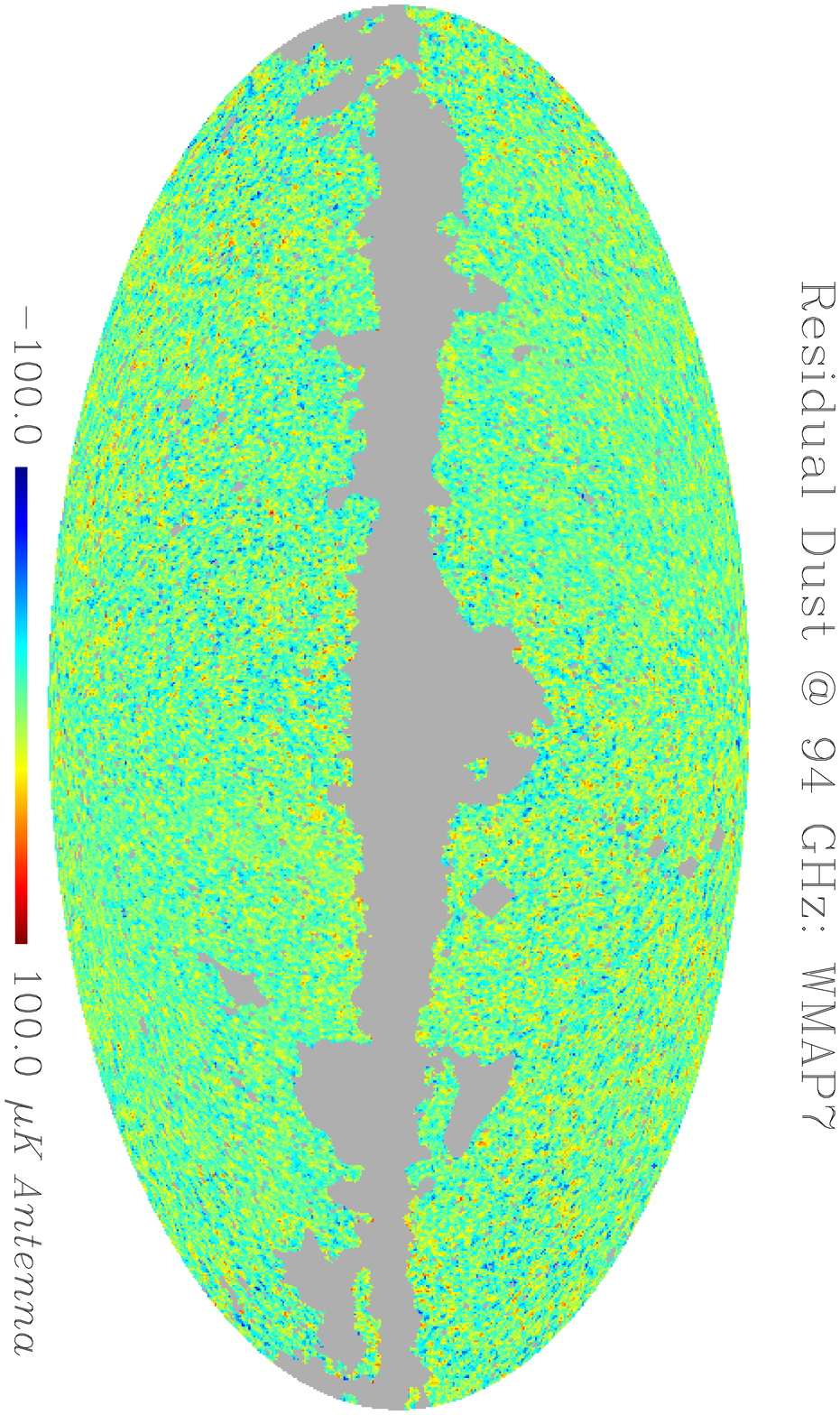}
\incgr[width=.5\columnwidth, angle=90]{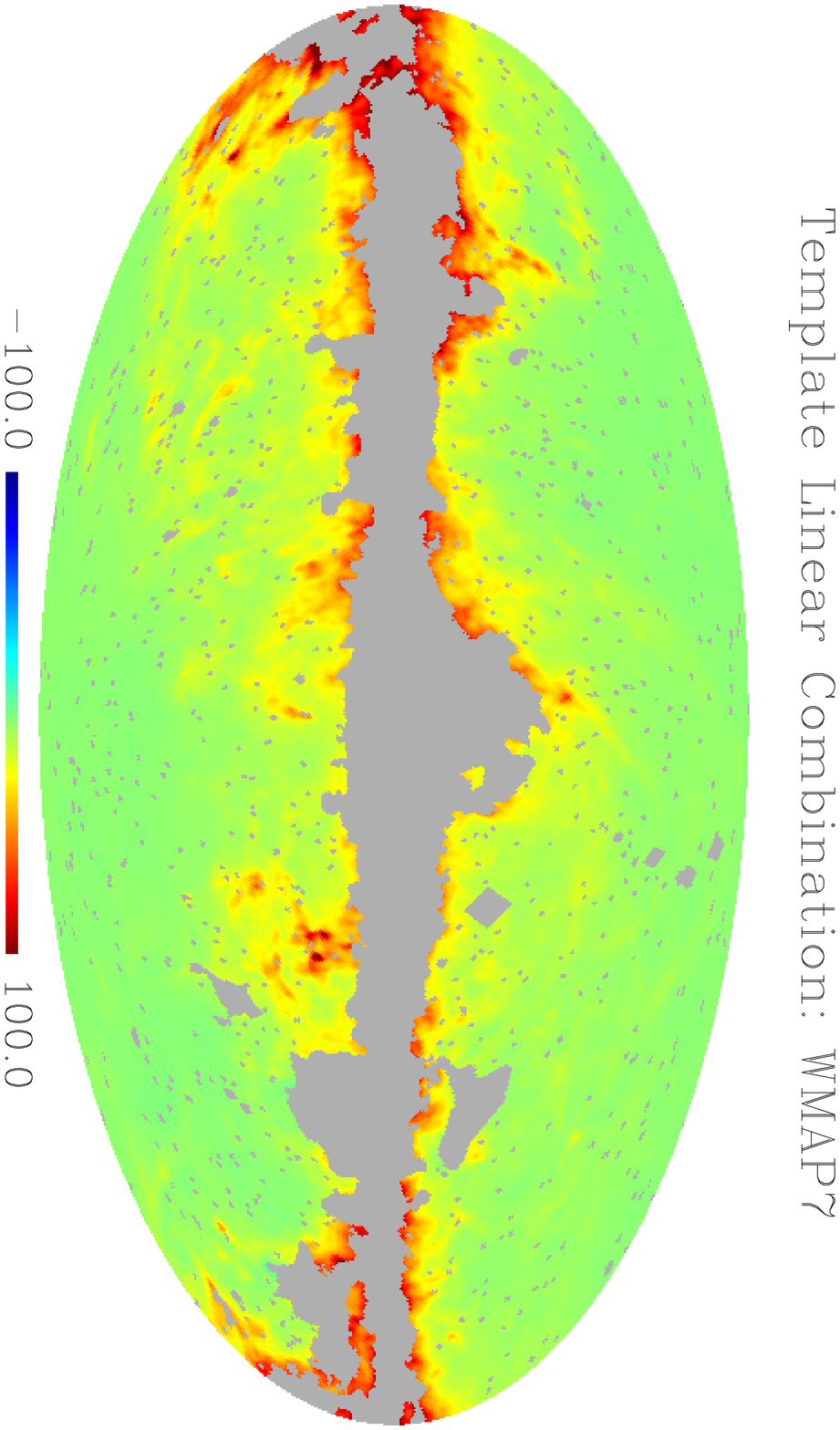}
\caption{Thermal dust amplitude maps (top) compared to the linear combination of templates (bottom) the difference (i.~e.~ residuals) is shown in the middle panel.}
\label{fig:wmap-pl_regression_fg2}
\end{center}
\end{figure}

The coefficients of the fit are quoted in Table~\ref{tab:wmap-pl_regression} and the resulting template and residuals in Figures~\ref{fig:wmap-pl_regression_fg1} and~\ref{fig:wmap-pl_regression_fg2}. Contrary to the usual technique, where the regression is performed on the WMAP maps, we regress out instead the \commander\ posterior average maps, which have noise contributed by the sampling procedure. Notice that the error we quote is obtained by integrating the posterior distribution over the other parameters, indices and component maps, and hence accounts for the uncertainty associated with the foreground fit as well as the instrumental noise. We also remind the reader that the input maps analyzed by \commander\ are noisier than the expected from simply smoothing them because we add regularizing noise required by the sampling algorithm (see Section~\ref{subsec:wmap}). These two factors increase the uncertainty on the template amplitudes. A detailed investigation is presented in Appendix~\ref{app:regression}.

\begin{figure}[!h]
\begin{center}
\incgr[width=.6\columnwidth, angle=90]{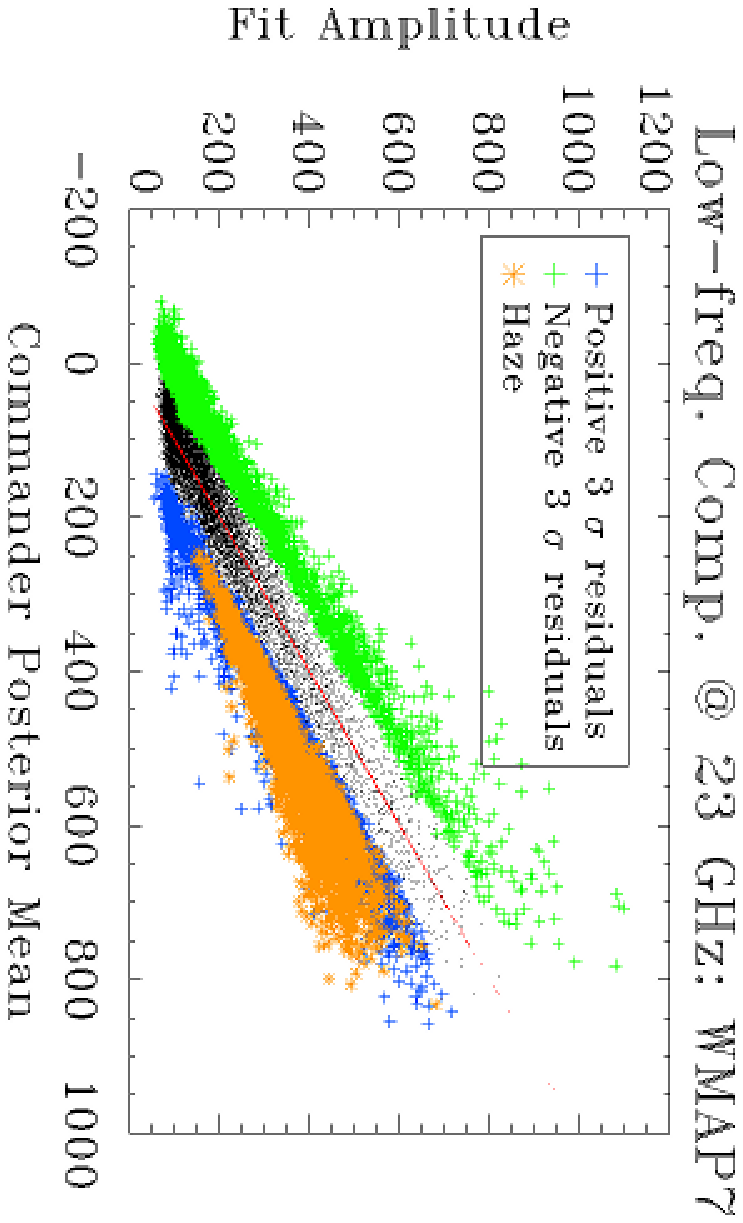}
\incgr[width=.6\columnwidth, angle=90]{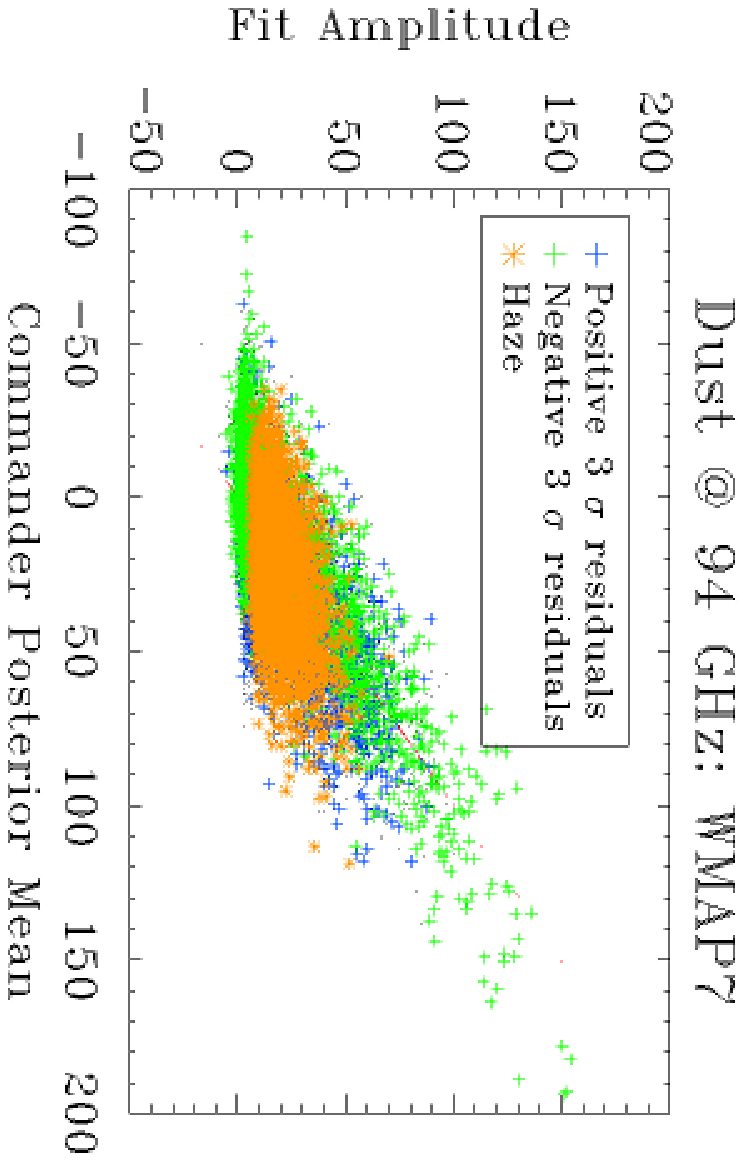}
\caption{For the low-frequency foreground (top) and dust amplitude (bottom) we show the pixel-to-pixel comparison between the best-fitting template combination (y-axis) and the \commander\ values (x-axis). The red line marks the y=x scaling. Blue and green dots denote residuals with a significance larger than $3\sigma$, positive and negative, respectively. Orange points are pixels corresponding to the Haze: $3\sigma$ positive excess within a disc of 36 degrees around the Galactic center.}
\label{fig:wmap-pl_scatter}
\end{center}
\end{figure}

As expected, the high frequency foreground component correlates with the FDS model of thermal dust only. The low-frequency component is mainly a combination of synchrotron emission and correlated dust, the latter possibly due to spinning dust. 
We note that the contribution of free-free emission is weak because the Galactic plane, where it is strongest, has been masked out. To further quantify the goodness-of-fit, we show scatter plots of the mean field \commander\ solution as function of the derived linear template shown in Figure~\ref{fig:wmap-pl_scatter}.
% ---------------------------------------------------------------------------------------------------
% ---
% ---------------------------------------------------------------------------------------------------

% ---------------------------------------------------------------------------------------------------
\begin{table}[!h]
\begin{center}
\caption{Regression coefficients of the \commander\ foreground amplitude maps. While the thermal dust correlates with the FDS model only, the low-frequency component is indeed a mixture of all three templates, synchrotron and dust emission being the strongest. Moreover, while the thermal dust residuals are featureless and consistent with noise, the low-frequency component residuals show a clear excess of power around the Galactic center.\label{tab:wmap-pl_regression}}
\begin{tabular}{ccccc}
	\tableline\tableline 
   \multicolumn{5}{c}{\bf WMAP 7-yr}\\ 
   	\tableline\tableline
  {\bf Dataset} & {\bf Haslam} & {\bf H$\alpha$} & {\bf FDS} & r\\ 
  	\tableline
   Low Freq.~ Comp. & $(3.6\pm1.2)\times10^{-6}$ & $6\pm6$ & $7\pm2$ & 0.94\\ 
   Thermal Dust & $(0.02 \pm 1.0)\times 10^{-6}$ & $-0.6 \pm 5$ & $1.1\pm1.8$ & 0.58\\ 
	\tableline
\end{tabular}
\end{center}
\end{table}
% ---------------------------------------------------------------------------------------------------

% ---------------------------------------------------------------------------------------------------
% ---------------------------------------------------------------------------------------------------
Although the correlation is clearly present and follows the $y=x$ line, looking at the residual map is very instructive. In the case of dust, the residuals are compatible with noise and this explains why the points cluster close to $y=0$ line. The situation is more intriguing for the low frequency component where the fit is not perfect and leaves an excess of power in the proximity of the Galactic center - the Galactic Haze - which has been advocated to be a distinct contribution. Fainter positive and negative regions exist as well, perhaps suggesting an overly-simplistic modeling of the dust component.

\section{Further Analysis}
\label{sec:discussion}

The most troublesome foreground is the dust, both thermal and spinning.  While the former remains week at WMAP frequencies outside the Galactic mask, 94 GHz being only slightly contaminated, the latter is poorly characterized and traced through the correlation between the lowest WMAP channels and the FDS dust model, under the assumption that thermal dust traces spinning dust reasonably well.  We now include this correlation in our foreground model by first regressing WMAP channels against Haslam, H$\alpha$ and FDS to obtain a spectral energy distribution for each foreground, and then use the resulting phenomenological dust SED with \commander to solve for the dust amplitude and a low-frequency component, described by a power law, together with the CMB.  We recognize that this spinning dust model is simple and unlikely to capture the full complexity of the dust emission.  However, more complicated models, like spatially varying spinning dust index or multiple component spinning dust, although physically motivated \citep[such as those proposed by][]{Hoang2010Apj,Hoang2011Apj}, would require larger frequency coverage for practical use.   We therefore adopt our simple model as an adequate description for the available WMAP data set.

\subsection{Low-frequency Component and Phenomenological Dust}
\begin{figure}[!h]
\begin{center}
\incgr[width=\columnwidth]{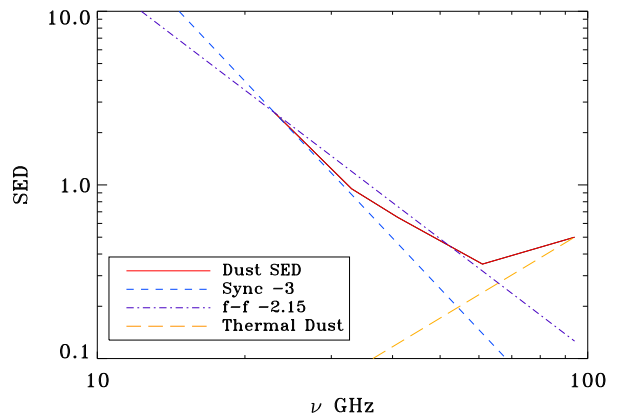}
\caption{SED for the phenomenological dust, compared to those of other components of interest: synchrotron, free-free and thermal dust emission.}
\label{fig:dust_sed}
\end{center}
\end{figure}

The first \commander\ run with this model returned a very steep spectral index for the low-frequency component, close to the limit of our prior for some regions in the Galactic plane known to have strong spinning dust emission. The $\chi^2$ of our solution in those regions was high, a sign that the foreground model fails. We then tuned the amplitude of our dust model, increasing it at low frequencies
until we obtained a shallower spectral index for the low frequency component and smaller $\chi^2$ values for the same regions. In practice, this means that we modify the SED template to increase the relative amount of spinning dust to thermal dust. The resulting spectral response is shown in Figure~\ref{fig:dust_sed}, together with those of the soft synchrotron ($\beta=-3$), free-free ($\beta=-2.15$) and thermal dust components ($\beta=1.7$). 
% ---------------------------------------------------------------------------------------------------
% ---------------------------------------------------------------------------------------------------

\begin{figure}[htb]
\begin{center}
%\incgr[width=.6\columnwidth, angle=90]{wmap_amp_1_rms.eps}
%\caption{Dust mean field map and corresponding rms at 33 GHz as resulting from \commander\ run.}
%\label{fig:fg_a2}
%\end{center}
%\end{figure}
%% ---
%\begin{figure}[h]
%\begin{center}
\incgr[width=.5\columnwidth, angle=90]{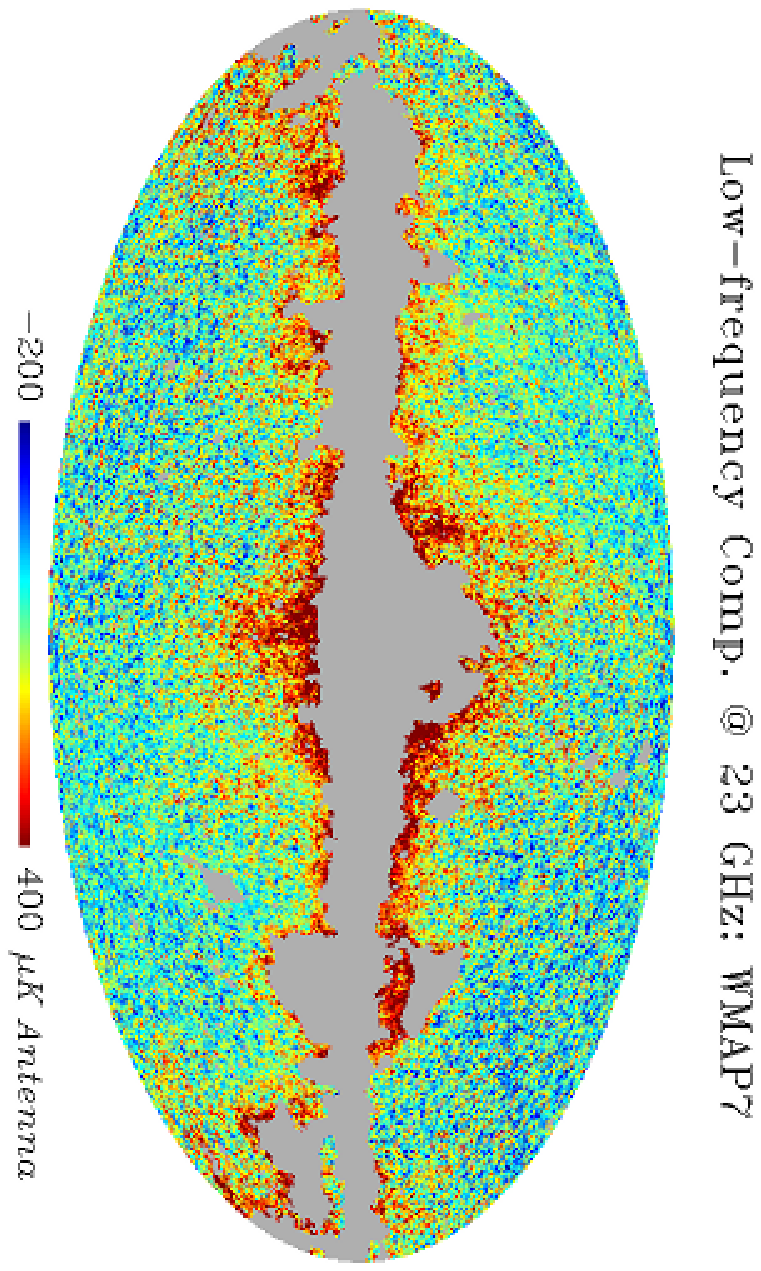}
\incgr[width=.5\columnwidth, angle=90]{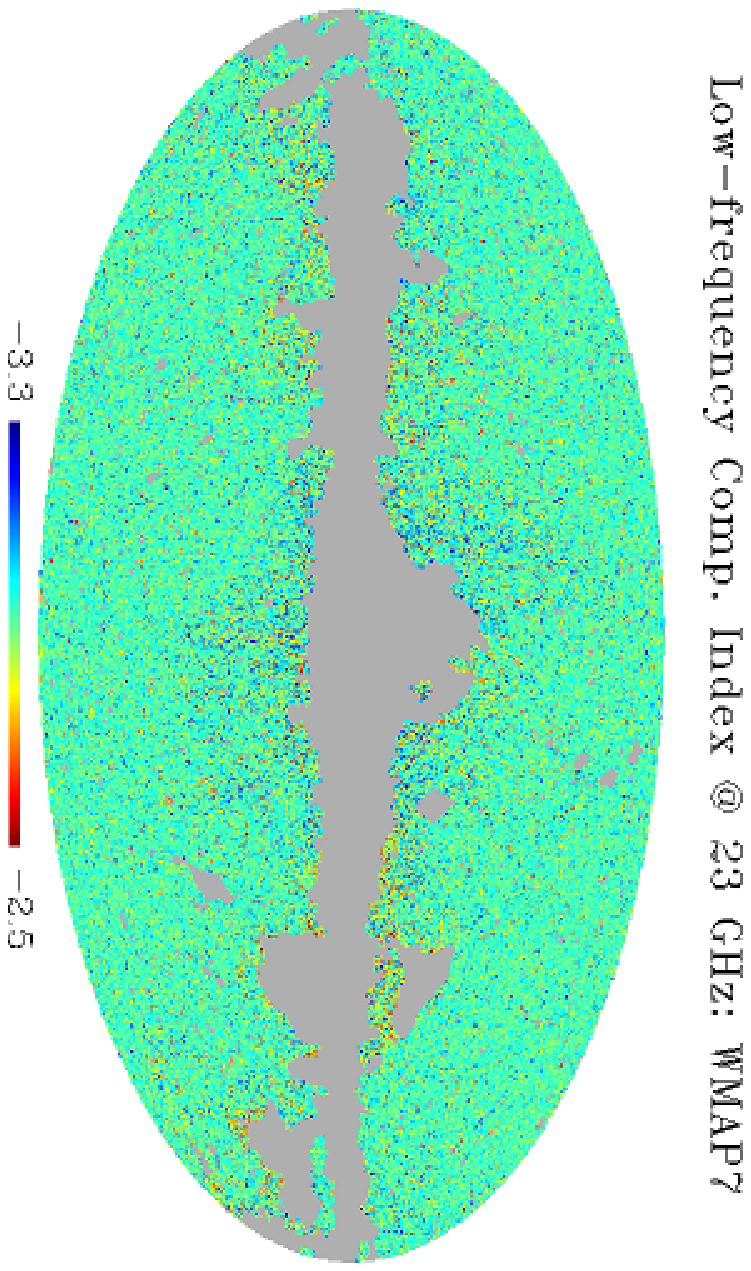}
\incgr[width=.5\columnwidth, angle=90]{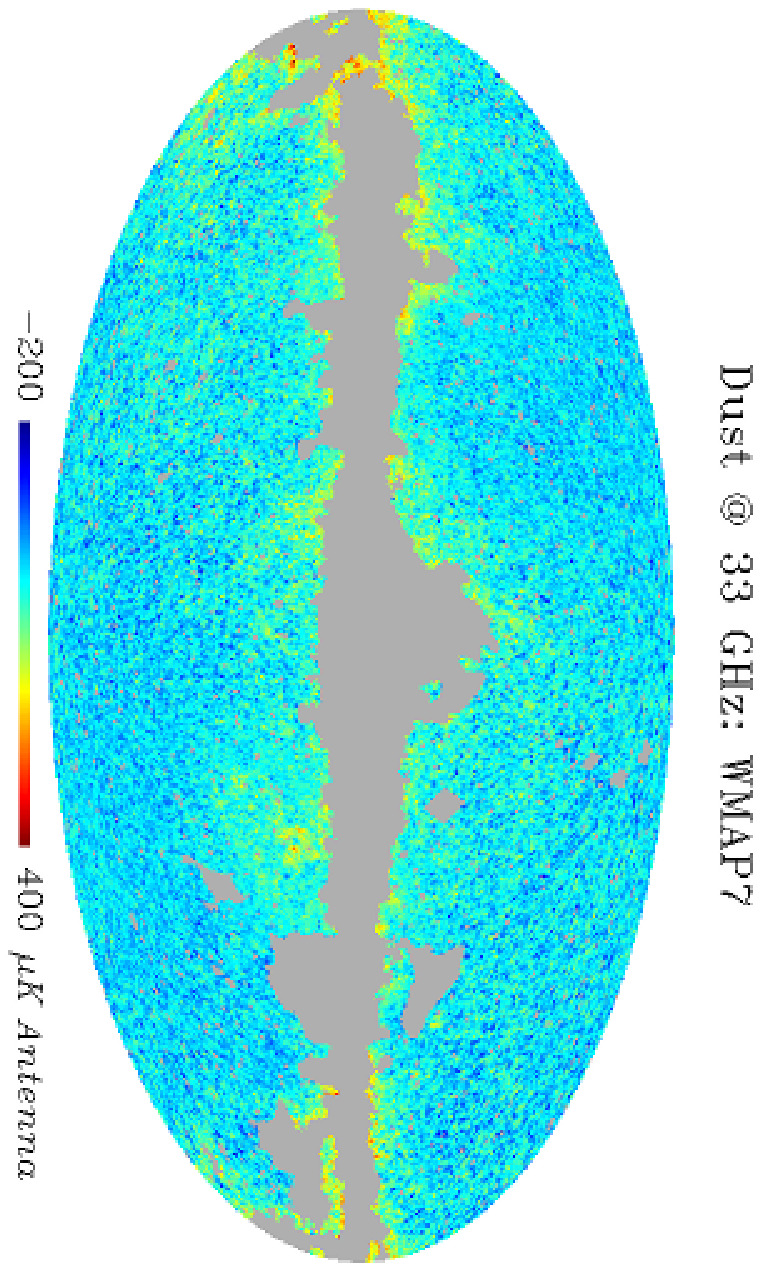}
%\incgr[width=.6\columnwidth, angle=90]{wmap_amp_0_rms.eps}
%\caption{Low-frequency component mean field map and corresponding rms at 23 GHz as resulting from \commander\ run.}
%\label{fig:fg_a1}
%\end{center}
%\end{figure}
%% ---
%\begin{figure}[h]
%\begin{center}
%\incgr[width=.6\columnwidth, angle=90]{wmap_ind_rms.eps}
%\incgr[width=.6\columnwidth, angle=90]{wmap_foreg_ind_23_unmask.eps}
\caption{Low-frequency component mean field amplitude (upper panel) and spectral index (middle panel) map at 23 GHz and dust-correlated emission (lower panel).}
\label{fig:fg_wmap}
\end{center}
\end{figure}

Figure~\ref{fig:fg_wmap} shows the mean field amplitude maps of the thermal/spinning dust, and of the low-frequency component along with its spectral index.
% The associated uncertainties on these maps are shown on the lower panel and are mainly driven by the low frequency component spectral index, as expected \citep{Eriksen2008ApJ676}. 
% ---------------------------------------------------------------------------------------------------
%\begin{figure*}[h]
%\begin{center}
%\incgr[width=.3\textwidth, angle=90]{wmap_foreg_amp_23.eps}
%\incgr[width=.3\textwidth, angle=90]{wmap_amp_0_rms.eps}
%\incgr[width=.3\textwidth, angle=90]{wmap_foreg_amp_94.eps}
%\incgr[width=.3\textwidth, angle=90]{wmap_amp_1_rms.eps}
%\incgr[width=.3\textwidth, angle=90]{wmap_foreg_ind_23.eps}
%\incgr[width=.3\textwidth, angle=90]{wmap_ind_rms.eps}
%\caption{
%  Mean field map (left column) and rms (right column) for hard
%  synchrotron (top), dust (middle) {\bf (GGD: should be shown at 94
%    GHz not 33 GHz.)} and synchrotron spectral index (bottom).
%}
%\label{fig:fg_maps}
%\end{center}
%\end{figure*}
% ---------------------------------------------------------------------------------------------------
At high Galactic latitudes, where the signal-to-noise is very low, the spectral index is consistent with the Gaussian prior, $\beta=-3.0\pm0.3$, $-2.9$ being the mean value, whereas closer to the plane the solution is driven by the data. In high latitude regions of strong synchrotron features, such as Loop I, the spectral index is noticeably softer than close to the plane, where spinning dust and free-free emission become more important. Finally, it is interesting to note that the dust map recovered at 94 GHz is remarkably similar to the FDS prediction.

\begin{figure}[h]
\begin{center}
%\incgr[width=.6\columnwidth, angle=90]{wmap-pl_indx.eps}
%\incgr[width=.6\columnwidth, angle=90]{wmap_indx.eps}
\incgr[width=.5\columnwidth, angle=90]{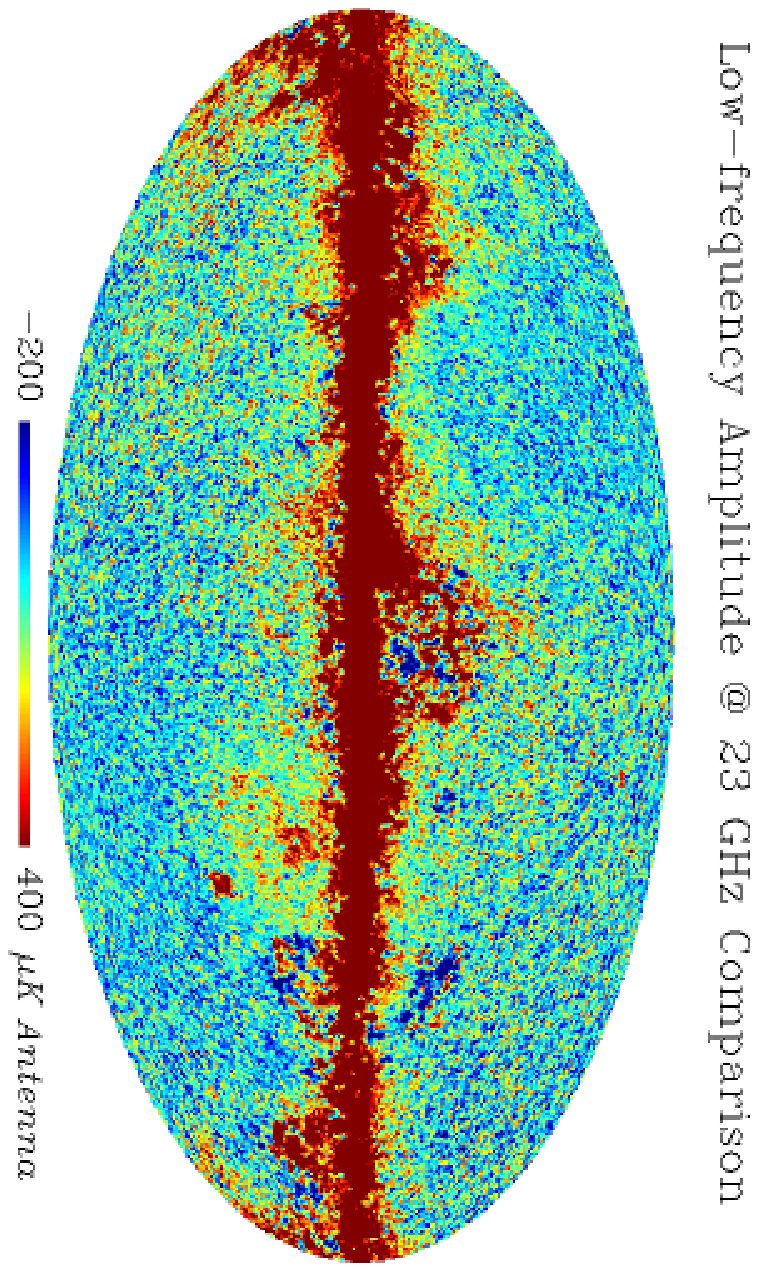}
\incgr[width=.5\columnwidth, angle=90]{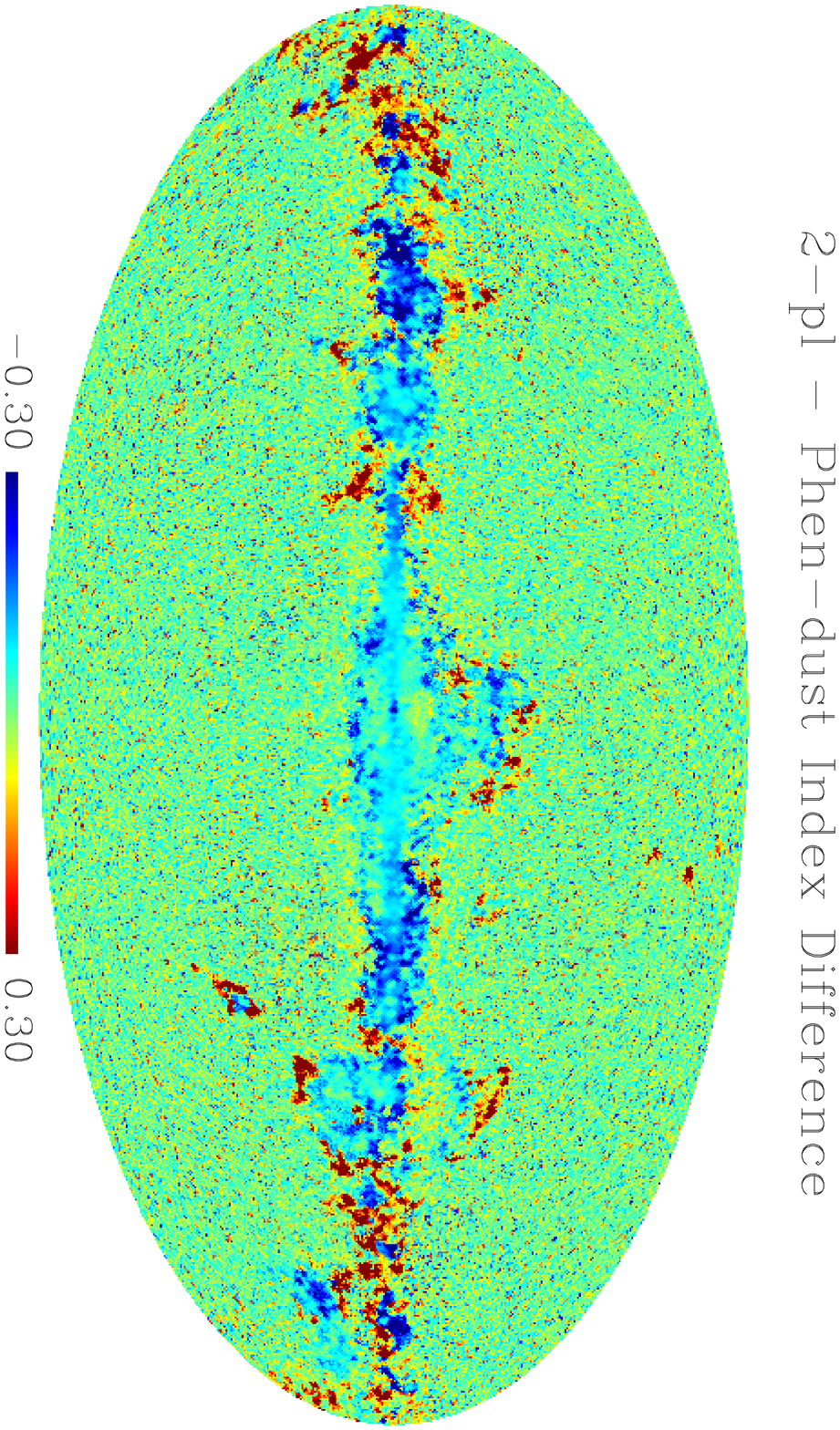}
%\caption{\commander\ posterior mean spectral indices of the low-frequency component derived assuming two different foreground models: two power-laws (upper panel) and power law and phenomenological description of dust (middle panel). The difference is shown in the lower panel}
%\label{fig:indx_comparison}
%\end{center}
%\end{figure}
%% ---------------------------------------------------------------------------------------------------
%\begin{figure}[h]
%\begin{center}
%%\incgr[width=.6\columnwidth, angle=90]{wmap-pl_low-f_amp.eps}
%%\incgr[width=.6\columnwidth, angle=90]{wmap_low-f_amp.eps}
%\incgr[width=.6\columnwidth, angle=90]{wmap_low-f_amp_comparison.eps}
\caption{Difference between two \commander\ posterior mean low-frequency component amplitudes (upper panel) and spectral indices (lower panel) assuming two different foreground models: two power-laws, and a power law with the phenomenological dust SED.}
\label{fig:comparison}
\end{center}
\end{figure}

As noted above, the low-frequency \commander\ solution represents the combination of several different emission mechanisms known to coexist at 23-41 GHz: synchrotron, free-free, and possible spinning dust residuals. This is confirmed by the low-frequency spectral index map that presents shallower values at high Galactic latitude, but clearly emphasizes regions in the Galactic plane with a very steep spectral index, likely to be spinning dust clouds. A comparison between the low-frequency component spectral index and amplitude for the two models is shown in Figure~\ref{fig:comparison}.  The evidence for dust-correlated emission is compelling, implying that our phenomenological dust SED is a good approximation of the sky.
% ---------------------------------------------------------------------------------------------------
% ---------------------------------------------------------------------------------------------------

For the most part, all of these emission mechanisms decrease in intensity above 23 GHz \citep[but see][]{dobler08b,dobler09}, and it is because of the noise in the data that a single power law results in a good $\chi^2$.  In an effort to disentangle these primary sources of emission, we again apply template regression to the \commander\ outputs using the Haslam 408 MHz map, the H$\alpha$ map, and the FDS map as tracers of synchrotron, free-free, and dust (thermal and spinning) emission, respectively. %(see Figure~\ref{fig:templates}).

%% ---------------------------------------------------------------------------------------------------
%\begin{figure}
%\begin{center}
%\incgr[width=.6\columnwidth]{wmap_scatter_0.eps}
%\incgr[width=.6\columnwidth, angle=90]{wmap_scatter_1.eps}
%\caption{Scatter plots of the amplitude solutions we derive. Top: FDS dust map versus out dust; bottom: linear combination of templates versus low frequency amplitude map. Only pixels outside the mask are displayed.}
%\label{fig:wscatter}
%\end{center}
%\end{figure}
%% ---------------------------------------------------------------------------------------------------

The regression coefficients are quoted in Table ~\ref{tab:regression} and indeed they confirm our success: outside the applied mask, dust is positively correlated with the FDS map only, whereas the low-frequency component map can be described as a linear combination of Haslam and H$\alpha$. The goodness of the regression is expressed by the coefficient.

%The goodness of the regression is shown in Figure~\ref{fig:wscatter}, respectively, for dust (upper panel) and synchrotron (lower panel). The straight red line marks $y=x$ in the plane.

\begin{figure}[htb]
\begin{center}
\incgr[width=.5\columnwidth, angle=90]{wmap_foreg_amp_23.eps}
\incgr[width=.5\columnwidth, angle=90]{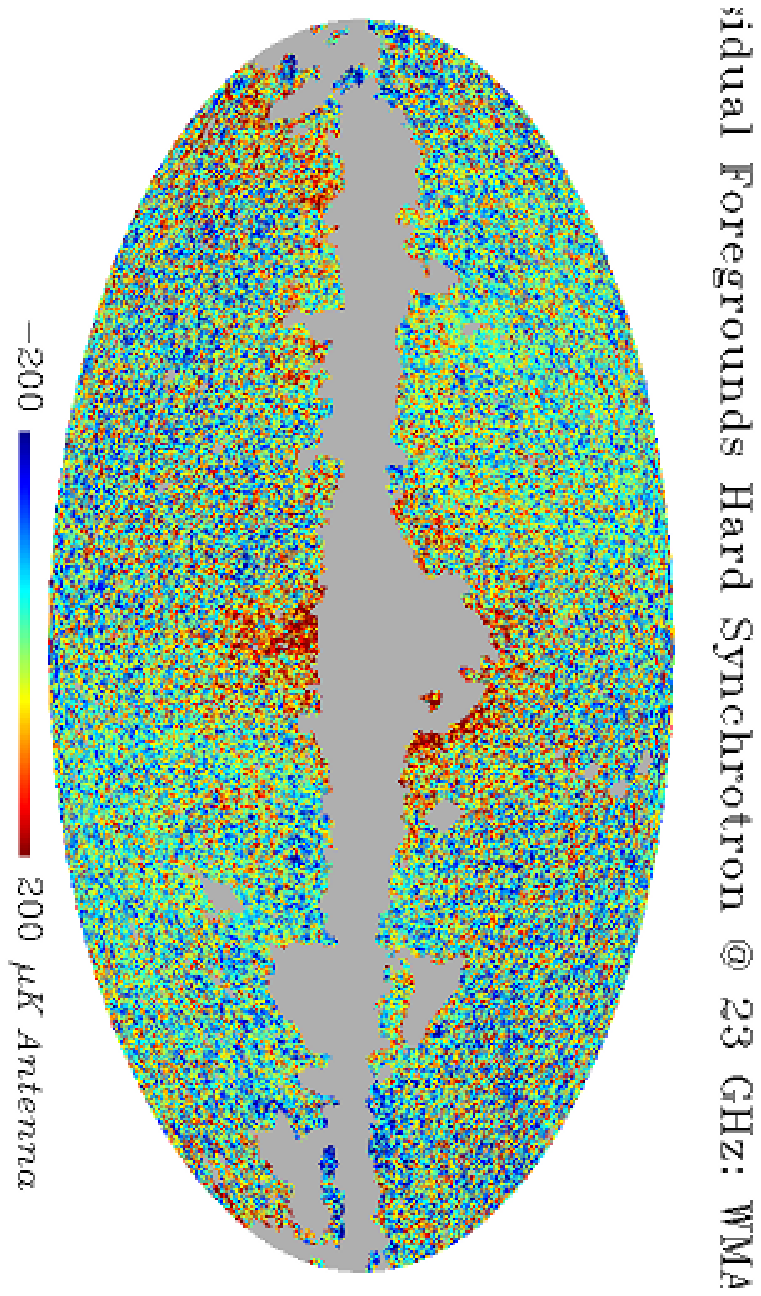}
\incgr[width=.5\columnwidth, angle=90]{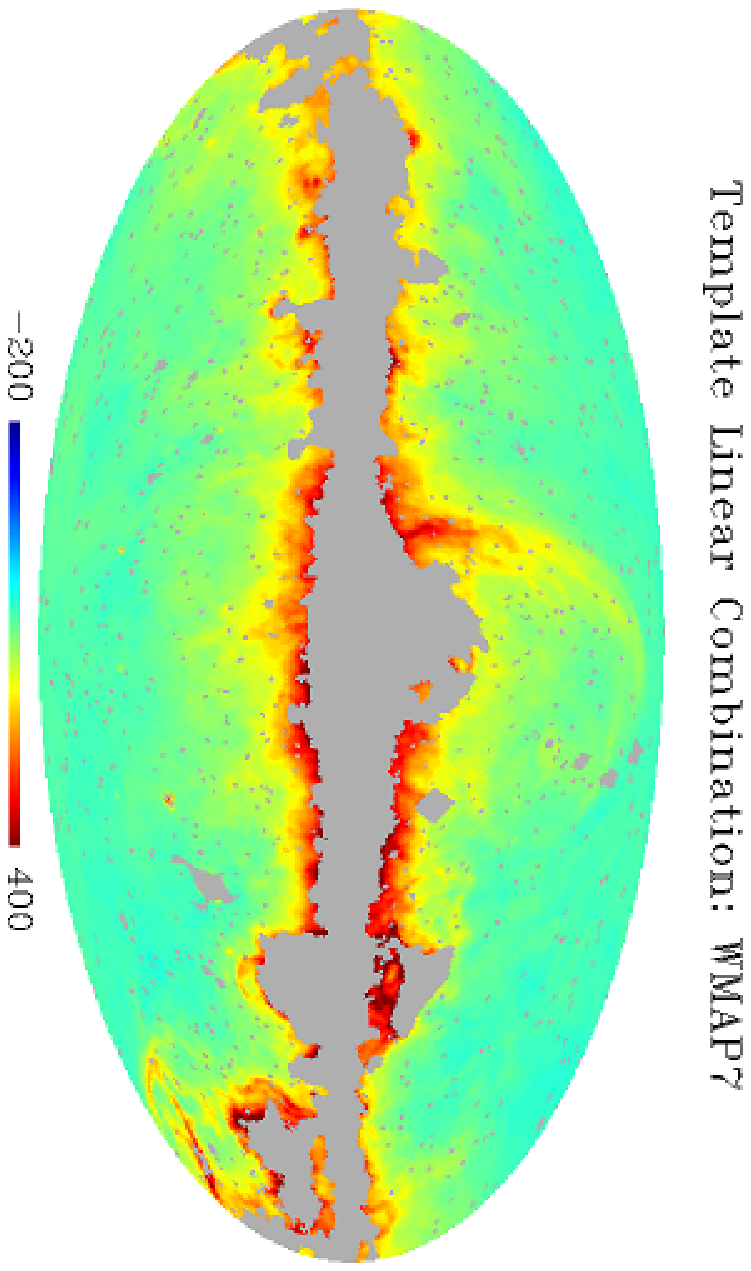}
\caption{WMAP 7-yr: Comparison between low-frequency foreground component amplitude solution and linear fit of the templates: an excess of power around the Galactic center is present.}
\label{fig:wfg_comp_fg1}
\end{center}
\end{figure}

 % ---------------------------------------------------------------------------------------------------
\begin{table}[htb]
\begin{center}
\caption{Regression coefficients for \commander\ foreground amplitudes: {\it Top}: WMAP 7-yr run; {\it Bottom} WMAP 7-yr and
  Haslam. The Galactic emission is decomposed into physical components,
  each of which shows a clear correlation with one template only.\label{tab:regression}}
\begin{tabular}{cccc}
  \tableline\tableline
  {\bf Dataset} & {\bf Haslam} & {\bf H$\alpha$} & {\bf FDS}\\ 
  \tableline
  \multicolumn{4}{c}{\bf WMAP 7-yr} \\ 
  \tableline 
   Low-freq. Comp.  &
  $(3.4\pm3.3)\times10^{-6}$ & $9\pm15$ & $0.6\pm6$ \\ Dust & $(0.1
  \pm 1.3)\times 10^{-6}$ & $-1 \pm 6$ & $2.3\pm2.3$ \\ 
\tableline
  \multicolumn{4}{c}{\bf  Haslam+WMAP 7-yr} \\ 
\tableline
  Soft Sync.  &
  $(5.9\pm0.4)\times10^{-6}$ & $-0.4\pm1.3$ &
  $(-0.6\pm7)\times10^{-1}$ \\ Low-freq. Comp. &
  $(-1.7\pm3.5)\times10^{-6}$ & $8\pm15$ & $0.4\pm6$ \\ Dust & $(-0.3
  \pm 1.2)\times 10^{-6}$ & $-0.25 \pm 6$ & $2.4\pm2.2$ \\
   \tableline
\end{tabular}
\end{center}
\end{table}
% ---------------------------------------------------------------------------------------------------
As in the former run with thermal dust described by a power law, the FDS template shows itself a remarkable tracer of the dust map recovered by \commander, with the residual showing no evidence for large scale deviations from the template. The same is true for the low frequency component with the \emph{significant} exception of the ``haze region'' around the Galactic center. Figure~\ref{fig:wfg_comp_fg1} shows the residuals after removing the template-correlated emission from the \commander\ mean field low-frequency foreground map.  
% ---------------------------------------------------------------------------------------------------
% ---
%\begin{figure}[h]
%\begin{center}
%\incgr[width=.6\columnwidth, angle=90]{wmap_foreg_amp_94.eps}
%\incgr[width=.6\columnwidth, angle=90]{wmap_fit_residuals_2.eps}
%\incgr[width=.6\columnwidth, angle=90]{wmap_fit_fg_2.eps}
%\caption{WMAP 7-yr: Comparison between dust amplitude solution and linear fit of the templates: residuals are compatible with noise.}
%\label{fig:wfg_comp_fg2}
%\end{center}
%\end{figure}
% ---------------------------------------------------------------------------------------------------
In other words, synchrotron emission elsewhere (in Loop I and in the Galactic plane) as well as free-free emission are traced well by our templates while the haze is definitely not, indicating that this large structure is not morphologically correlated with any of the templates. We find that the haze emission is present in the data and contributes significantly to the total emission towards the Galactic center (and particularly in the south).

Our findings are in agreement with what has been described by several authors and in particular by \cite{Haze_Dobler2008ApJ...680.1222D}, who claim the presence of an additional foreground component characterized by a spectral behaviour harder than a synchrotron component, but compatible with neither free-free (because the spectrum is too soft) nor spinning dust emission (because of lack of a thermal feature).  Previous analyses were based on presubtraction of CMB cleaned maps from the data, which is problematic due to the fact that no CMB estimator is completely clean of foregrounds. This leads to a bias in the inferred spectrum of the Galactic emissions. To the extent that it is possible with five bands, we have attempted to reduce this systematic by simultaneously solving for the cleaned CMB map and the foreground maps while also determining a spectral index on the sky, within a Bayesian framework, where the goodness of the fit is controlled pixel-by-pixel through a $\chi^2$ evaluation.

\subsection{WMAP 7-year Data combined with Ancillary Data}

Our results in the previous section indicate that the haze represents either a new component not present in the Haslam map or a variation of the spectral index associated with Haslam 408 MHz data. In order to assess this, we include the 408 MHz Haslam map together with WMAP data and run \commander\ on \emph{six} bands covering a factor of $\sim$250 in frequency. This allows us to fit for an additional component and separate soft synchrotron from other emission mechanisms, the former being described by a power law with fixed spectral index $\beta=-3$, the latter by a spatially varying spectral index.  We expect the soft contribution to be mainly driven by the Haslam map. Our foreground model in this second \commander\ run becomes: CMB, dust described by a spatially constant spectrum which accounts for both spinning and thermal emission, soft synchrotron with fixed spectral response and an additional low frequency component with spatially varying spectral index:
% ----------------------------------------------------------------------------------------------------
\be
\begin{split}
  T_\nu(p) &= M_\nu + \sum_{d=x,y,z} D_{\nu,d}(p) + \Big(\frac{\nu}{\nu_0}\Big)^\beta(p)A_{\rm low freq}(p) \\
   & + \Big(\frac{\nu}{\lambda_0}\Big)^{-3}A_{\rm soft\,synch}(p) + s(\nu) A_{\rm dust}(p),
\end{split}
\label{eq:fg_h}
\ee
% ----------------------------------------------------------------------------------------------------
where $\nu_0=22.8$~GHz and $\lambda_0=408$~MHz, and $s(\nu)$ is the fixed effective dust 
spectrum, normalized to $\mu_0=33$~GHz.

The mean posterior CMB we obtain with this \commander\ model is slightly different compared to the 5-band case showing less power close to the Galactic plane. This suggests that the improved flexibility of our model has resulted in better component separation. The overall $\chi^2$ of these 6-band run does not change, since we add one map and we fit for one more foreground component. It is interesting to notice that we retrieve a value of the monopole in the Haslam map of $\simeq3.2 {\rm K}$.

\begin{figure}[htb]
\begin{center}
\incgr[width=.5\columnwidth, angle=90]{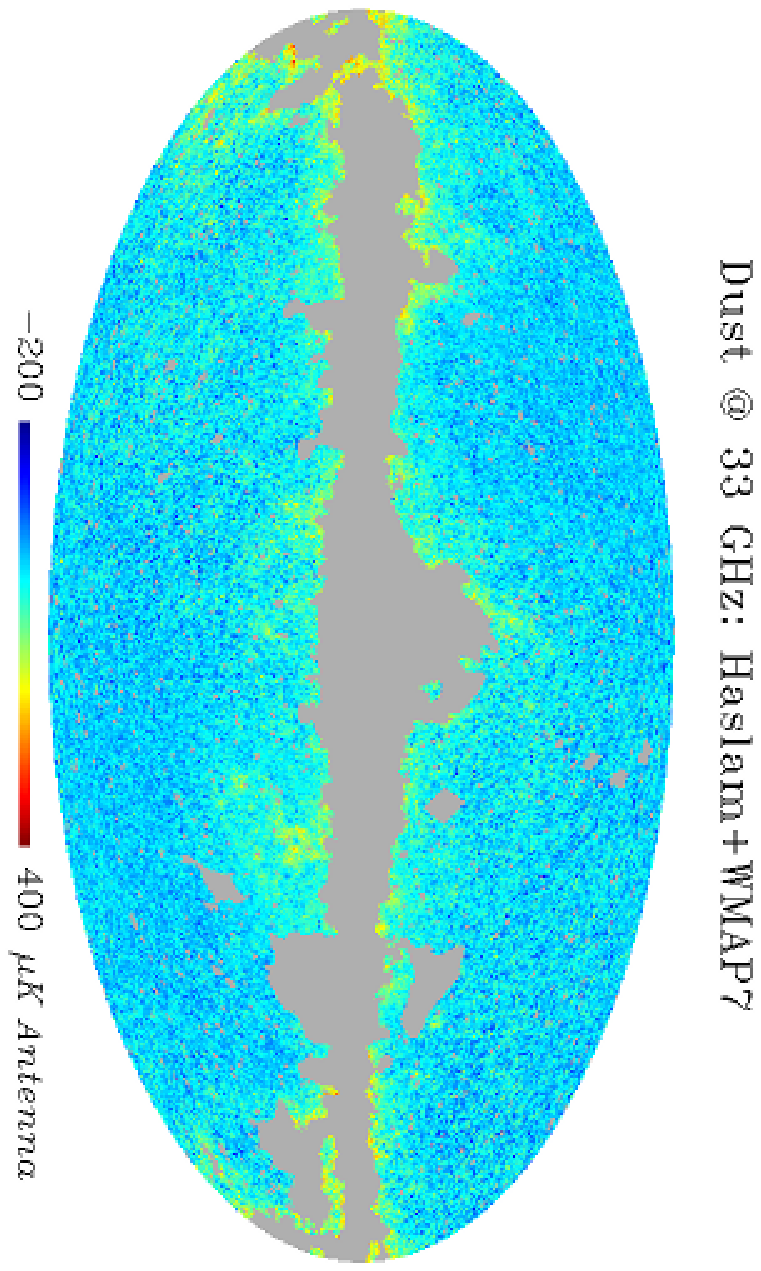}
\incgr[width=.5\columnwidth, angle=90]{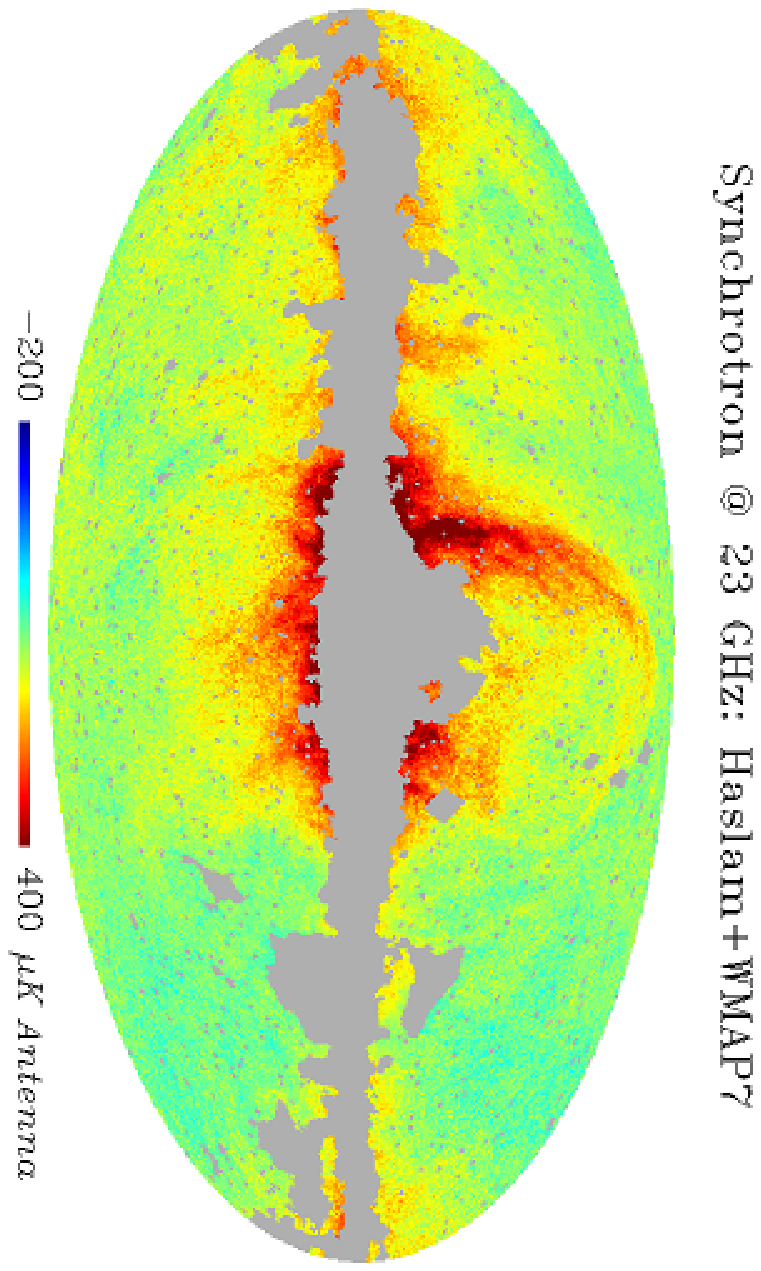}
\incgr[width=.5\columnwidth, angle=90]{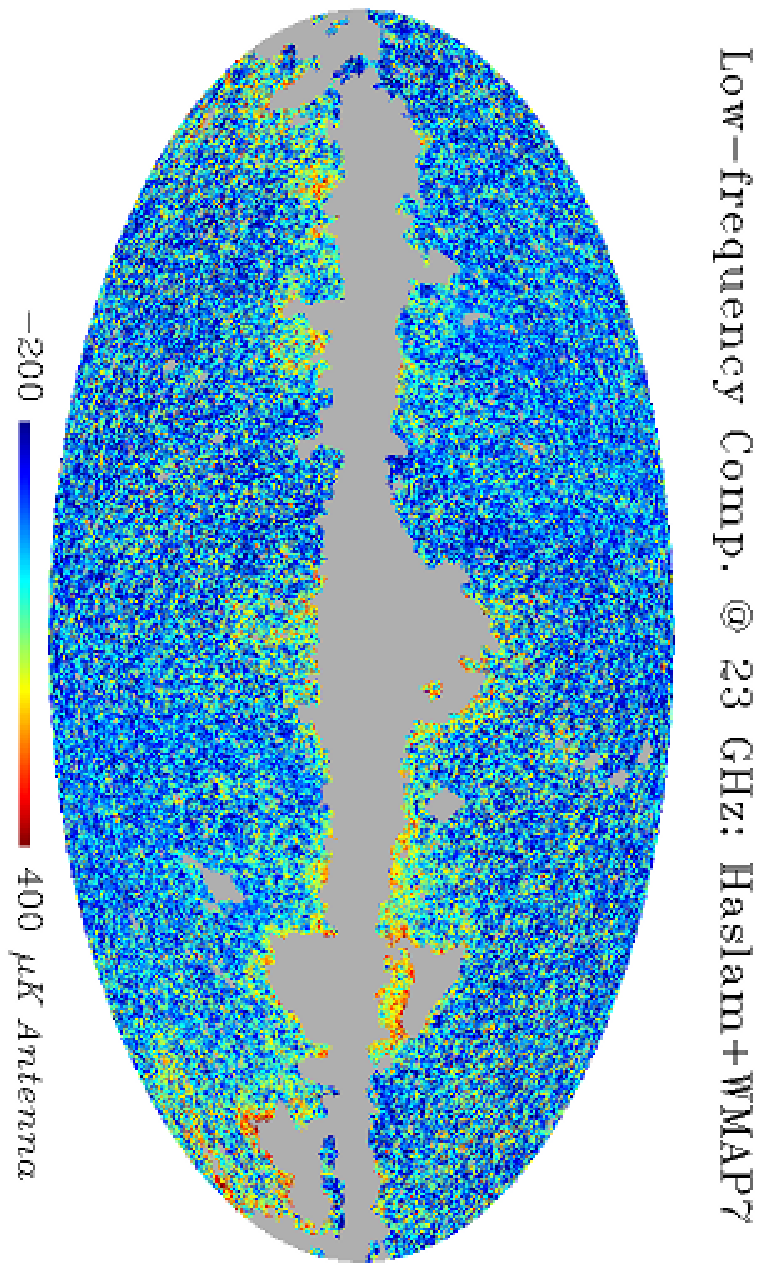}
\caption{WMAP 7-yr+Haslam: \commander\ posterior mean foreground amplitude maps.}
\label{fig:hfg_comp}
\end{center}
\end{figure}

In Figure~\ref{fig:hfg_comp} we report the mean field map of the foreground amplitudes: by visual inspection, the three amplitude maps look strongly correlated with the foreground templates. %(Figure~\ref{fig:templates}). To quantify this we again produce a scatter plot for each pair which is displayed in Figure ~\ref{fig:fg_hcomp}. 
The coefficients of the regression are summarized in Table ~\ref{tab:regression}.

% ---------------------------------------------------------------------------------------------------
%% ---
\begin{figure}[htb]
\begin{center}
\incgr[width=.5\columnwidth, angle=90]{haslam_foreg_amp_23.eps}
\incgr[width=.5\columnwidth, angle=90]{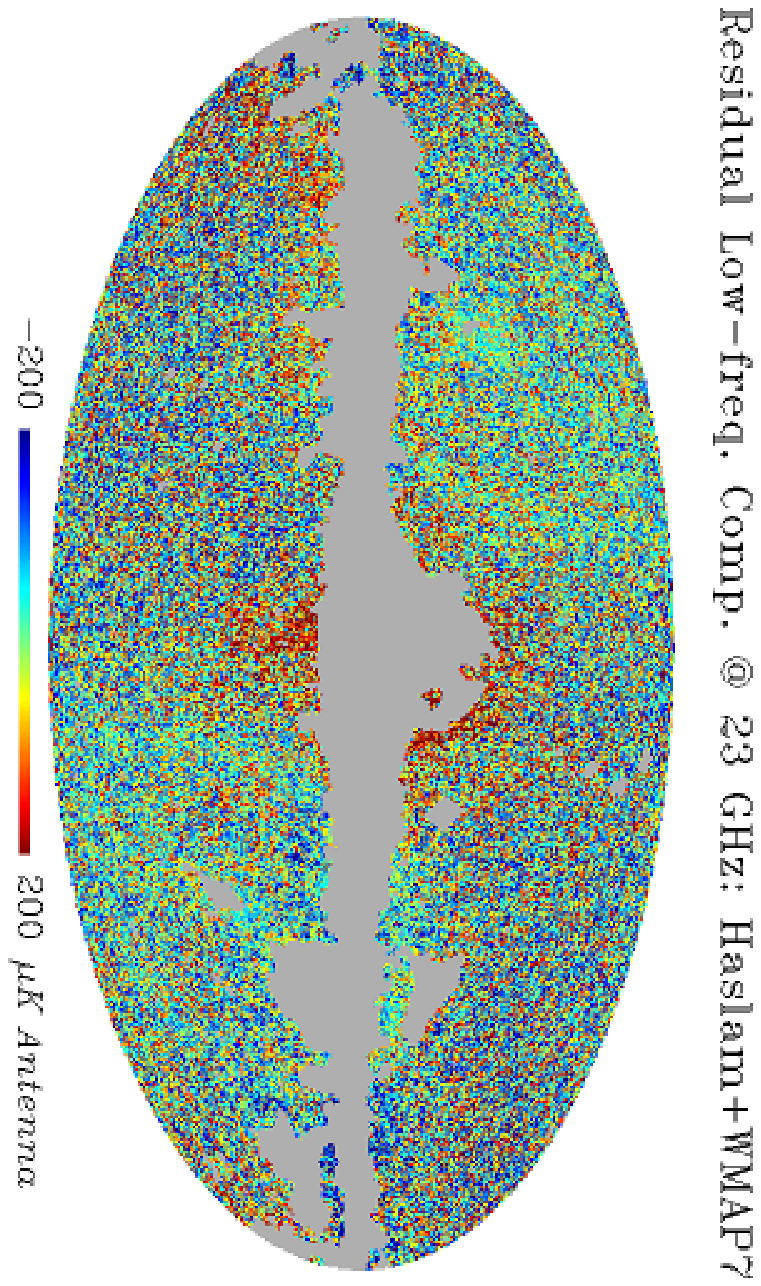}
\incgr[width=.5\columnwidth, angle=90]{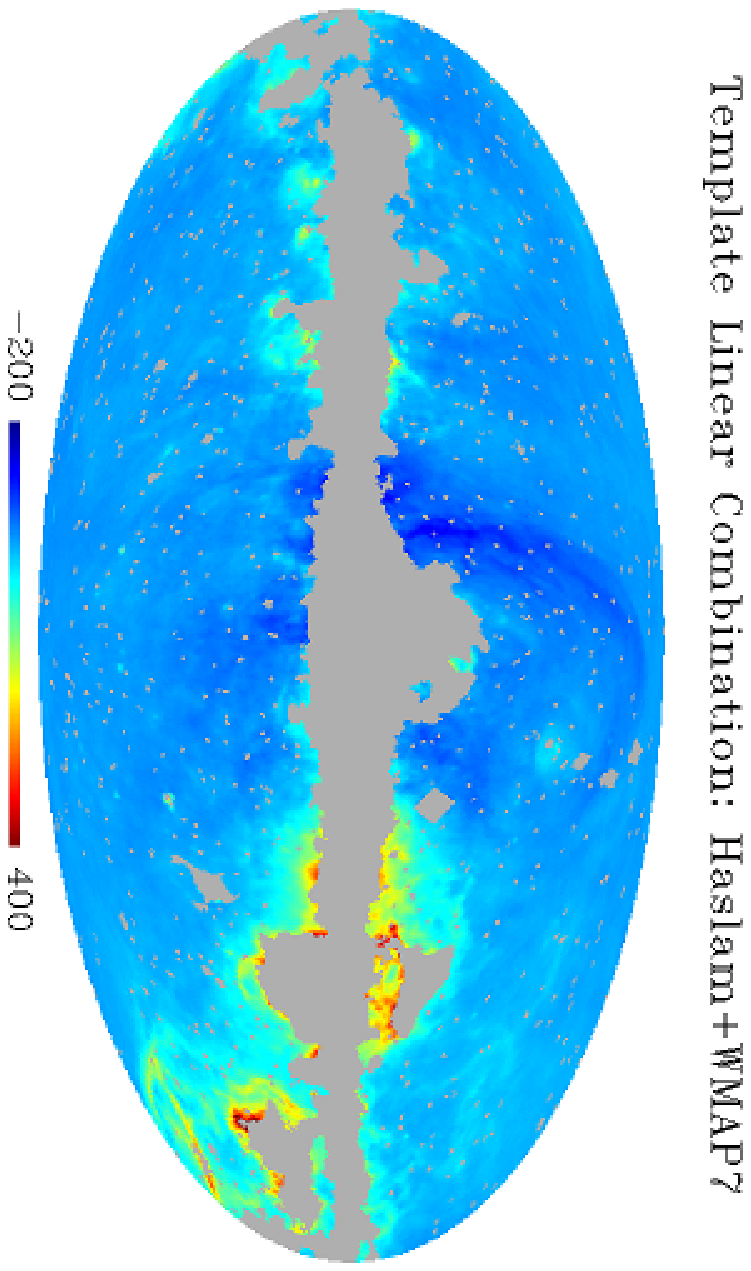}
\caption{WMAP 7-yr+Haslam:
  Comparison between low-frequency component amplitude solution and linear fit of the templates. The Haslam-uncorrelated low frequency component shows an excess of power which is not present in any of the templates.}
\label{fig:hfg_regression}
\end{center}
\end{figure}
% ---------------------------------------------------------------------------------------------------

The residuals obtained after regressing out the templates from the \commander\ solution obtained for this more flexible foreground model are shown in Figure~\ref{fig:hfg_regression}.  The residuals are quite striking and demonstrate that the soft synchrotron component (i.e., emission with a spectral index of $\nu^{-3.0}$) is almost perfectly correlated with Haslam and is not present, while the low frequency component clearly has a $\nu^{-2.15}$ free-free component that is strongly correlated with H$\alpha$ and the haze component, which is softer. This may suggest that the haze is not merely a spectral index variation of emission present in the Haslam map. In fact, the low frequency component is anti-correlated with Haslam (driven primarily by the small negative residuals of Loop I). It is plausible that the prior we set for the spectral index of the soft component, $\beta=-3$, is not steep enough to characterize the Spur ($\beta=-3.05$), although at high Galactic latitudes a flatter behavior is measured ($\beta=-2.9$). Driven by this consideration, we ran \commander\ setting the spectral index of the synchrotron component to $\beta=-2.9$ and $\beta=-3.05$: the former resulted into more negative residuals in the synchrotron component, the latter reduced without completely removing them.

The haze is a relatively weak residual compared to the soft synchrotron component, thus a spectral index change is required to generate it \citep{Mertsch2010JCAP}. If this were interpreted as a simple variation of the spectral index of Haslam rather than a separate component not visible in Haslam, then this would imply that the spectral index is harder from 23-60 GHz than from 0.408-23 GHz. i.e., the spectrum would be concave up. However, cooling of electrons always generates spectra which are concave down, because energy losses of cosmic-ray electrons are proportional to the electron energy squared.  Thus, power-law injection of electrons results in concave down spectra as they cool. In other words, a different population of electrons with a harder spectral index must be present: a simple variation of the spectral index of the same electron population is not what we measured, since the amplitude of the soft component that we retrieved is strongly correlated with the Haslam map at 408 MHz. It is true that the assumption of a power-law for the synchrotron component can be too simplistic, and a (negative) curvature might be present, as a result of a break in the spectral behaviour. However, given the number of frequencies available, sparse between 0.408 and 23 GHz and relatively narrow range of the WMAP data, allowing an extra parameter to be determined is not possible.

Again, the dust component is only strongly correlated with the FDS map, even more so than in the previous model, again likely indicating a better foreground separation.

Lastly, we point out that the spectral index map of the haze clearly shows that
its spectral behaviour is not compatible with free-free, but it is driven by the prior ($\beta=-2.5\pm0.3$). Unfortunately, this may suggest that the regularizing noise we added is too high to enable us to be definitive about the spectral index of the haze.

\begin{figure}[htb]
\begin{center}
\incgr[width=.5\columnwidth, angle=90]{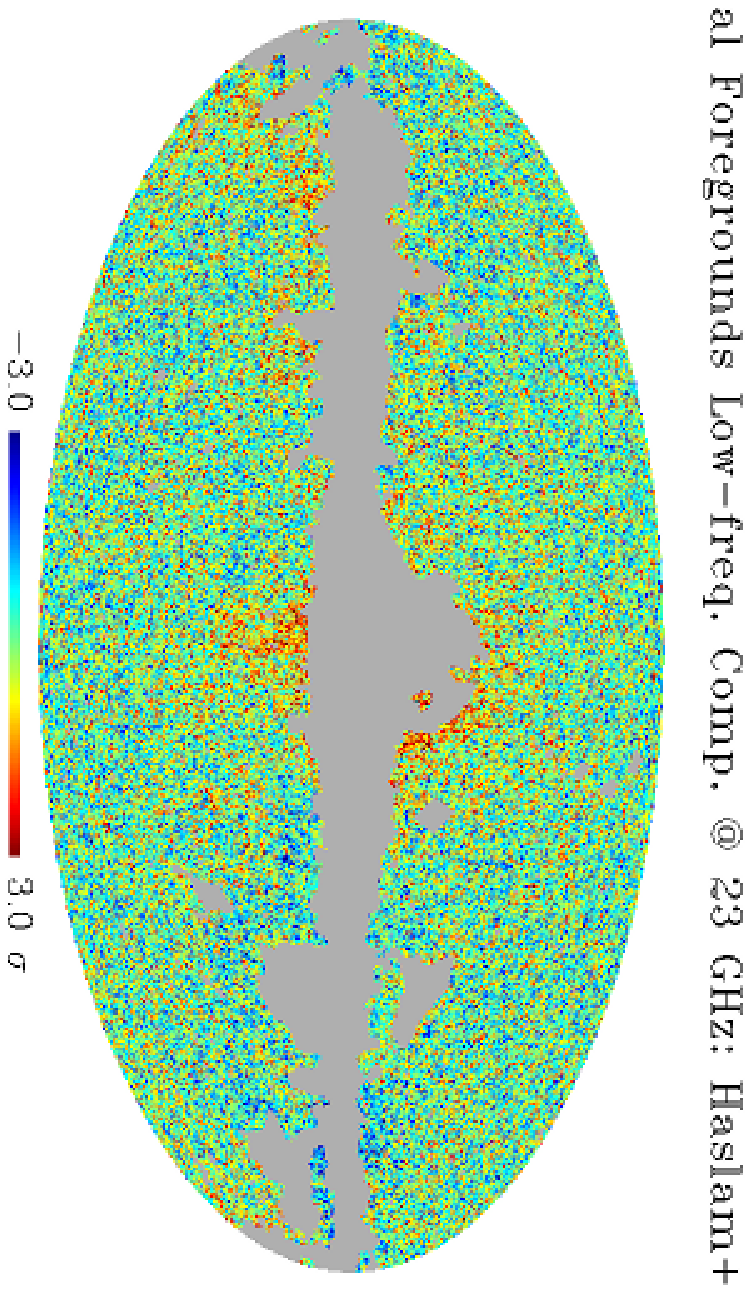}
\incgr[width=.5\columnwidth, angle=90]{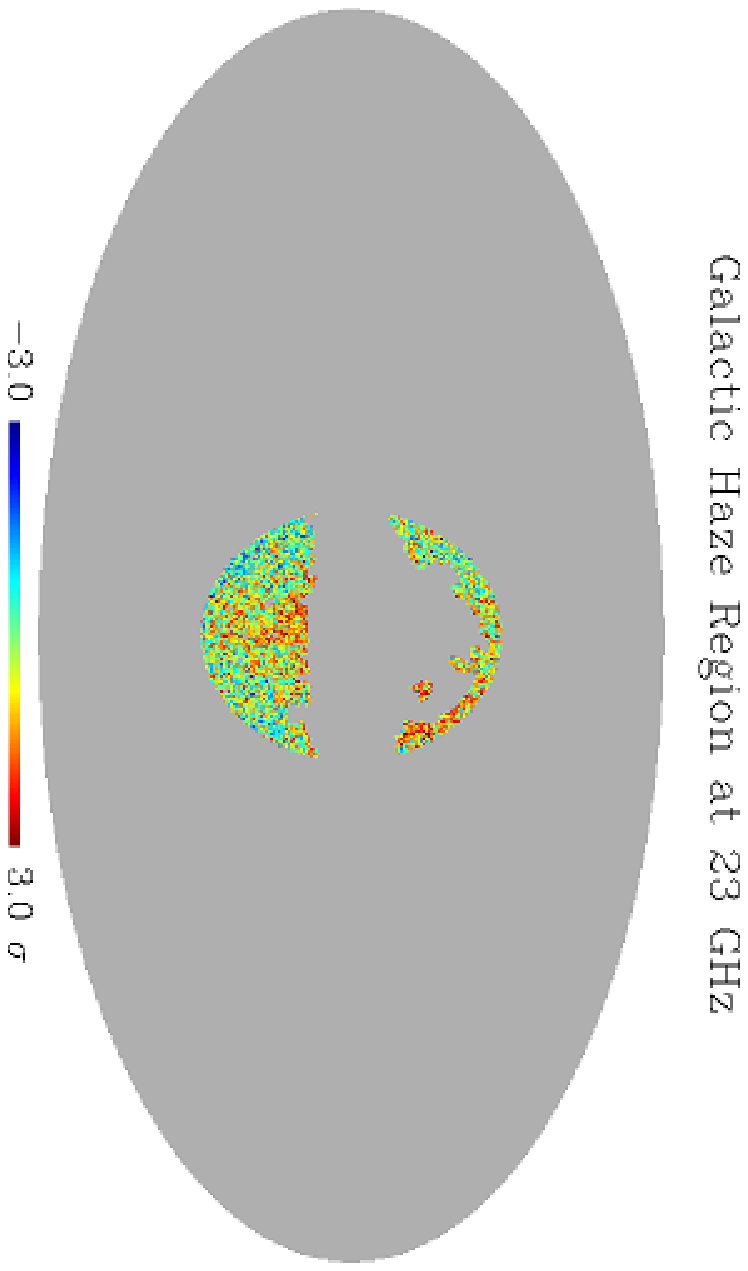}
\caption{Significance of the residuals for the soft synchrotron component (top panel) and pixel selected for the study of the correlation on a small region around the Galactic center.}
\label{fig:haze_spot}
\end{center}
\end{figure}

Up until now we have explored template fits in which the spectra are taken to be uniform across the sky.  This is obviously not the case, and so to demonstrate the effects of this assumption we break the sky into regions and perform individual fits on these.  Examples of this technique can be found in \cite{Hildebrandt2007,Haze_Dobler2008ApJ...680.1222D,dobler11b,Ghosh2012MNRAS}. In practice, it means computing the correlation coefficients given in Eq.~\ref{eq:regression_coef} on a subset of pixels for the chosen regions. Here we will not perform such a detailed analysis, but restrict ourselves only to the Haze region around the Galactic center. The idea behind this approach is that if the Haze corresponds to a region of global variation of the spectral index of the synchrotron component, the correlation between our residual map and the Haslam map computed on the Haze pixels would be high. We can select the brightest pixels based on the rms map we derived from \commander: in Figure \ref{fig:haze_spot} we show the significance of the residuals and the pixels above $3\sigma$ which we chose within 36 degrees of the Galactic center.
% ---------------------------------------------------------------------------------------------------
% ---------------------------------------------------------------------------------------------------
The correlation coefficients we obtained for the Haze are $3.3\pm3.5 \times 10^{-6}$, $0.1\pm 21$ and $3\pm7$ respectively for Haslam, H$\alpha$ and FDS templates, showing no evidence for strong correlation. It is interesting to notice that while the amplitudes for free-free and dust emission remain stable with the threshold, this is not the case of the Haslam map, whose correlation decreases with the lower bound of the chosen significance cut. We tried 2.0, 2.5, 3.0, 3.5 $\sigma$ finding more and more correlation: this is not surprising since the number of selected pixels decreases and the statistics degrades. This is a due to the regularizing noise.

This approach will be more effective when larger frequency coverage becomes be available.

\begin{figure}[htb]
\begin{center}
\incgr[width=.5\columnwidth]{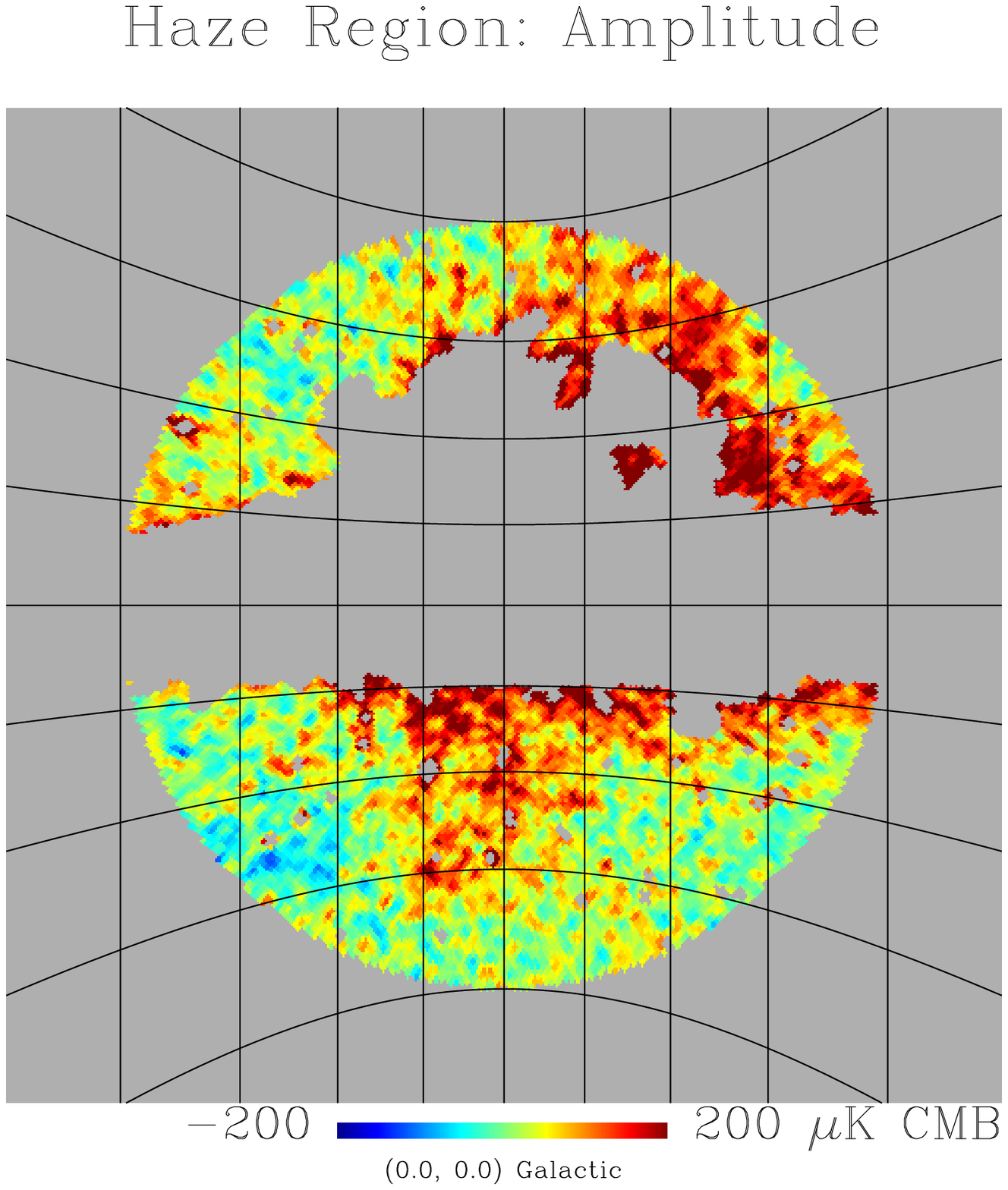}
\incgr[width=.5\columnwidth]{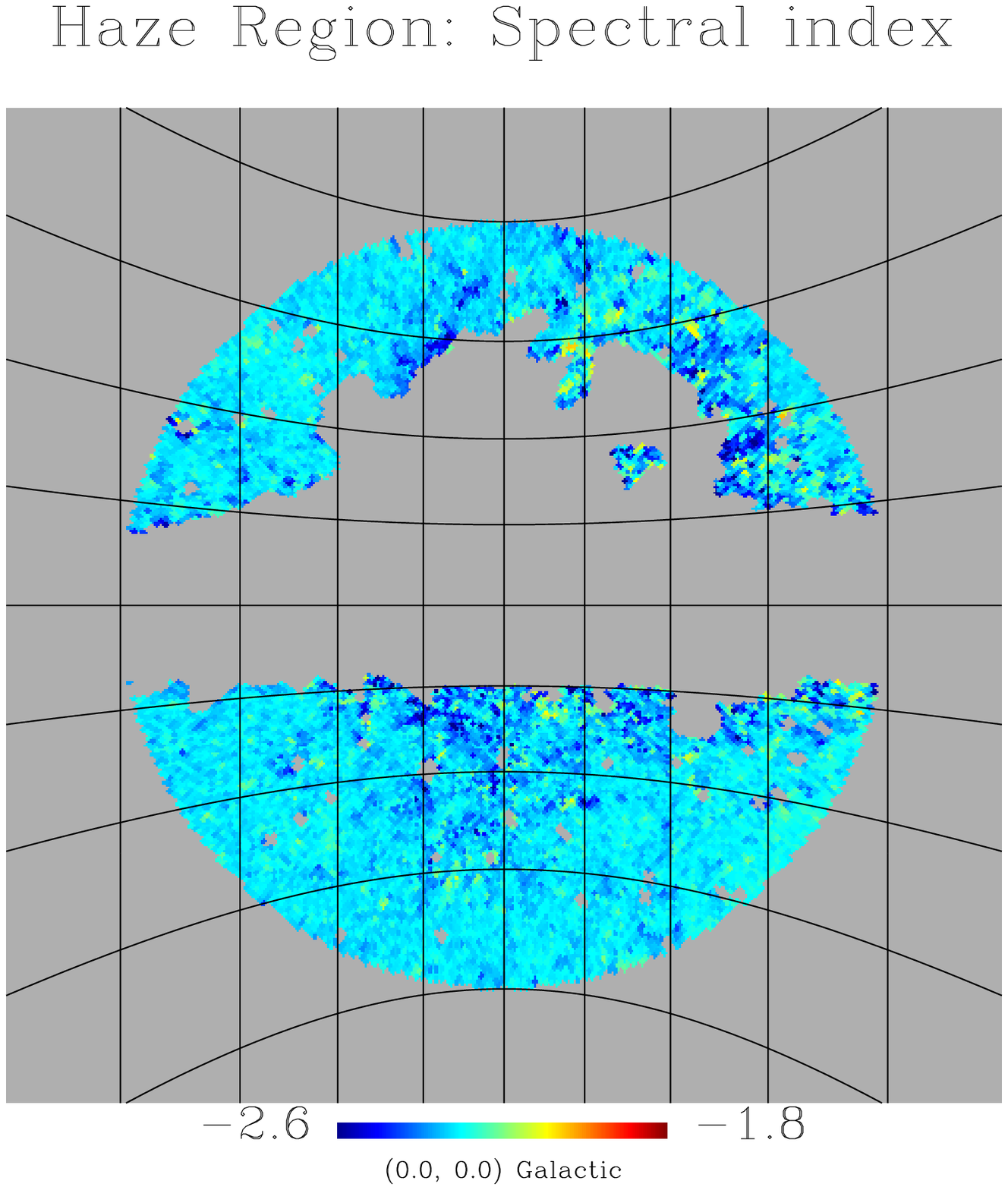}
%\incgr[width=.6\columnwidth]{haslam_haze_pix.eps}
\caption{Haze amplitude and spectral index obtained after regressing \commander\ posterior mean amplitudes found without adding regularization noise.}
\label{fig:haze_pars}
\end{center}
\end{figure}

To improve the constraints on the spectral index of the hard low-frequency component, we re-ran commander without adding noise to the input maps and adopting a different prior, $\beta=-2.15\pm0.3$, with the idea that the higher signal-to-noise ratio would allow the data to drive the solution. (In order to speed converge, we also dropped the sample of $C_l$s).  Except for regions with  strong free-free emission, we now observe a mean spectral index of $-2.3\pm0.27$ for $|b_{\rm Gal}|>30^\circ$, whereas we find $-2.4\pm0.22$ for the haze region. Though not conclusive, this does support the hypothesis of a harder spectral index for the haze component proposed by \cite{dobler11b}. The regression coefficients for this case are  $0.4\pm2\times10^{-7}$, $9\pm7$ and $0.08\pm2$ for the Haslam, H$\alpha$ and FDS template respectively, which show a clear improvement of the component identification. Figure \ref{fig:haze_pars} shows the posterior mean amplitude and spectral index of the low-frequency component with a 35$^\circ$ latitude around the Galactic center.
% ---------------------------------------------------------------------------------------------------
% ---------------------------------------------------------------------------------------------------

\section{Conclusions}
\label{sec:conclusions}
We applied the Gibbs sampling technique implemented in the \commander\ software to WMAP 7-year data aiming at characterizing, simultaneously, CMB (map and angular power spectrum) and foreground emission. Our analysis improves previous work in a number of aspects. We pushed the analysis to higher angular resolution, \~1\deg, which enabled us to directly compare the foreground solutions to available templates, resulting in excellent agreement. This represents a success on its own but also confirms the power of the Bayesian approach, which is likely to perform better when the larger Planck data set becomes available.

Higher angular resolution requires better modeling of the noise in the maps after the smoothing procedure: we added scan-modulated regularizing noise estimated via Monte Carlo simulations. Despite the limited number of frequencies available, we did not directly use foreground templates in the Gibbs sampling run, but rather let the parameters of the model vary across the sky. This allowed us to distinguish regions in the sky where a specific foreground mechanism dominates on the basis of the its spectral behaviour. The presence of a strongly dust correlated emission at low frequencies, explained by invoking spinning dust grains, emerges not only through the amplitude map but also in the spatial variation of the spectral index of the low frequency component, which mainly results from a mixture of synchrotron emission and dust-correlated emission. The number of available frequencies still limits us, since it forces us to use a constant spectral index for the dust.

Regressing \commander\ solutions against foreground templates is a complementary way to disentangle the various emission mechanisms. It is by applying this procedure that we confirm the presence of excess of signal localized around the Galactic center, the microwave haze. A hint of such a emission has been found in the Fermi data \citep{Haze_Dobler2010ApJ...717..825D,dobler11}. The study of this component has stimulated a rich production, both pointing out possible systematics in the applied regression procedures \citep[,see for instance][and]{Linden2010ApJ,Mertsch2010JCAP} and proposing possible physical mechanisms \citep{Haze_Bottino2010MNRAS.402..207B,dobler11,guo11,Guo2011,biermann10,Crocker2011,Mertsch2011}. The WMAP team remains skeptical about the presence of the haze \citep{gold11}, their main counter argument being the lack of evidence for such a signal in the polarization maps \citep[though][showed that such a signal would likely not be detectable in the WMAP data given the noise]{dobler11b}.  Regarding the spatial correspondence between the WMAP haze and the Fermi
haze/bubbles, we note that the increased noise in our output maps makes a direct comparison difficult.  Comparison with earlier WMAP
data was presented by \cite{Haze_Dobler2010ApJ...717..825D,Su2010}, and more recently with WMAP 7-year data by \cite{dobler11b}.  We find, as in those studies, that the morphological correlation between the two is reasonable at $|b_{\rm Gal}|<30^\circ$, where the signal in microwaves is most unambiguous.

We emphasize that our analysis overcomes the problem of CMB subtraction and the circularity argument arising from the removal of an internal linear combination of the channels used in previous analyses. Moreover, we addressed the issue of possible contamination of spinning dust as an explanation of the excess signal by including a correlation between the low frequency emission and thermal dust. We computed the spectral energy distribution of the dust as described by the FDS model, performing a template fitting of the WMAP channels as a preprocessing step. We input the resulting SED to \commander, solving for an amplitude map, together with a low-frequency component. The resulting dust map is highly correlated with the FDS model, and the low-frequency component results from a sum of synchrotron and free-free emission only. When regressing this map against foreground templates, the Galactic Haze is still present. One criticism to our approach could be the simplicity of the spinning dust model, but since one of our goals was to be independent of external data sets within a completely Bayesian framework, adding more complexity to the dust model is not possible given the available degrees of freedom.

We also argued that the haze is not an artifact of the spectral variation of the synchrotron component across the sky. To this end, we included the Haslam map at 408 MHz in the \commander\ run and expanding the foreground model: dust, combing both thermal and spinning contribution, synchrotron emission with fixed spectral response $\beta=-3$ and an additional component with a free spectral index, which would describe free-free, and any other contribution.  We found that we could completely separate synchrotron and dust emission, and that we were left with an amplitude map that can be easily characterized by its spectral index. Free-free emission clearly shows up in the Galactic plane, together with the Haze which seems to have a harder spectrum than the synchrotron component. Unfortunately, outside the Galactic plane, the index map is quite noisy and the haze does not have a distinctive signature compared to the rest of the sky, although it is definitely not compatible with free-free emission.

Our analysis has improved previous knowledge of the foreground and shed new light on the nature of the Galactic Haze. We addressed at least three criticisms often attached to haze studies: coherent CMB removal, contamination from spinning dust, and the spatial variation of the synchrotron component spectral index. The excess of signal seems to be stable with respect to them. However, we have made assumptions on the dust spectral behaviour, forced by the number of frequencies available, and  we did not rule out the possibility of the presence of curvature in the synchrotron component spectral index or a model with a broken power law. %This, together with the use of a more complex foreground model, will be lifted when Planck data will be public, allowing a detailed description of the thermal dust, through a gray-body model with spatially varying emissivity and temperature.
Planck data will address these remaining issues and enable a more complete foreground model.

\section*{Acknowledgements}
Many of the results in this paper have been derived using the HEALPix \citep{Gorski2005Healpix} software and analysis package. We acknowledge use of the Legacy Archive for Microwave Background Data Analysis (LAMBDA). Support for LAMBDA is provided by the NASA Office of Space Science. JGB gratefully acknowledges support by the {\em Institut Universitaire de France}. GD has been supported by the Harvey L.~Karp discovery award.

\bibliography{Biblio}

\begin{thebibliography}{53}
\expandafter\ifx\csname natexlab\endcsname\relax\def\natexlab#1{#1}\fi

\bibitem[{Bennett {et~al.}(2003)}]{Bennett:2003ca}
Bennett, C. {et~al.} 2003, \apjs, 148, 97

\bibitem[{{Biermann} {et~al.}(2010){Biermann}, {Becker}, {Caceres}, {Meli},
  {Seo}, \& {Stanev}}]{biermann10}
{Biermann}, P.~L., {Becker}, J.~K., {Caceres}, G. {et~al.} 2010, \apjl, 710, L53

\bibitem[{{Bottino} {et~al.}(2010){Bottino}, {Banday}, \&
  {Maino}}]{Haze_Bottino2010MNRAS.402..207B}
{Bottino}, M., {Banday}, A.~J., \& {Maino}, D. 2010, \mnras, 402, 207

\bibitem[{{Casassus} {et~al.}(2008){Casassus}, {Dickinson}, {Cleary},
  {Paladini}, {Etxaluze}, {Lim}, {White}, {Burton}, {Indermuehle}, {Stahl}, \&
  {Roche}}]{casassus08}
{Casassus}, S., {Dickinson}, C., {Cleary}, K. {et~al.} 2008, \mnras, 391, 1075

\bibitem[{{Cholis} {et~al.}(2009){Cholis}, {Dobler}, {Finkbeiner},
  {Goodenough}, \& {Weiner}}]{Haze_Cholis2009PhRvD..80l3518C}
{Cholis}, I., {Dobler}, G., {Finkbeiner}, D.~P., {Goodenough}, L., \& {Weiner},
  N. 2009, \prd, 80, 123518

\bibitem[{{Chu} {et~al.}(2005){Chu}, {Eriksen}, {Knox}, {G{\'o}rski}, {Jewell},
  {Larson}, {O'Dwyer}, \& {Wandelt}}]{Chu2005PhRvD71}
{Chu}, M., {Eriksen}, H.~K., {Knox}, L. {et~al.} 2005, \prd, 71, 103002

\bibitem[{{Crocker} \& {Aharonian}(2011)}]{Crocker2011}
{Crocker}, R.~M. \& {Aharonian}, F. 2011, Physical Review Letters, 106, 101102

\bibitem[{{Crocker} {et~al.}(2011){Crocker}, {Jones}, {Aharonian}, {Law},
  {Melia}, {Oka}, \& {Ott}}]{crocker11}
{Crocker}, R.~M., {Jones}, D.~I., {Aharonian}, F. {et~al.} 2011, \mnras, 413, 763

\bibitem[{{Cumberbatch} {et~al.}(2009){Cumberbatch}, {Zuntz}, {Kamfjord
  Eriksen}, \& {Silk}}]{Haze_Cumberbatch2009arXiv0902.0039C}
{Cumberbatch}, D.~T., {Zuntz}, J., {Kamfjord Eriksen}, H.~K., \& {Silk}, J.
  2009, arXiv:0902.0039

\bibitem[{{Dickinson} {et~al.}(2009){Dickinson}, {Eriksen}, {Banday}, {Jewell},
  {G{\'o}rski}, {Huey}, {Lawrence}, {O'Dwyer}, \&
  {Wandelt}}]{Dickinson2009ApJ705}
{Dickinson}, C., {Eriksen}, H.~K., {Banday}, A.~J. {et~al.} 2009, \apj, 705, 1607

\bibitem[{{Dobler}(2011)}]{dobler11b}
{Dobler}, G. 2012, \apj, 750, 17

\bibitem[{{Dobler} {et~al.}(2011){Dobler}, {Cholis}, \& {Weiner}}]{dobler11}
{Dobler}, G., {Cholis}, I., \& {Weiner}, N. 2011, \apj, 741, 25

\bibitem[{{Dobler} {et~al.}(2009){Dobler}, {Draine}, \&
  {Finkbeiner}}]{dobler09}
{Dobler}, G., {Draine}, B., \& {Finkbeiner}, D.~P. 2009, \apj, 699, 1374

\bibitem[{{Dobler} \&
  {Finkbeiner}(2008{\natexlab{a}})}]{Haze_Dobler2008ApJ...680.1222D}
{Dobler}, G. \& {Finkbeiner}, D.~P. 2008{\natexlab{a}}, \apj, 680, 1222

\bibitem[{{Dobler} \& {Finkbeiner}(2008{\natexlab{b}})}]{dobler08b}
---. 2008{\natexlab{b}}, \apj, 680, 1235

\bibitem[{{Dobler} {et~al.}(2010){Dobler}, {Finkbeiner}, {Cholis}, {Slatyer},
  \& {Weiner}}]{Haze_Dobler2010ApJ...717..825D}
{Dobler}, G., {Finkbeiner}, D.~P., {Cholis}, I., {Slatyer}, T., \& {Weiner}, N.
  2010, \apj, 717, 825

\bibitem[{{Draine} \& {Lazarian}(1998)}]{draine1998}
{Draine}, B.~T. \& {Lazarian}, A. 1998, \apj, 508, 157

\bibitem[{{Erickson}(1957)}]{Erickson1957}
{Erickson}, W.~C. 1957, \apj, 126, 480

\bibitem[{{Eriksen} {et~al.}(2008{\natexlab{a}}){Eriksen}, {Dickinson},
  {Jewell}, {Banday}, {G{\'o}rski}, \& {Lawrence}}]{Eriksen2008ApJ672L}
{Eriksen}, H.~K., {Dickinson}, C., {Jewell}, J.~B. {et~al.} 2008{\natexlab{a}}, \apjl, 672, L87

\bibitem[{{Eriksen} {et~al.}(2006){Eriksen}, {Dickinson}, {Lawrence},
  {Baccigalupi}, {Banday}, {G{\'o}rski}, {Hansen}, {Lilje}, {Pierpaoli},
  {Seiffert}, {Smith}, \& {Vanderlinde}}]{Eriksen2006ApJ641}
{Eriksen}, H.~K., {Dickinson}, C., {Lawrence}, C.~R. {et~al.} 2006, \apj, 641, 665

\bibitem[{{Eriksen} {et~al.}(2007{\natexlab{a}}){Eriksen}, {Huey}, {Banday},
  {G{\'o}rski}, {Jewell}, {O'Dwyer}, \& {Wandelt}}]{Eriksen2007ApJ665L}
{Eriksen}, H.~K., {Huey}, G., {Banday}, A.~J. {et~al.} 2007{\natexlab{a}}, \apjl, 665,
  L1

\bibitem[{{Eriksen} {et~al.}(2007{\natexlab{b}}){Eriksen}, {Huey}, {Saha},
  {Hansen}, {Dick}, {Banday}, {G{\'o}rski}, {Jain}, {Jewell}, {Knox}, {Larson},
  {O'Dwyer}, {Souradeep}, \& {Wandelt}}]{Eriksen2007ApJ656}
{Eriksen}, H.~K., {Huey}, G., {Saha}, R. {et~al.} 2007{\natexlab{b}}, \apj, 656, 641

\bibitem[{{Eriksen} {et~al.}(2008{\natexlab{b}}){Eriksen}, {Jewell},
  {Dickinson}, {Banday}, {G{\'o}rski}, \& {Lawrence}}]{Eriksen2008ApJ676}
{Eriksen}, H.~K., {Jewell}, J.~B., {Dickinson}, C. {et~al.} 2008{\natexlab{b}}, \apj, 676, 10

\bibitem[{{Eriksen} {et~al.}(2004){Eriksen}, {O'Dwyer}, {Jewell}, {Wandelt},
  {Larson}, {G{\'o}rski}, {Levin}, {Banday}, \& {Lilje}}]{Eriksen2004ApJS155}
{Eriksen}, H.~K., {O'Dwyer}, I.~J., {Jewell}, J.~B. {et~al.} 2004, \apjs, 155, 227

\bibitem[{{Fern{\'a}ndez-Cerezo} {et~al.}(2006){Fern{\'a}ndez-Cerezo},
  {Guti{\'e}rrez}, {Rebolo}, {Watson}, {Hoyland}, {Hildebrandt},
  {Rubi{\~n}o-Mart{\'{\i}}n}, {Mac{\'{\i}}as-P{\'e}rez}, \& {Sosa
  Molina}}]{Sergi2006}
{Fern{\'a}ndez-Cerezo}, S., {Guti{\'e}rrez}, C.~M., {Rebolo}, R. {et~al.} P. 2006, \mnras,
  370, 15

\bibitem[{{Finkbeiner}(2003)}]{Finkbeiner2003ApJS..146..407F}
{Finkbeiner}, D.~P. 2003, \apjs, 146, 407

\bibitem[{{Finkbeiner}(2004)}]{Haze_Finkbeiner2004ApJ...614..186F}
---. 2004, \apj, 614, 186

\bibitem[{{Finkbeiner} {et~al.}(1999){Finkbeiner}, {Davis}, \&
  {Schlegel}}]{Finkbeiner1999ApJ524}
{Finkbeiner}, D.~P., {Davis}, M., \& {Schlegel}, D.~J. 1999, \apj, 524, 867

\bibitem[{{Ghosh} {et~al.}(2012){Ghosh}, {Banday}, {Jaffe}, {Dickinson},
  {Davies}, {Davis}, \& {Gorski}}]{Ghosh2012MNRAS}
{Ghosh}, T., {Banday}, A.~J., {Jaffe}, T. {et~al.} 2012, \mnras, 422, 3617

\bibitem[{{Gold} {et~al.}(2011){Gold}, {Odegard}, {Weiland}, {Hill}, {Kogut},
  {Bennett}, {Hinshaw}, {Chen}, {Dunkley}, {Halpern}, {Jarosik}, {Komatsu},
  {Larson}, {Limon}, {Meyer}, {Nolta}, {Page}, {Smith}, {Spergel}, {Tucker},
  {Wollack}, \& {Wright}}]{gold11}
{Gold}, B., {Odegard}, N., {Weiland}, J.~L. {et~al.} 2011, \apjs, 192, 15

\bibitem[{{G{\'o}rski} {et~al.}(2005){G{\'o}rski}, {Hivon}, {Banday},
  {Wandelt}, {Hansen}, {Reinecke}, \& {Bartelmann}}]{Gorski2005Healpix}
{G{\'o}rski}, K.~M., {Hivon}, E., {Banday}, A.~J. {et~al.} 2005, \apj, 622, 759

\bibitem[{{Guo} \& {Mathews}(2011)}]{Guo2011}
{Guo}, F. \& {Mathews}, W.~G. 2011, arXiv:1103.0055

\bibitem[{{Guo} {et~al.}(2011){Guo}, {Mathews}, {Dobler}, \& {Oh}}]{guo11}
{Guo}, F., {Mathews}, W.~G., {Dobler}, G., \& {Oh}, S.~P. 2011, arXiv:1110.0834

\bibitem[{{Haslam} {et~al.}(1982){Haslam}, {Salter}, {Stoffel}, \&
  {Wilson}}]{Haslam1982A&AS47}
{Haslam}, C.~G.~T., {Salter}, C.~J., {Stoffel}, H., \& {Wilson}, W.~E. 1982,
  \aaps, 47, 1

\bibitem[{{Hildebrandt} {et~al.}(2007){Hildebrandt}, {Rebolo},
  {Rubi{\~n}o-Mart{\'{\i}}n}, {Watson}, {Guti{\'e}rrez}, {Hoyland}, \&
  {Battistelli}}]{Hildebrandt2007}
{Hildebrandt}, S.~R., {Rebolo}, R., {Rubi{\~n}o-Mart{\'{\i}}n}, J.~A. {et~al.} 2007, \mnras, 382, 594

\bibitem[{{Hoang} {et~al.}(2010){Hoang}, {Draine}, \&
  {Lazarian}}]{Hoang2010ApJ}
{Hoang}, T., {Draine}, B.~T., \& {Lazarian}, A. 2010, \apj, 715, 1462

\bibitem[{{Hoang} {et~al.}(2011){Hoang}, {Lazarian}, \&
  {Draine}}]{Hoang2011ApJ}
{Hoang}, T., {Lazarian}, A., \& {Draine}, B.~T. 2011, \apj, 741, 87

\bibitem[{{Hooper} {et~al.}(2007){Hooper}, {Finkbeiner}, \&
  {Dobler}}]{Haze_Hooper2007PhRvD..76h3012H}
{Hooper}, D., {Finkbeiner}, D.~P., \& {Dobler}, G. 2007, \prd, 76, 083012

\bibitem[{{Jarosik} {et~al.}(2010){Jarosik}, {Bennett}, {Dunkley}, {Gold},
  {Greason}, {Halpern}, {Hill}, {Hinshaw}, {Kogut}, {Komatsu}, {Larson},
  {Limon}, {Meyer}, {Nolta}, {Odegard}, {Page}, {Smith}, {Spergel}, {Tucker},
  {Weiland}, {Wollack}, \& {Wright}}]{Jarosik:2010}
{Jarosik}, N., {Bennett}, C.~L., {Dunkley}, J. {et~al.} 2011, \apjs, 192, 14

\bibitem[{{Jewell} {et~al.}(2004){Jewell}, {Levin}, \&
  {Anderson}}]{Jewell2004ApJ609}
{Jewell}, J., {Levin}, S., \& {Anderson}, C.~H. 2004, \apj, 609, 1

\bibitem[{{Jewell} {et~al.}(2009){Jewell}, {Eriksen}, {Wandelt}, {O'Dwyer},
  {Huey}, \& {G{\'o}rski}}]{Jewell2009ApJ697}
{Jewell}, J.~B., {Eriksen}, H.~K., {Wandelt}, B.~D. {et~al.} 2009, \apj, 697, 258

\bibitem[{{Kogut} {et~al.}(2007){Kogut}, {Dunkley}, {Bennett}, {Dor{\'e}},
  {Gold}, {Halpern}, {Hinshaw}, {Jarosik}, {Komatsu}, {Nolta}, {Odegard},
  {Page}, {Spergel}, {Tucker}, {Weiland}, {Wollack}, \& {Wright}}]{kogut07}
{Kogut}, A., {Dunkley}, J., {Bennett}, C.~L. {et~al.} 2007, \apj, 665, 355

\bibitem[{{Komatsu} {et~al.}(2010){Komatsu}, {Smith}, {Dunkley}, {Bennett},
  {Gold}, {Hinshaw}, {Jarosik}, {Larson}, {Nolta}, {Page}, {Spergel},
  {Halpern}, {Hill}, {Kogut}, {Limon}, {Meyer}, {Odegard}, {Tucker}, {Weiland},
  {Wollack}, \& {Wright}}]{Komatsu:2010wmap7}
{Komatsu}, E., {Smith}, K.~M., {Dunkley}, J. {et~al.} 2011, \apjs, 192, 18

\bibitem[{{Larson} {et~al.}(2011){Larson}, {Dunkley}, {Hinshaw}, {Komatsu},
  {Nolta}, {Bennett}, {Gold}, {Halpern}, {Hill}, {Jarosik}, {Kogut}, {Limon},
  {Meyer}, {Odegard}, {Page}, {Smith}, {Spergel}, {Tucker}, {Weiland},
  {Wollack}, \& {Wright}}]{Larson2010}
{Larson}, D., {Dunkley}, J., {Hinshaw}, G. {et~al.} 2011, \apjs, 192, 16

\bibitem[{{Larson} {et~al.}(2007){Larson}, {Eriksen}, {Wandelt}, {G{\'o}rski},
  {Huey}, {Jewell}, \& {O'Dwyer}}]{Larson2007ApJ656}
{Larson}, D.~L., {Eriksen}, H.~K., {Wandelt}, B.~D. {et~al.} 2007, \apj, 656, 653

\bibitem[{{Linden} \& {Profumo}(2010)}]{Linden2010ApJ}
{Linden}, T. \& {Profumo}, S. 2010, \apjl, 714, L228

\bibitem[{{Mertsch} \& {Sarkar}(2010)}]{Mertsch2010JCAP}
{Mertsch}, P. \& {Sarkar}, S. 2010, \jcap, 10, 19

\bibitem[{{Mertsch} \& {Sarkar}(2011)}]{Mertsch2011}
---. 2011, Physical Review Letters, 107, 091101

\bibitem[{{O'Dwyer} {et~al.}(2004){O'Dwyer}, {Eriksen}, {Wandelt}, {Jewell},
  {Larson}, {G{\'o}rski}, {Banday}, {Levin}, \& {Lilje}}]{ODwyer2004ApJ617L}
{O'Dwyer}, I.~J., {Eriksen}, H.~K., {Wandelt}, B.~D. {et~al.} 2004, \apjl, 617, L99

\bibitem[{{Park} {et~al.}(2007){Park}, {Park}, \&
  {Gott}}]{Park2007ApJ...660..959P}
{Park}, C.-G., {Park}, C., \& {Gott}, III, J.~R. 2007, \apj, 660, 959

\bibitem[{{Planck Collaboration} {et~al.}(2011){Planck Collaboration}, {Ade},
  {Aghanim}, {Arnaud}, {Ashdown}, {Aumont}, {Baccigalupi}, {Balbi}, {Banday},
  {Barreiro}, \& et~al.}]{planck11XX}
{Planck Collaboration} 2011, \aap, 536, A20

\bibitem[{{Rudjord} {et~al.}(2009){Rudjord}, {Groeneboom}, {Eriksen}, {Huey},
  {G{\'o}rski}, \& {Jewell}}]{Rudjord2009ApJ692}
{Rudjord}, {\O}., {Groeneboom}, N.~E., {Eriksen}, H.~K. {et~al.} 2009, \apj, 692, 1669

\bibitem[{{Schlegel} {et~al.}(1998){Schlegel}, {Finkbeiner}, \&
  {Davis}}]{schlegel98}
{Schlegel}, D.~J., {Finkbeiner}, D.~P., \& {Davis}, M. 1998, \apj, 500, 525

\bibitem[{{Su} {et~al.}(2010){Su}, {Slatyer}, \& {Finkbeiner}}]{Su2010}
{Su}, M., {Slatyer}, T.~R., \& {Finkbeiner}, D.~P. 2010, \apj, 724, 1044

\bibitem[{{Wandelt} {et~al.}(2004){Wandelt}, {Larson}, \&
  {Lakshminarayanan}}]{Wandelt2004PhRvD70}
{Wandelt}, B.~D., {Larson}, D.~L., \& {Lakshminarayanan}, A. 2004, \prd, 70,
  083511

\end{thebibliography}

\appendix

\section{The CMB Map and Power Spectrum Posterior}
\label{app:commander}
Here we review the basic concept behind Gibbs sampling. Let us first focus on the case of one frequency map and no foregrounds.  The data model for this case is
\begin{equation}
\Bd = \Bs + \Bn,
\end{equation}
where $\Bd$ is the data, $\Bs$ the CMB sky signal, and
$\Bn$ instrumental noise. We assume both the CMB signal and noise to be Gaussian random fields
with vanishing mean and covariance matrices $\BS$ and $\BN$, respectively. The CMB sky can be written in spherical harmonics as $\Bs = \sum_{\ell, m} a_{\ell m} Y_{\ell m}$, with the CMB covariance matrix then fully characterized by the angular power spectrum $C_{\ell}$ according to $\textrm{C}_{\ell m, \ell' m'} = \langle a_{\ell m}^* a_{\ell^\prime m^\prime}\rangle = C_{\ell} \delta_{\ell \ell^\prime} \delta_{m m'}$. The noise matrix $\BN$ is left unspecified for now, but we note that for white noise it is diagonal in pixel space, $N_{ij} = \sigma_{i}^2 \delta_{ij}$, for pixels $i$ and $j$ and noise variance $\sigma^2_{i}$.

Our goal is to sample from the posterior density for both the sky signal $\Bs$ and the power
spectrum $C_{\ell}$, given by
\begin{align}
P(\Bs, C_{\ell}|\Bd) &\propto P(\Bd|\Bs, C_{\ell}) P(\Bs, C_{\ell}) \\
&\propto P(\Bd|\Bs, C_{\ell}) P(\Bs|C_{\ell}) P(C_{\ell}),
\end{align}
In what follows we assume the prior $P(C_{\ell})$ is uniform.
Since we have assumed Gaussianity, the joint posterior distribution may be written as
\begin{equation}
P(\Bs, C_{\ell}|\Bd) \propto 
e^{-\frac{1}{2} (\Bd-\Bs)^t \BN^{-1} (\Bd-\Bs)}
\prod_{\ell}\frac{e^{-\frac{2\ell+1}{2}
    \frac{\sigma_{\ell}}{C_{\ell}}}}{C_{\ell}^{\frac{2\ell+1}{2}}}
P(C_{\ell}),
\label{eq:cmb_posterior}
\end{equation}
where we have defined the quantity $\sigma_{\ell} \equiv \frac{1}{2\ell+1} \sum_{m=-\ell}^{\ell}
|a_{\ell m}|^2$ as the angular power spectrum of the full-sky CMB signal.

For the case here with the CMB signal assumed to be a Gaussian field, one can integrate over the
CMB sky signal and analytically solve for the marginalized posterior $P(C_{\ell}|\Bd)$.  However,
evaluating the posterior numerically for any specific angular power spectrum is computationally prohibitive
as it involves the computation of the inverse and determinant of very large matrices.  We therefore sample from the
posterior using a Gibbs sampling algorithm.

\subsection{Gibbs Sampling}
One procedure to sample from the joint density $P(\Bs,C_{\ell}|\Bd)$, as proposed by \citet{Jewell2004ApJ609} and
\citet{Wandelt2004PhRvD70}, is to alternately sample from the respective conditional densities
\begin{align}
\Bs^{i+1} &\leftarrow P(\Bs | C_{\ell}^i, \Bd) \\
C_{\ell}^{i+1} &\leftarrow P(C_{\ell} | \Bs^{i+1}, \Bd).
\end{align}
Here $\leftarrow$ indicates sampling from the distribution on the
right-hand side. After some ``burn-in" period, the joint samples $(\Bs^i, C_{\ell}^i)$ will be
distributed from the joint posterior. Thus, the problem is reduced to that
of sampling from the two \emph{conditional} densities $P(\Bs |
C_{\ell}, \Bd)$ and $P(C_{\ell} | \Bs, \Bd)$.

The conditional density $P(C_{\ell} | \Bs, \Bd)$ in this case is independent of the data,
$P(C_{\ell} | \Bs, \Bd) = P(C_{\ell} | \Bs)$, simply because the underlying CMB sky signal provides
all the information needed to estimate the ensemble angular power spectrum $C_{\ell}$.
Under the assumption of Gaussianity and isotropy, this conditional is given by the inverse Gamma distribution, $S_\ell$:
\begin{equation}
P(C_{\ell} | \Bs) \propto \frac{e^{-\frac{1}{2}
    \Bs_{\ell}^{t}\BS_{\ell}^{-1}\Bs_{\ell}}}{\sqrt{|\BS_{\ell}|}} =    \frac{e^{-\frac{2\ell+1}{2} \frac{\sigma_{\ell}}{C_{\ell}}}}{C_{\ell}^{\frac{2\ell+1}{2}}}.
\end{equation}
In order to sample from this conditional density we first draw $2\ell-1$ normal random variates $\rho_\ell^k$,
compute the sum $\rho_\ell^2=\sum_{k=1}^{2\ell-1}|\rho_\ell^k|^2$, and finally set
\be
\mathcal{C}_\ell=\frac{\sigma_\ell}{\rho_\ell^2},
\ee
giving a sample distributed according to the inverse Gamma distribution.

The conditional density for the sky map given the angular power spectrum and data follows directly from
the form of the joint Bayes posterior in Equation \ref{eq:cmb_posterior}, and given by
\begin{align}
P(\Bs | C_{\ell}, \Bd) &\propto P(\Bd|\Bs, C_{\ell}) P(\Bs|C_{\ell}) \\
&\propto e^{-\frac{1}{2} (\Bd-\Bs)^t \BN^{-1} (\Bd-\Bs)} 
\,\,e^{-\frac{1}{2}\Bs^t\BS^{-1}\Bs} \\
&\propto e^{-\frac{1}{2} (\Bs-\hat{\Bs})^t (\BS^{-1} + \BN^{-1}) (\Bs-\hat{\Bs})},
\end{align}
where we have defined the so-called
mean-field map (or Wiener filtered data) $\hat{\Bs} = (\BS^{-1}
+ \BN^{-1})^{-1} \BN^{-1} \Bd$. Thus, $P(\Bs | C_{\ell}, \Bd)$ is a Gaussian distribution with mean equal to $\hat{\Bs}$ and a covariance matrix equal to $(\BS^{-1} +
\BN^{-1})^{-1}$.  In order to sample from this conditional, we first
generate two independent white noise maps $\omega_{0}$ and $\omega_{1}$, and solve
\begin{equation}
\left[\BS^{-1} + \BN^{-1}\right] \Bs = \BN^{-1}\Bd + \BS^{-\frac{1}{2}} \omega_0 + \BN^{-\frac{1}{2}} \omega_1,
\label{eq:lin_sys}
\end{equation}
The resulting map $\Bs$ is exactly a sample from the conditional
$P(\Bs | C_{\ell}, \Bd)$.  The addition of the white noise maps simply  reflects our uncertainty in the true but unobserved CMB sky - there are many CMB maps
that are consistent with the data and power spectrum estimate, and we are simply making a random
choice from this set of maps.

The discussion so far was limited to a single band and no modeling of instrumental response. The generalization is straightforward and can be found in \cite{Eriksen2007ApJ656}, together with a discussion on the actual numerical implementation which reduces the round-off errors. Here we quote the main result, which generalizes Eq.~\ref{eq:lin_sys} to:
\begin{equation}
\begin{split}
\left[\BS^{-1} + \sum_{\nu} \BA_{\nu}^t \BN_{\nu}^{-1}\BA_{\nu}\right]
\Bs = & \\ \sum_{\nu} \BA_{\nu}^t \BN_{\nu}^{-1}\Bd_{\nu}
+ \BS^{-\frac{1}{2}} &\omega_0 + \sum_{\nu} \BA_{\nu}^t
\BN_{\nu}^{-\frac{1}{2}} \omega_{\nu},
\end{split}
\label{eq:lin_sys2}
\end{equation}
where $\BA_\nu$ describes the beam response of the detector $\nu$ and $\BN$ its noise properties. Note that we now draw one white noise map for each frequency band,
$\omega_{\nu}$. The sampling procedure for $P(C_{\ell}|\Bs)$ is
unchanged.

\subsection{The foreground sampler}
\label{sec:fg_sampling}

The Gibbs sampling algorithm can be easily extended to include foregrounds
described by a parametric model, $F(\theta)$. Including the foreground model parameters into the joint posterior, the sampling procedure generalizes to:
\begin{align}
\Bs^{i+1} &\leftarrow P(\Bs | C_{\ell}^i, \theta^i, \Bd) \nonumber \\
C_{\ell}^{i+1} &\leftarrow P(C_{\ell} | \Bs^{i+1}, \Bd) \nonumber \\
\theta^{i+1} &\leftarrow P(\theta | C_\ell^{i+1}, \Bs^{i+1}, \Bd). \nonumber
\end{align}
We note that for the foreground models of interest, we do not have an algorithm to produce an exact sample
from the conditional density $P(\theta | C_\ell^{i+1}, \Bs^{i+1}, \Bd)$ but we instead generalize Gibbs sampling to
an MCMC algorithm, the details of which can be found in \citet{Eriksen2008ApJ676}).

The parametric family of data models including foregrounds implemented in \commander\ are of the form
\be
\begin{split}
\Bd_\nu & = \BA_\nu \Bs + \sum_{m=1}^M a_{\nu,m} \Bt_m  +\\
& + \sum_{n=1}^{N} b_{n} f_n(\nu) \mathbf{f}_n + \sum_{k=1}^{K} \mathbf{c}_{k} \; \mathbf{g}_{k}(\nu; \theta_k)+ \Bn_{\nu}.
\end{split}
\ee
We may identify three main classes of foregrounds:
\bit
\item[i)] $\Bt_m$ are M templates multiplied by an amplitude at every frequency, $a_{\nu,m}$;
\item[ii)] $\Bf_n$ are N templates whose spectral behaviour is known and described by the function $f_n(\nu)$; we allow for an overall rescaling, $b_n$, for each of them;
\item[iii)] K foregrounds are described by a map of coefficients, $\Bc_k$, multiplied by the spectral response which is function of the frequency and the parameters of the foreground model, $\mbf\theta_k$. 
\eit
We remind the reader that the bold face notation means an array of size $N_{\rm pix}$. An example of the first class of foregrounds is given by monopole and dipole residuals; a special case of the second class is a free-free template (Figure~\ref{fig:templates}, top panel) whose spectral behaviour is known and follows the relation $(\nu/\nu_0)^{-2.15}$; for the third type we may quote synchrotron emission described by an amplitude we solve for at the reference frequency, e.~g.~ $\mu_0=23$~GHz, and a spectral response given by $(\nu/\mu_0)^{\mbf\beta}$.

It is instructive to compute the degrees of freedom for such a foreground model applied to WMAP data. We solve three maps, $\Bs$, $\mbf\beta$ and $\BA_{\rm synch}$, 5 monopoles and dipoles, and two overall amplitudes if we describe free-free emission and thermal dust by means of the second class of foreground models. In total we look for $3 \, N_{\rm pix} + 22$ parameters. This is already pretty close to the maximum number of parameters allowed by the WMAP, $\leq 5 \, N_{\rm pix}$. We could ask for one more map, either dust or free-free, assuming a single power law and fixing the spectral index, or allowing a curvature term in the synchrotron model. Since a simple power law for thermal dust and free-free emission has been shown to be consistent with the data in the frequency range spanned by WMAP \citep{Dickinson2009ApJ705}, this may be used to disentangle the spinning dust contribution from thermal dust.

We notice that to solve for the spectral index of a given component, the instrumental beam response of each channel must be taken into account. Up to now, \commander\ is able to work with maps at the same angular resolution only. Smoothing all frequency maps and ancillary data to a common angular scale is then a necessary pre-processing step.

As discussed in \cite{Eriksen2008ApJ676}, it turns out to be more efficient  to sample all the map amplitudes at once, followed by the spectral response parameter, and finally the angular power spectrum, following the iterative scheme:
\begin{align}
&\{\Bs^{i+1}, a_m, b_n, \Bc_k \}^{i+1} \leftarrow P(\Bs, a_m, b_n, \Bc_k | C_{\ell}^i, \theta^i, \Bd) \nonumber\\
&\theta^{i+1} \leftarrow P(\theta | C_\ell^i, \Bs^{i+1}, a_m^{i+1}, b_n^{i+1}, \Bc_k^{i+1}, \Bd) \nonumber\\
&C_{\ell}^{i+1} \leftarrow P(C_{\ell} | \Bs^{i+1}, \Bd).
\end{align}

Sampling from the conditional density $P(\Bs, a_m, b_n, \Bc_k | C_{\ell}^i, \theta^i, \Bd)$ is a generalization of Eq.~\ref{eq:lin_sys2}, whereas the sampling of the angular power spectrum remains unchanged, since $\mathcal{C}_\ell$ are functions of the sky signal only. Sampling of the non-linear degrees of freedom is through a standard inversion sampler: first compute the conditional probability density $P(x|\theta)$, where $x$ is the currently sampled  parameter and $\theta$ denotes the set of all other parameters in the model, assuming the likelihood to be independent pixel by pixel: $-2{\rm ln}\mathcal{L}(x)=\chi^2=\sum_\nu(d_\nu-s_\nu(X,\theta))^2/\sigma_\nu^2$. Then, the corresponding cumulative distribution is computed, $F(x|\theta)=\int_{-\infty}^xP(y|\theta)dy$, and the value of the $x$ variable is chosen by drawing a uniformly distributed random number, $u$, and reading $F(x|\theta)=u$.

This concludes the review on the implementation of Gibbs sampling as implemented in the computer code \commander.

\section{Noise impact on the regression procedure}
\label{app:regression}

We have observed that the regression coefficients we found when fitting foreground templates to \commander\ posterior mean amplitudes are consistent with those discussed in other works, but they have much larger errors. We argued in the text that this is the result of the sampling procedure and has two causes: i) our input maps are noisier because of the additional noise term and ii) our uncertainties on the foreground amplitudes take into account the error on the other parameters of the model: CMB, other foreground amplitudes and spectral indices. In this respect, our errors are more conservative.

\begin{figure}[!h]
\begin{center}
\incgr[width=.2\textwidth, angle=90]{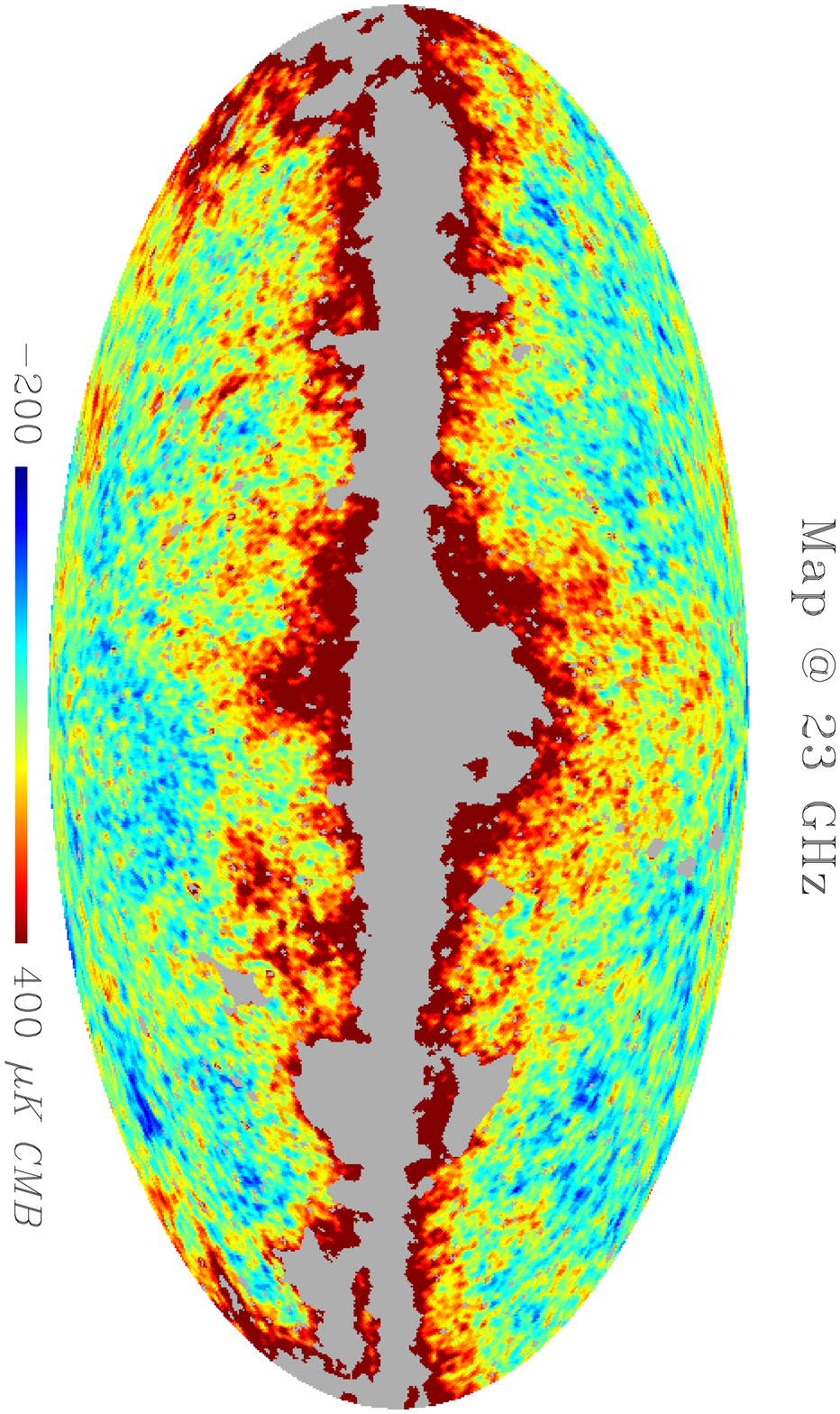}
\incgr[width=.2\textwidth, angle=90]{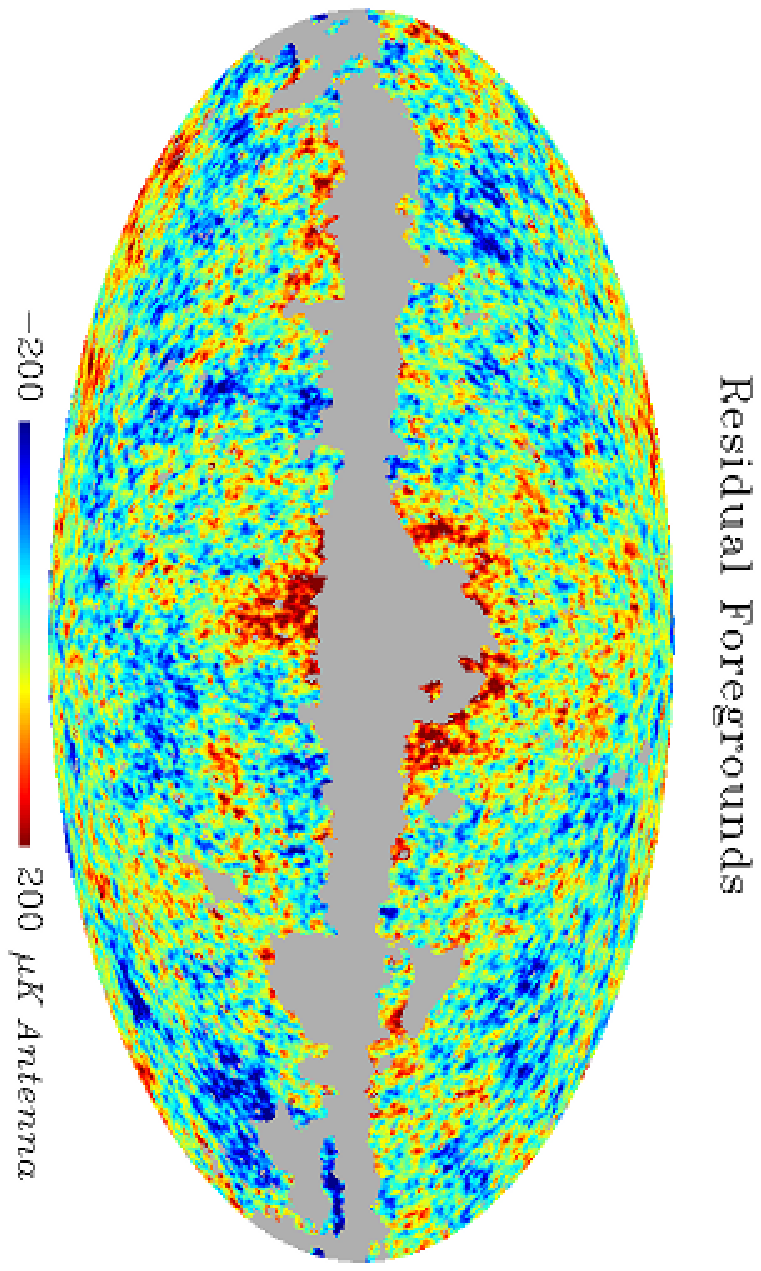}
\incgr[width=.2\textwidth, angle=90]{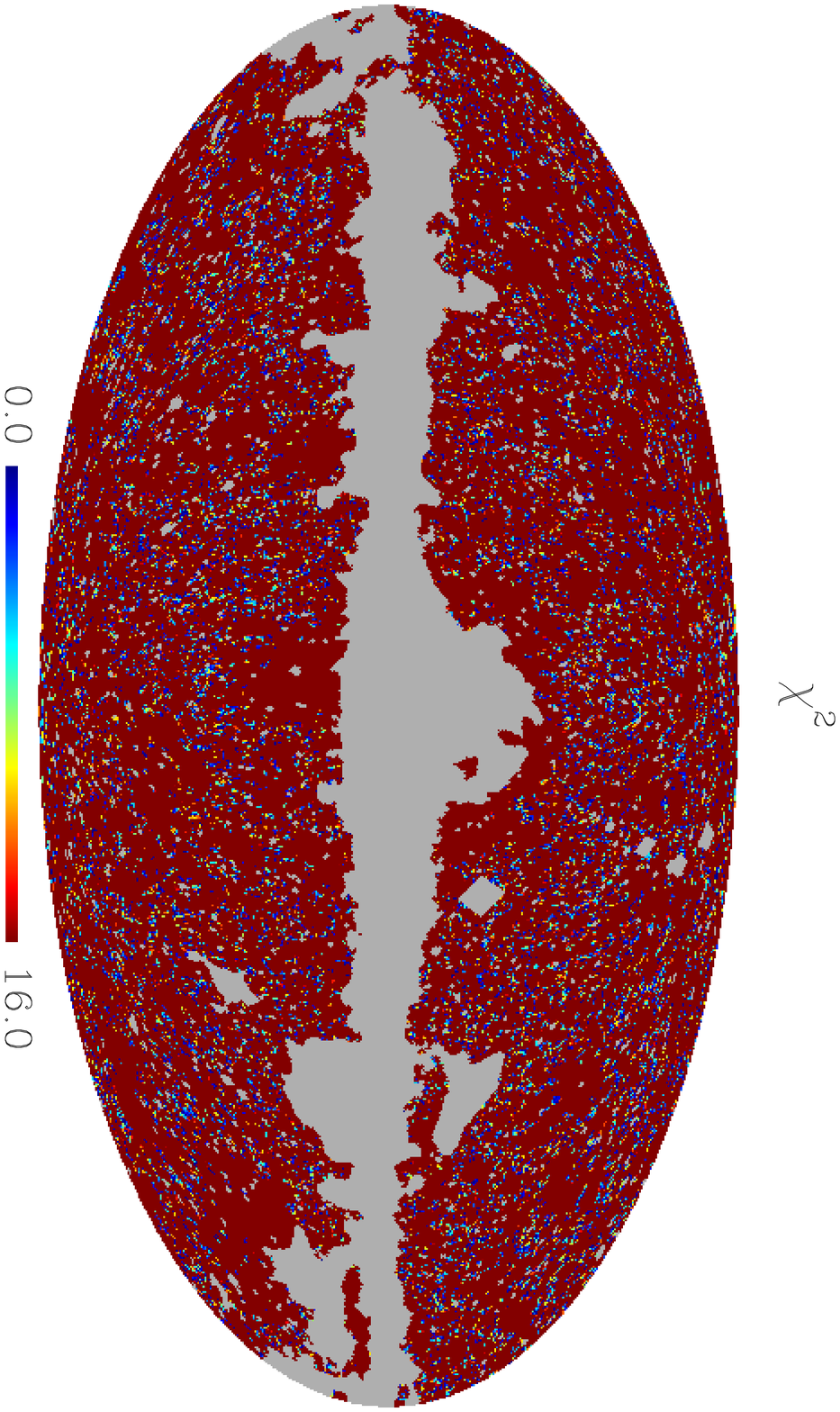}
\incgr[width=.2\textwidth, angle=90]{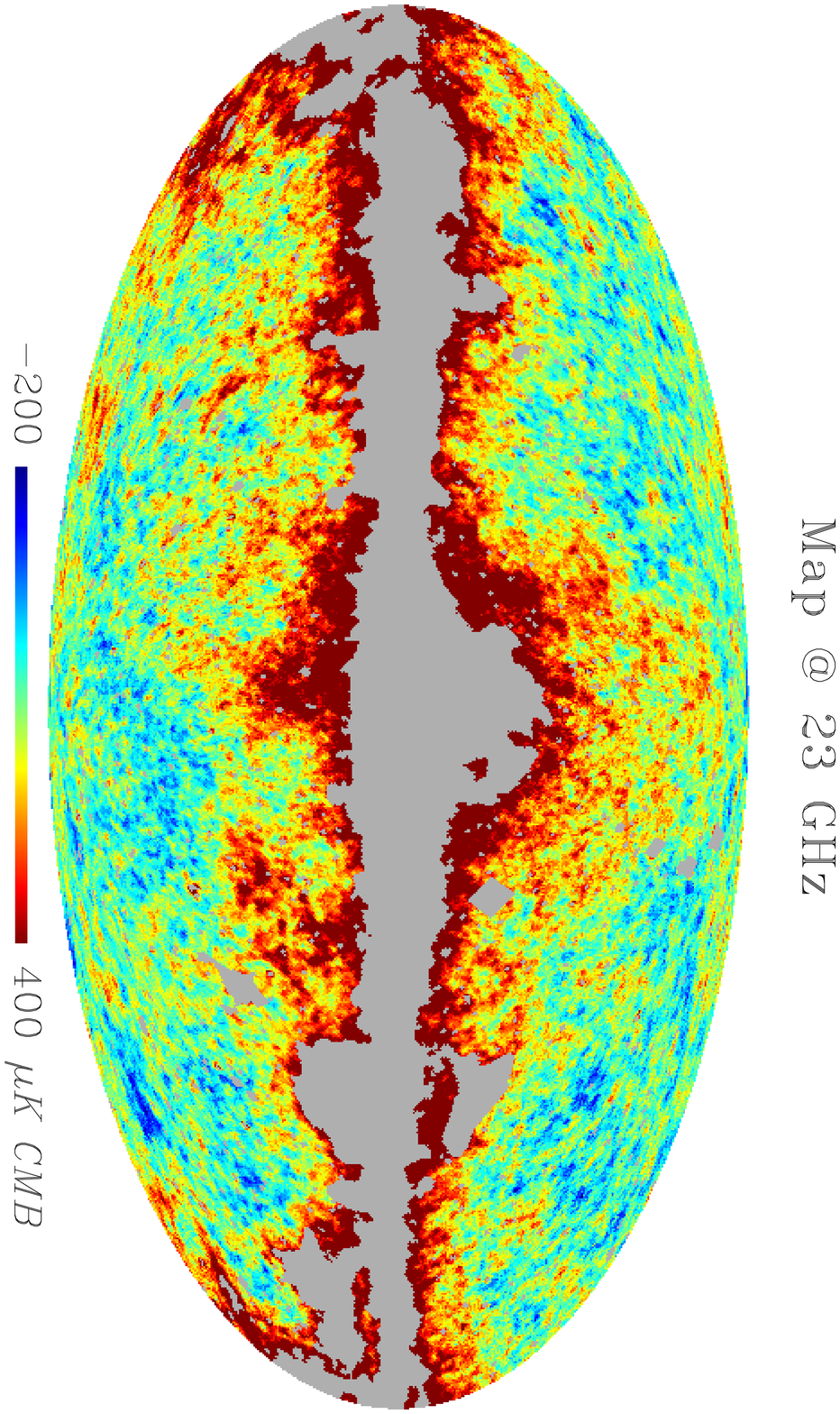}
\incgr[width=.2\textwidth, angle=90]{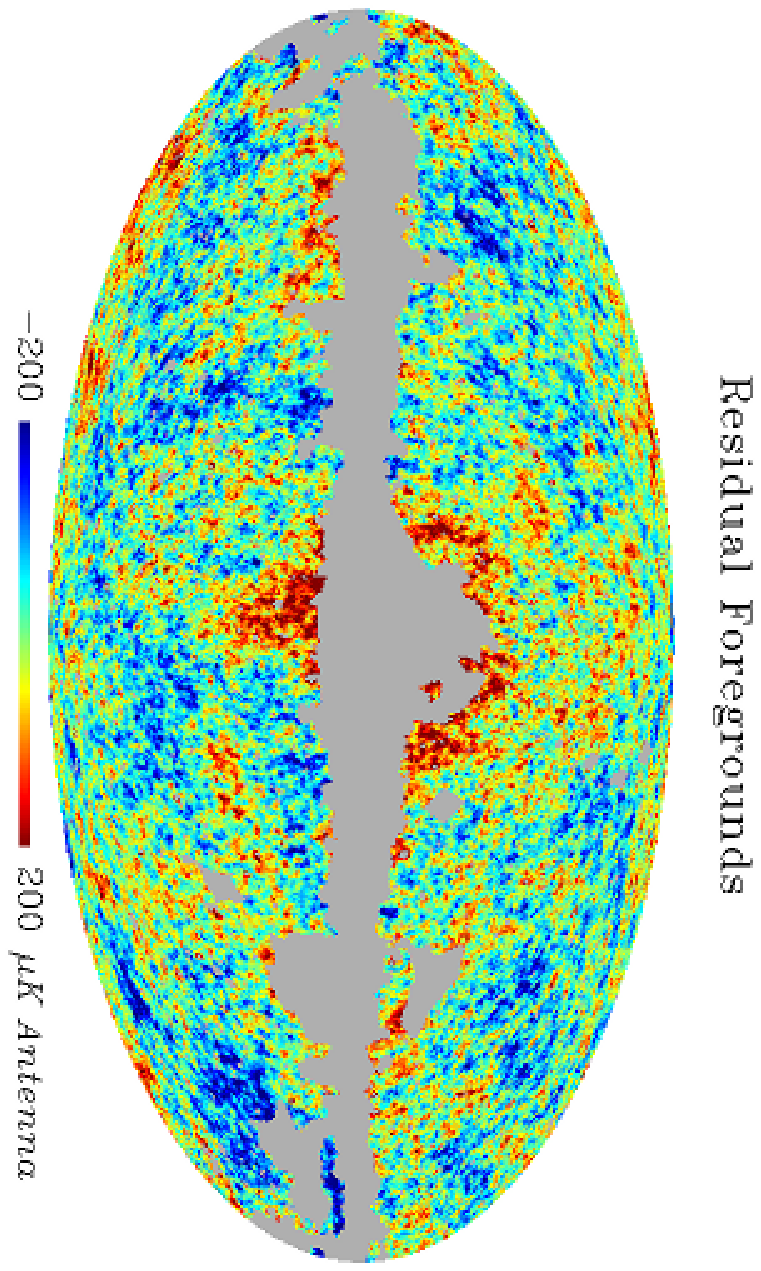}
\incgr[width=.2\textwidth, angle=90]{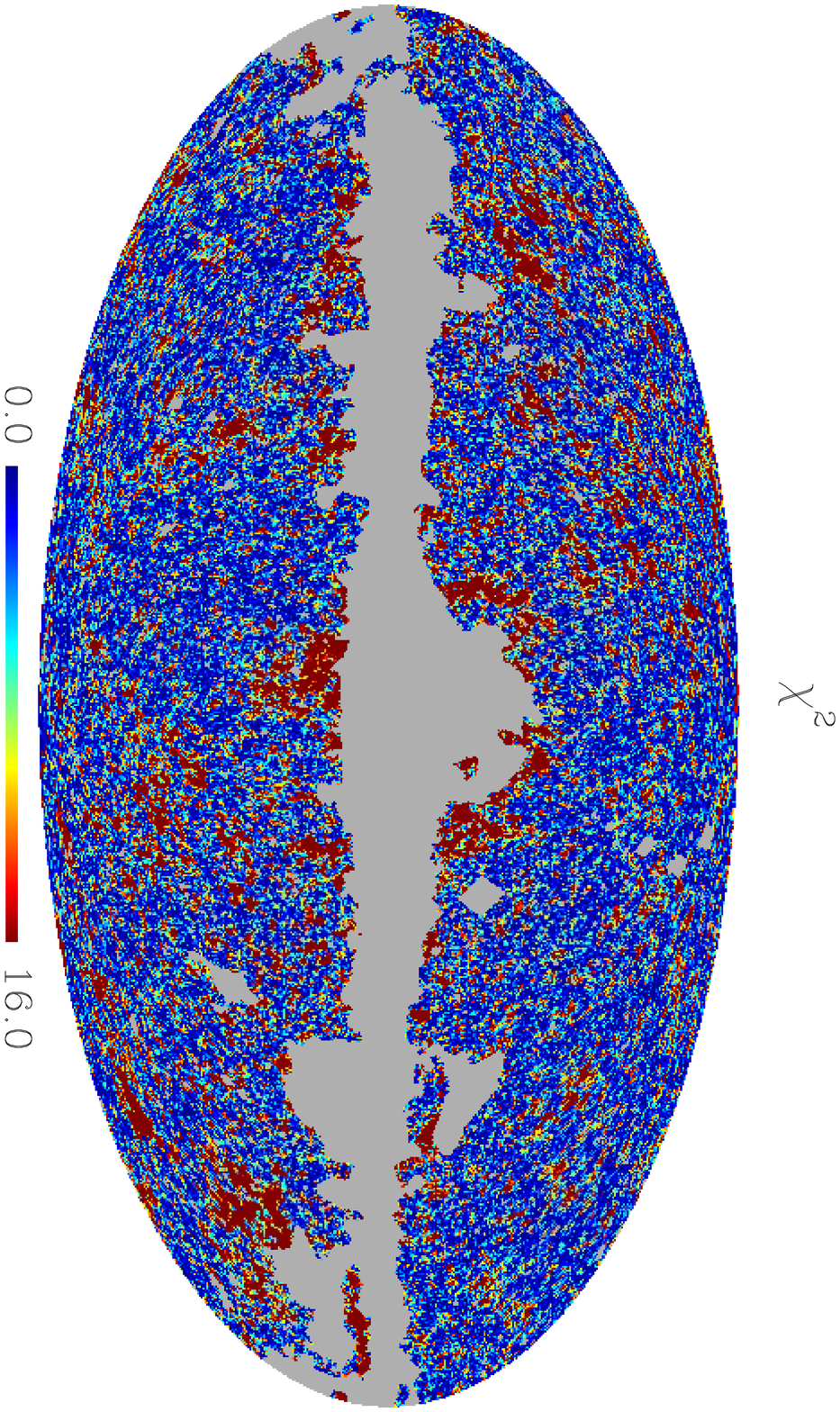}
\incgr[width=.2\textwidth, angle=90]{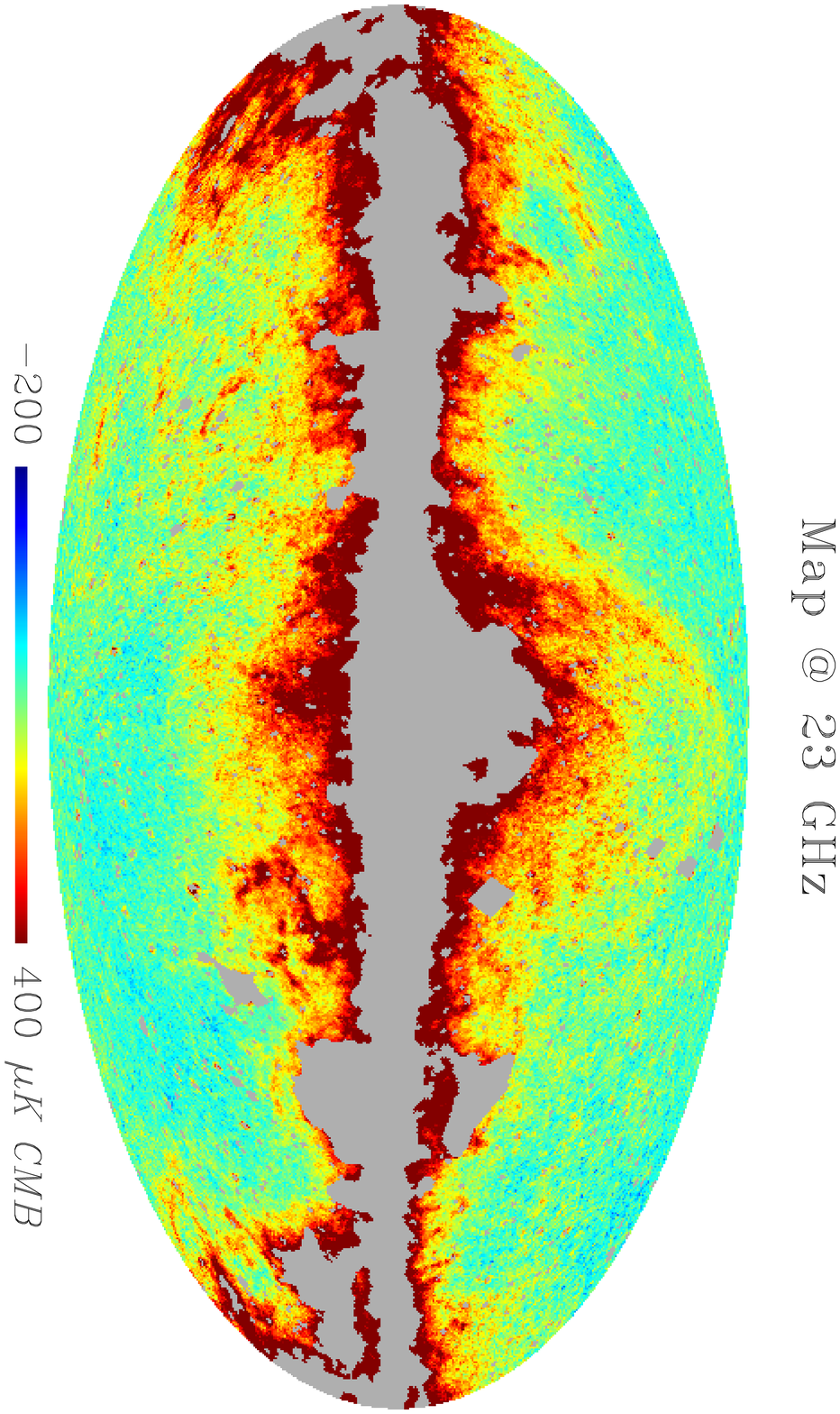}
\incgr[width=.2\textwidth, angle=90]{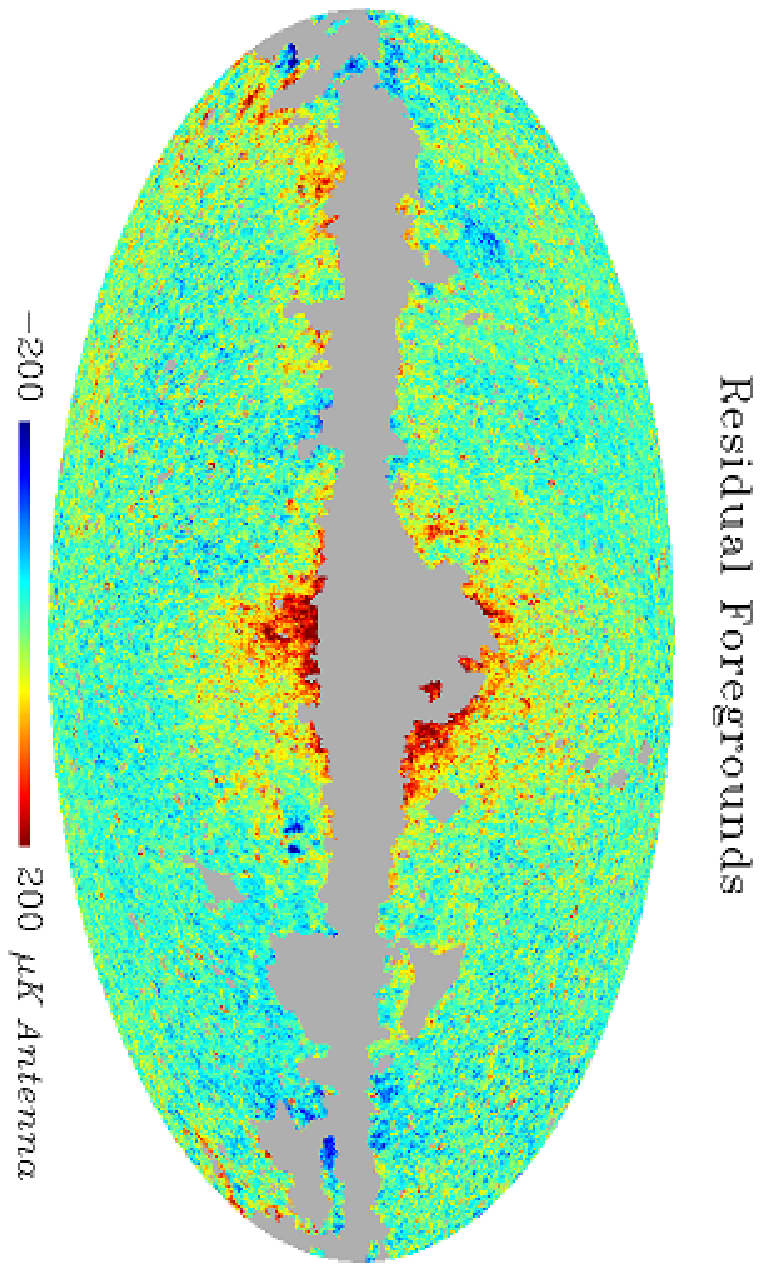}
\incgr[width=.2\textwidth, angle=90]{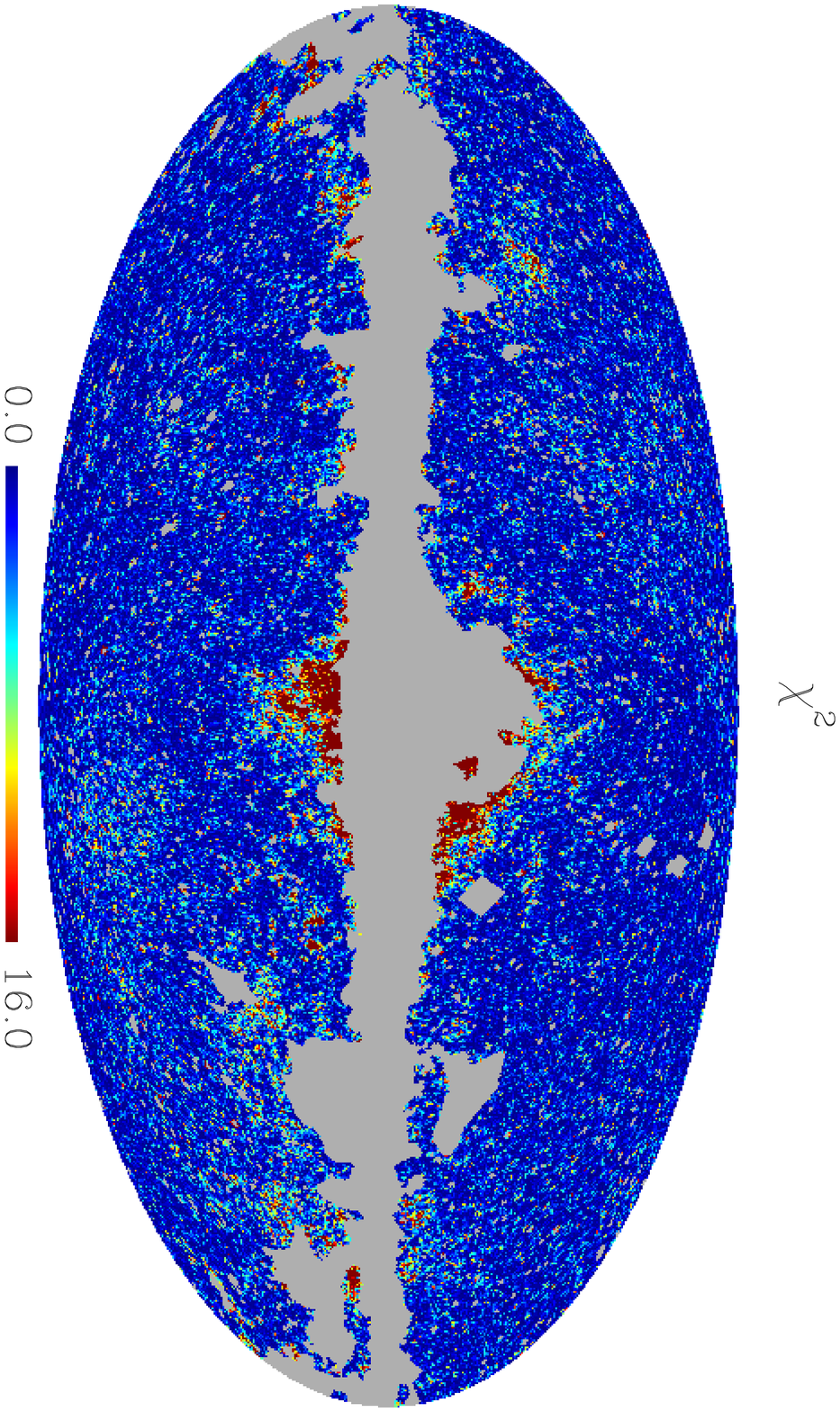}
\caption{Example of three regressions performed on the WMAP K-band smoothed to 60 arcminute (top), \commander\ input K-band (middle row) and output foreground amplitude (bottom).}
\label{fig:reg_maps}
\end{center}
\end{figure}

To clearly show this, we perform the same regression described in Equations~\ref{eq:Amp} and \ref{eq:regression_coef} on two different maps: 1) smoothed WMAP K-band and 2) \commander\ input K-band, which have to be compared to \commander\ output amplitude performance. The three maps are shown in Figure~\ref{fig:reg_maps}, together with the fit residuals and the corresponding $\chi^2$.

Table~\ref{tab:chk_regression} compares the regression coefficients for the signal at 23 \GHz. Consistently, the amplitudes are the same but the error bars increase dramatically due to the noise added to the maps. In particular we move from 10-30$\sigma$ detection to 1-3$\sigma$, which is driven by the scaling factor applied to the K-band (16, see Section~\ref{subsec:wmap}).

This comparison suggests that a lower level of noise added to the input maps is useful to better characterize diffuse foreground emission. We will further investigate this issue in a forthcoming work.
% ---------------------------------------------------------------------------------------------------
\begin{table}[!h]
\begin{center}
\caption{Regression coefficients of the \commander\ foreground amplitude maps compared to those obtained when regressing smoothed WMAP channel and \commander\ input WMAP maps\label{tab:chk_regression}}
\begin{tabular}{ccccc}
	\tableline\tableline 
   \multicolumn{5}{c}{\bf WMAP 7-yr}\\ 
   	\tableline\tableline
  {\bf Dataset} & {\bf Haslam} & {\bf H$\alpha$} & {\bf FDS} & r\\ 
  	\tableline
   Smoothed K-band & $(3.7\pm0.17)\times10^{-6}$ & $6.4\pm.6$ & $7.0\pm.25$ & 0.91\\ 
   \commander\ Input K-band & $(3.7 \pm 1.1)\times 10^{-6}$ & $6 \pm 5$ & $7\pm2$ & 0.85\\ 
   \commander\ output @ 23GHz & $(3.6 \pm 1.2)\times 10^{-6}$ & $6 \pm 6$ & $7\pm2$ & 0.94\\ 
	\tableline
\end{tabular}
\end{center}
\end{table}
% ---------------------------------------------------------------------------------------------------

% ---------------------------------------------------------------------------------------------------
% ---------------------------------------------------------------------------------------------------

\end{document}